\documentclass[journal]{IEEEtran}
\usepackage{blindtext, graphicx,lipsum}
\usepackage[dvipsnames]{xcolor}
\usepackage{cuted,float}
\newlength\figureheight 
\newlength\figurewidth 
\usepackage{cite}
\usepackage{graphicx}
\usepackage{amsthm,amsmath,amssymb,mathtools,bm,etoolbox,float}
\usepackage{tkz-euclide}
\usepackage{multicol}
\usepackage{tikz}
\newlength\fheight
\newlength\fwidth
\usepackage{tkz-euclide}
\usepackage[utf8]{inputenc}
\usepackage{pgfplots} 
\usepackage{pgfgantt}
\usepackage{pdflscape}
\pgfplotsset{compat=newest} 
\pgfplotsset{plot coordinates/math parser=false}
\pgfplotsset{every  tick/.style={black,},ylabel style={font=\tiny},xlabel style={font=\tiny},tick label style={font=\tiny},legend style= {font=\scriptsize},
minor x tick num=1,minor y tick num=1,xminorticks=true,yminorticks=true,}

\def\endthebibliography{%
  \def\@noitemerr{\@latex@warning{Empty `thebibliography' environment}}%
  \endlist
}
\usepackage{amsthm}
\usetikzlibrary{arrows,plotmarks}
\usepackage{tikz-3dplot}
\usepackage{balance}
\newtheorem{remark}{Remark}
\newtheorem{theorem}{Theorem}
\newtheorem{corollary}{Corollary}
\usepackage{algorithm,algorithmic}
\usepackage{caption,subcaption}
\begin{document}

\title{On The Joint Effects of HPA Nonlinearities and IQ Imbalance On Mixed RF/FSO Cooperative Systems}

\author{Elyes~Balti,~\IEEEmembership{Student~Member,~IEEE,}
        and~Brian~K.~Johnson,~\IEEEmembership{Senior~Member,~IEEE}}%

\maketitle

\begin{abstract}
In this work, we provide a framework analysis of dual-hop hybrid Millimeter Wave Radio Frequency (RF)/Free Space Optical (FSO) MIMO relaying system. The source is equipped with multiple antennas and employs conjugate beamforming while the destination consists of multiple apertures with selection combining. The system also consists of a relay operating at amplify-and-forward mode. The RF channels are subject to Nakagami-m fading while the optical links experience the M\'alaga distribution. In addition, we introduce the impairments to the relay and receiver. In fact, the relay is impaired by the High Power Amplifier (HPA) nonlinearities while the receiver suffers from the In phase and Quadrature Imbalance. Moreover, we assume two types of HPA nonlinearities impairments called Soft Envelope Limiter (SEL) and Traveling Wave Tube Amplifier (TWTA). Closed-forms of the outage probability, the bit error probability, and the ergodic capacity are derived. Capitalizing on these performances, we derive the high SNR asymptotes to unpack insightful metrics such as the diversity gain. We also address the impacts of some key factors on the system performance such as the impairments, the interferers, the number of antennas and apertures and the pointing errors, etc. Finally, the analytical expressions are confirmed by Monte Carlo simulation.
\end{abstract}

\begin{IEEEkeywords}
Millimeter wave, MIMO, Soft Envelope Limiter, Traveling Wave Tube Amplifier, IQ Imbalance, Interference.
\end{IEEEkeywords}

\IEEEpeerreviewmaketitle

\section{Introduction}
Millimeter Waves (MmWaves) technology has been emerged as a promising solution to address many issues of cellular networks, specially, the bandwidth shortage. MmWaves refer to the spectrum band from 28 to 300 GHz\footnote{Although a rigorous definition of mmWaves frequencies would place them between 30 and 300 GHz, industry has loosely defined them to include the spectrum from 10 to 300 GHz.} and it offers huge amount of bandwidth to increase the capacity of the cellular networks \cite{surv}. MmWaves can also be served as a practical solution of the shared users systems consisting of licensed and secondary users. Due to the limited spectrum in Microwave band, only limited spectrum holes are left for secondary users to get access to the network and communicate in an efficient way. In this context, mmWaves effectively solves this shortcoming and get both licensed and secondary users to communicate reliably. MmWaves can be also introduced in hybrid Radio Frequency (RF) and Free Space Optical (FSO) fifth-generation (5G) systems. In addition, mmWaves applications become practical as relative commercial products have been developed such as IEEE 802.11 ad, 5G New Radio (NR) mmWaves prototype, and Wireless Gigabit Alliance (WiGig) \cite{rheath,5gnr,product}. 

Wireless optical communications also known as FSO is considered as the key stone for the next generation of wireless communication since it has recently gained enormous attention for the vast majority of the most well-known networking applications such as fiber backup, disaster recoveries and redundant links \cite{1}. The main advantages of employing the FSO is to reduce the power consumption and provide higher bandwidth. Moreover, FSO becomes as an alternative or a complementary to the RF communication as it overcomes the problems of the spectrum scarcity and its license access to free frequency band. In this context, many previous attempts have leveraged some of these advantages by introducing the FSO into classical systems to be called Mixed RF/FSO systems \cite{c1,c2,c3,j1,j2,j3,j4,asym,eT}. This new system architecture reduces not only the interferences level but also it offers full duplex Gigabit Ethernet throughput and high network security \cite{4}. 

\subsection{Literature}
Because of the higher frequencies, MmWaves are only justified and applied for short range communications. In fact, the main dilemma for MmWaves are the severe pathloss and the distance between the transmitter (TX) and the receiver (RX) that both affect the link reliability \cite{pathloss1,pathloss2}. A common technique to overtake this shortcoming is to employ multiple antennas at both sides to create an array gain to compensate for the pathloss severity. For cellular system, the user equipment (UE) suffers from intra-cell and inter-cell interference of the other users \cite{cell,cell1,cell2}. By employing sectorized array or high directional antenna array, created by massive number of antennas, the interference can be reduced \cite{antenna}. Due to the high bandwidth, most of MmWaves systems are noise-limited resulting in low signal-to-noise-ratio (SNR) range. For this reason, it is straightforward that MmWaves systems cannot support high order of modulations since the SNR cannot increase to relative higher values \cite{modulation1,modulation2,modulation3}. Furthermore, MmWaves are mainly employed to densify the network cells mainly in microcells where high achievable rate is the main requirement for such cell size. However, for macrocells, it is recommended to employ Sub-6 GHz where the power signal power is mainly required for long-distant communications and hence MmWaves are not a good candidate for such cell size \cite{macrocell,microcell,v2xdiversity}.

The optical signal can be detected following different schemes and the most widely used are the heterodyne and intensity-modulation and direct detection (IM/DD) \cite{muratbook}. Although previous work have shown that the heterodyne configuration outperforms IM/DD, it is still hard to be implemented in the system. As a result, recent work have focused on IM/DD with on-off keying (OOK) due to its cost effective and easy implementation, however, this scheme requires an adaptive threshold for the demodulation \cite{muratbook}. To address this shortcoming, the subcarrier intensity modulation (SIM) has been suggested as an alternative to IM/DD with OOK since this technique states that the RF signal is premodulated before the laser modulation \cite{djordjevic,zedini}.

It is true that the FSO contributes in densifying the number of users, the cellular networks still suffer from low signal coverage in some areas mainly located in forests and mountains where the optical signal cannot travel to such long distances and it is also heavily absorbed by the intermediate objects due to its high frequency. Although the literature has shown the superiority of the mixed RF/FSO systems over the classical RF systems, they still suffer from the reliability scarcity and power efficient coverage. 
To overtake this difficulty, previous research attempts have proposed cooperative relaying techniques hybridized with the mixed RF/FSO systems since it improves not only the capacity of the wireless system but also it offers high Quality of Service (QoS). Recently, this new efficient system model has attracted considerable attention in particular using various relaying schemes. The most common used relaying techniques are Decode-and-Forward \cite{j2}, Amplify-and-Forward \cite{c1,c2,c3}, and Quantize-and-Encode \cite{10}. Regarding the system with multiple relays, activating all relays to simultaneously forward the communication is not recommended because the problem of synchronization at the reception always occurs with optical communications. To solve this problem, only one relay is allowed to transmit the signal. In this case, a relay selection protocol is required to select this candidate relay. In the literature, there are many protocols previously proposed such as opportunistic relay selection \cite{j1}, distributed switch and stay, max-select protocol \cite{12}. Unlike these protocols which require the knowledge of the total CSIs of the channels, Krikidis \textit{et al.} have proposed partial relay selection (PRS) in \cite{prs} which requires the CSI of only one channel (source-relay or relay-destination). Unlike the slow time-varying channels, the rapid time-varying channels are characterized by high time-varying CSIs. In this case, the CSI used for relay selection is different from the CSI used for signal transmission, so the CSI is outdated due to the slow feedback coming from the relays. Unlike \cite{prs} where the PRS is assumed with perfect CSI estimation, outdated CSI of Rayleigh and Nakagami-m fading is assumed in \cite{c1,c2,16}.  Besides, research attempts have introduced the full-duplex relaying as it has the potential to double the spectral efficiency. Due to the self-interference which substantially degrades the performance, related works proposed beamforming designs to cancel the self-interference and improve the achievable rate \cite{fd1,fd2}. Furthermore, relaying as well as the intelligent reflecting surfaces (IRS) have been introduced in the context of physical layer security of vehicular adhoc networks (VANET) to protect the legitimate receiver from the eavesdropping attacks by transmitting friendly jammers and/or artificial noise in order to maximize the secrecy capacity \cite{pls1,pls2,neji1,neji2,maalej}.

In spite of these considerable contributions in the area of mixed RF/FSO systems, they assumed ideal system without hardware impairments. In practice, however, the hardware (source, relays, destination) are susceptible to impairments, e.g., high power amplifier (HPA) nonlinearities \cite{17}, phase noise \cite{19} and In phase and Quadrature (IQ) imbalance \cite{20}. Due to its low quality and price, the relay suffers from the nonlinear PA impairment which is caused primarily by the nonlinear amplification of the signal that may cause a distortion and a phase rotation of the signal. The most well-known nonlinear power amplifier model are Traveling Wave Tube Amplifier (TWTA), Soft Envelope Limiter (SEL) \cite{22} and Ideal Soft Limiter Amplifier (ISLA) \cite{23}. Maletic \textit{et al.} \cite{22} concluded that the SEL has less severe impact on the system performance than the TWTA model. Furthermore, there are few attempts \cite{c1,c2,j2} considering mixed RF/FSO system affected by a general model of impairments but they did not specify the type/nature of the hardware impairments. Qi \textit{et. al} \cite{31} concluded that the impairments have deleterious effects on the system by limiting its performance in terms of outage probability, symbol error rate and channel capacity especially, in the high Signal-to-Noise Ratio (SNR) regime. In fact, previous work \cite{32} demonstrated that the impaired systems have a finite capacity limit at high SNR while there are floors for both the outage probability and the symbol error rate \cite{33}.

\subsection{Related Work}
The existing work of the mixed RF/FSO systems cover various permutations of the system parameters. The authors in \cite{55,56} consider dual-hop hybrid RF/FSO system employing AF with fixed gain (FG). Particularly, Zedini \textit{et. al} in \cite{55} derive the outage probability, the bit error rate (BER) and the ergodic capacity assuming that the RF and FSO follows Nakagami-m and unified G$^2$, respectively. Besides, Al-Quwaiee \textit{et. al} in \cite{56} present the same performance as the aforementioned work but they assume that the RF and FSO channels experience Rayleigh and Double Generalized Gamma fading, respectively. On the other side, \cite{58,59} develop asymmetric dual-hop mixed RF/FSO systems with variable gain (VG). Ansari \textit{et. al} in \cite{58} derive novel closed-forms of the outage probability, BER and the average capacity where the RF and FSO links experience Rayleigh and unified G$^2$ while Yang \textit{et. al} in \cite{59} derive the same performance achieved by \cite{58} but they assume transmit diversity at the source and selection combining at the receiver. In addition the RF links are subject to Nakagami-m while the FSO fading is modeled by M{\'a}laga distribution. Further work \cite{c1,c2} assume mixed RF/FSO multiple relays systems with outdated CSI and they extend their work compared to the previous attempts by considering non-ideal hardware suffering from an aggregate model of hardware impairments. Although, the aforementioned work have considered many permutations of the system parameters, they did not consider more realistic and practical RF/FSO system including both the spatial diversity brought by the multiple relays and a particular model of the HPA nonlinearities rather than assuming a general model of impairments. 

\subsection{Contribution}
In this paper, we introduce the hardware and interference constraints on the proposed system. Specifically, the relay suffers from the nonlinear HPA as well as the interference, while the destination is susceptible to IQ imbalance. The nonlinear HPA models studied in this work are SEL, and TWTA hardware imperfections. Moreover, we consider Fixed Gain (FG) relaying and we assume that the optical signal can be detected following either heterodyne or IM/DD while a subcarrier signal is used to modulate the intensity of an optical carrier (representing SIM technique). The arrangement of this work follows these steps:
 \begin{itemize}
    \item Present a detailed description of the system architecture and the different models of impairments and we then take into account a macroscopic analysis and study the impacts of the hardware impairments and the interference on the system performance.
    \item Specify the statistics of the RF and the optical channels in terms of the probability density function (PDF), the cumulative distribution function (CDF) and the high order moments.
    \item  After calculating the end-to-end Signal-to-Interference-plus-Noise-and-Distortion Ratio (SINDR), we present the analytical formulations of the end-to-end outage probability, the probability of error, the ergodic capacity, the upper bounds, and the asymptotic high SNR. Capitalizing on these expressions, we will derive quantitative summaries and valuable engineering insights to draw meaningful conclusions and observations of the proposed system.
\end{itemize}

\subsection{Structure}
This paper is organized as follows: Section II describes the system models, the HPA nonlinearities, the IQ imbalance, and the interference. The target performances in terms of the outage probability, the probability of error, and the ergodic capacity analysis are presented in Sections III, IV, and V, respectively. Numerical results and their discussions are given in Section VI. The final Section discusses the summary of this work.

\subsection{Notation}
For the sake of organization, we provide some useful notations to avoid the repetition. $f_{\rm h}(\cdot)$ and $F_{\rm h}(\cdot)$ denote the PDF and CDF of the random variable $h$, respectively. The Generalized Gamma distribution with parameters $\alpha, \beta$ and $\gamma$ is given by $\mathcal{GG} (\alpha,\beta,\gamma)$. Moreover, the Gamma distribution with parameters $\alpha$ and $\beta$ is given by $\mathcal{G} (\alpha,\beta)$. In addition, the Gaussian distribution of parameter $\mu, \sigma^2$ is denoted by $\mathcal{N}(\mu,\sigma^2)$. The operator $\mathbb{E}[\cdot]$ stands for the expectation while $\mathbb{P}[\cdot]$ denotes the probability measure. The symbol $\backsim$ stands for "distributed as".

\section{System Model}
\begin{figure}[t]
    \centering
    \includegraphics[width=\linewidth]{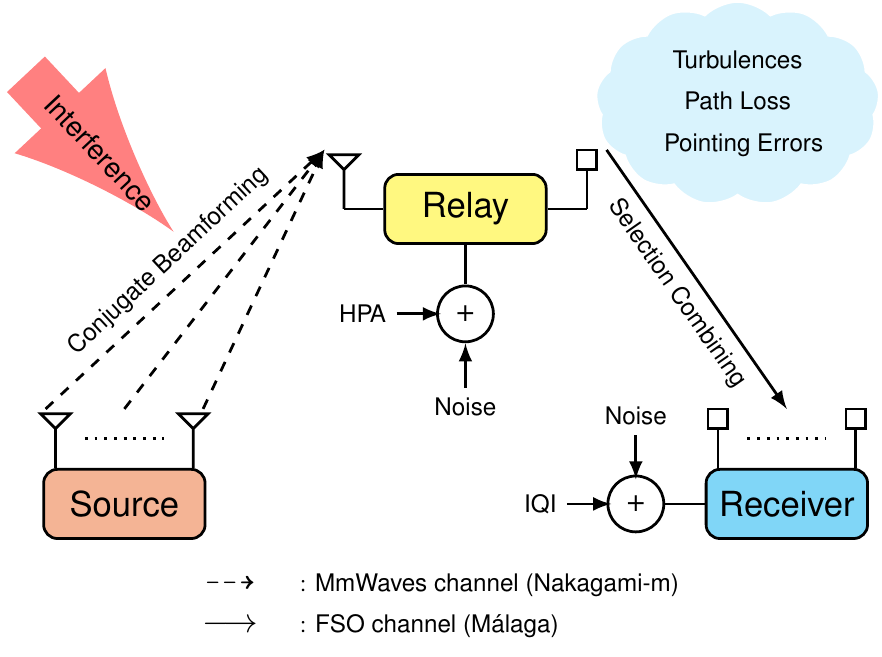}
    \caption{Dual-hop MIMO MmWaves and FSO Relaying System. The RF signaling of the first hop is affected by the interference while the FSO beam in the second hop experiences the atmospheric turbulences, pathloss and pointing errors. We observe that the relay amplifier is subject to the nonlinearities effect caused by the high power amplification while the receiver suffers from IQ imbalance impairment.}
    \label{system}
\end{figure}
The system consists of a source (S) equipped with $N$ antennas, and a destination ($D$) consisting of $M$ apertures. $S$ can communicate with RX only through an intermediary relay ($R$) which is considered as RF to FSO converter. $R$ is equipped with one receive antenna for RF signal reception and one transmit aperture for FSO  transmission. The links between $S$ and $R$ are RF nature, while the channels between $R$ and $D$ are FSO. RF and FSO links are impaired by the interferers and the pointing errors, respectively. To combat the effects of multipath fading and the atmospheric turbulences, we assume transmit diversity at $S$ and selection combining at $D$. Additionally, due to the low quality of the hardware, the relay is susceptible to the effect of HPA nonlinearities and the common models used in this work are SEL and TWTA, while the receiver suffers from the IQ imbalance during the signal reception. 
\subsection{MmWaves Channel Model}
The received RF signal at $R$ is given by
\begin{align}
    y_{\text{SR}} = \sqrt{\frac{A_{\text{SR}}P_{s}}{N}}\sum_{n=1}^{N} h_{n} x + \sum_{k=1}^{M_{ R}}f_{k}d_{k} + n_{\text{SR}}
\end{align}
where $A_{\text{SR}}$ is the average power gain, $P_s$ is the total transmitted power from $S$ equally splitted among the $N$ channels, x is the modulation symbol, $h_n$ is the $n$-th channel between $S$ and $R$, $d_k$ is the modulation symbol of the $k$-th interferer with an average power $\mathbb{E}[|d_k|^2] = P_{R_k}$, $f_{k}$ is the fading coefficient between the $k$-th interferer and $R$, $M_R$ is the number of interferer affecting the RF channels, and $n_{\text{SR}} \backsim \mathcal{N} (0,~\sigma^2_{\text{SR}})$. \textcolor{black}{Note that the total average power of the interference is $P_r = \sum_{k=1}^{M_R}P_{R_k}$.}

The average power gain of of the RF links is given by \cite[Eq.~(7)]{mmwavej}
\begin{equation}
A_{\text{SR}}\text{[dB]} = G_{\text{TX}} + G_{\text{RX}} - 20\log_{10}\left(\frac{4\pi L_{\text{SR}}}{\lambda_1} \right) - (\alpha_{\text{ox}} + \alpha_{\text{rain}}) L_{\text{SR}}
\end{equation}
where $G_t$ and $G_r$ are the transmit and receive antenna gains, respectively, $\lambda_1$ is the wavelength of the RF links, $\alpha_{\text{ox}}$ and $\alpha_{\text{rain}}$ are the attenuations caused by the oxygen absorption and rain, respectively, and $L_{\text{SR}}$ is the link distance between $S$ and $R$.\\
The noise variance in the RF links is given by \cite{mmwavec}
\begin{equation}
    \sigma_{\text{SR}}^2 \text{[dBm]} = B + N_0 + N_f  
\end{equation}
where $B$ is the mmWaves bandwidth, $N_0$ is the noise power spectral density and $N_f$ is the noise figure at the receiver.
\begin{table}[H]
\renewcommand{\arraystretch}{.8}
\caption{MmWaves Sub-System \cite{mmwavej,mmwavec}}
\label{table1}
\centering
\begin{tabular}{|l|c|c|}
\hline
\hline
\bfseries Parameter &\bfseries Symbol &\bfseries  Value\\
\hline
Carrier Frequency & $f_c$& 28 GHz\\
\hline
Bandwidth & $B$& 850 MHz\\
\hline
Transmit Antenna Gain& $G_{\text{TX}}$& 44 dBi\\
\hline
Receive Antenna Gain & $G_{\text{RX}}$& 44 dBi\\
\hline
 Oxygen Attenuation& $\alpha_{\text{ox}}$ & 15.1 dB/km\\
\hline
Noise Power Spectral Density& $N_0$& -144 dBm/MHz\\
\hline
Receive Noise Figure& $N_f$& 5 dB\\
\hline\hline
\end{tabular}
\end{table}
Since $\| \mathbf{h} \|^2$ is the sum of $N$ Nakagami-m independent and identically distributed (i.i.d) random variables with parameter $m_{\text{SR}}$, the PDF of $\gamma_{\text{SR}}$ is given by
\begin{equation}
    f_{\gamma_{\text{SR}}}(\gamma) = \frac{1}{\Gamma(Nm_{\text{SR}})\overline{\gamma}^{Nm_{\text{SR}}}_{\text{SR}}}\gamma^{Nm_{\text{SR}}-1} e^{-\frac{\gamma}{\overline{\gamma}_{\text{SR}}}}
\end{equation}
where $\Gamma(\cdot)$ is the Gamma function and \textbf{h} is the vector form of $h_n$. If $m_{\text{SR}}$ is integer, the CDF is given by
\begin{equation}
F_{\gamma_{\text{SR}}}(\gamma) = 1 - e^{-\frac{Nm_{\text{SR}}\gamma}{\overline{\gamma}_{\text{SR}}}}\sum_{n=1}^{Nm_{\text{SR}}-1}\frac{1}{n!}\left( \frac{Nm_{\text{SR}}\gamma}{\overline{\gamma}_{\text{SR}}} \right)^n.  
\end{equation}
\subsection{FSO Channel Model}
\subsubsection{\textcolor{black}{Atmospheric Turbulences and Signal Models}}
Once the RF signal reaches $R$, it is amplified based on the instantaneous Channel State Information (CSI) and then forwarded to $D$. The received signal at $D$ is given by
\begin{equation}
y_{\text{RD}} = (\eta\mu I_m)^{\frac{r}{2}}Gy_{\text{SR}} + n_{\text{RD}}   
\end{equation}
where $\mu$ is the electrical-to-optical conversion coefficient, $\eta$ is the receiver responsivity, $I_m$ is the optical irradiance for $m = 1 \ldots M$,
$n_{\text{RD}} \backsim \mathcal{N} (0,~\sigma^2_{\text{RD}})$, \textcolor{black}{$r$ = 1 and $r$ = 2, respectively, denote the modes of heterodyne and Intensity Modulation and Direct Detection (IMDD)}, and $G$ is the relaying gain. Since selection combining is assumed at $D$, the aperture with the highest channel gain is selected to collect the received signal. The selection is achieved based on the following
\begin{equation}
    I_{\tt{max}} = \max(I_1,~I_2,\ldots,~I_M).
\end{equation}
After removing the DC bias, $D$ recovers the original information by demodulating the signal. Note that the optical irradiance are i.i.d random variables. The  FSO  fading  involves  three  components  which are  the  turbulence-induced  fading  ($I_a$),  the  atmospheric  pathloss  ($I_\ell$)  and  the  pointing errors ($I_p$). The $m$-th  channel  gain $I_m$ can  be  written  as  follows
\begin{align}
    I_m = I_a I_p I_\ell.
\end{align}
Since $I_{\rm a}$ follows the $\mathcal{M}$ distribution, the PDF can be given by \cite[Eq.~(19)]{j2}
\begin{equation}\label{ird}
f_{I_a}(I_a) = A \sum_{n=1}^{\beta}a_nI_a^{\frac{\alpha+n}{2}-1}K_{\alpha-n}\left(2\sqrt{\frac{\alpha\beta I_a}{g\beta+\Omega^{\prime}}}  \right)    
\end{equation}
where $K_{\nu}(\cdot)$ is the modified Bessel function of the second kind with order $\nu$. The parameters $g, \Omega^{'}, \alpha$, and $\beta$ are detailed in \cite{j2}. In addition, the parameter $A$, and $a_n$ are defined as
\begin{equation}
    A = \frac{2\alpha\frac{\alpha}{2}}{g^{1+\frac{\alpha}{2}}\Gamma(\alpha)}\left(\frac{g\beta}{g\beta+\Omega^{\prime}}\right)^{\beta+\frac{\alpha}{2}}.
\end{equation}
\begin{equation}
    a_n = {\beta -1 \choose n-1} \frac{(g\beta+\Omega^{'})^{1-\frac{n}{2}}}{(n-1)!}\left(\frac{\Omega^{'}}{g}\right)^{n-1}\left(\frac{\alpha}{\beta}\right)^{\frac{n}{2}}.
\end{equation}
\subsubsection{\textcolor{black}{Truncation Analysis}}
\textcolor{black}{We rewrite the PDF (\ref{ird}) in a more suitable form for mathematical tractability in the derivation. Defining $\xi = \beta/(g\beta+\Omega^{\prime})$, we can rewrite $A$ and $a_n$ as $A = \Xi\left(\alpha\xi \right)^{\alpha/2}$ and $a_n = \omega_n\left(\alpha \xi \right)^{n/2}$, where $\Xi$ and $\omega_n$ are defined as
\begin{equation}
\Xi = \frac{2\left(g\xi\right)^\beta}{g\Gamma(\alpha)}    
\end{equation}
\begin{equation}
    \omega_n = {\beta-1 \choose n-1}\frac{\beta^{1-n}}{\xi(n-1)!}\left( \frac{\Omega^{\prime}}{g} \right)^{n-1}
\end{equation}
Then, we express the PDF as
\begin{equation}
f_{I_a}(I_a) = \Xi\sum_{n=1}^\beta \omega_n  \left(\alpha\xi \right)^{\frac{\alpha+n}{2}}I_a^{\frac{\alpha+n}{2}-1}K_{\alpha-n}\left(2\sqrt{\alpha \xi I_a} \right)  
\end{equation}
The modified Bessel function can be written in terms of generalized power series representation as
\begin{equation}\label{bessel}
K_{\nu}(x) = \frac{\pi}{2\sin(\pi\nu)}\sum_{p=0}^{\infty}\left[\frac{(0.5x)^{2p-\nu}}{\Gamma(p-\nu+1)p!} - \frac{(0.5x)^{2p+\nu}}{\Gamma(p+\nu+1)p!}  \right]    
\end{equation}
where $\nu \notin \mathbb{Z}$ and $|x| < \infty$. Injecting (\ref{bessel}) in (\ref{ird}), we get
\begin{equation}\label{exp}
\begin{split}
f_{I_a}(I_a) =& \Xi \sum_{n=1}^\beta \frac{\pi\omega_n}{2\sin[\pi(\alpha-n)]}\sum_{p=0}^\infty\left[  \rho_p(\alpha,n)I_a^{p+n-1} \right.\\&\left.- \rho_p(n,\alpha)I_a^{p+\alpha-1} \right]    
\end{split}
\end{equation}
where $(\alpha-n)\notin \mathbb{Z}$ or, equivalently, $\alpha \notin \mathbb{Z}$, and
\begin{equation}\label{rho}
\rho_p(x,y) = \frac{\left(\alpha \xi \right)^{p+y}}{\Gamma(p-x+y+1)p!}
\end{equation}
In practice, some finite value $L$ is used for calculating the
upper limit in the summation term of (\ref{exp}). Let $\Hat{f}_{I_a}(I_a)$ denotes this truncated function. It is given by
\begin{equation}
\begin{split}
\Hat{f}_{I_a}(I_a) =& \Xi \sum_{n=1}^\beta \frac{\pi\omega_n}{2\sin[\pi(\alpha-n)]}\sum_{p=0}^L\left[  \rho_p(\alpha,n)I_a^{p+n-1} \right.\\&\left.- \rho_p(n,\alpha)I_a^{p+\alpha-1} \right]
\end{split}
\end{equation}
The elimination of the infinite terms after the first $L+1$ terms results in a truncation error defined as $\mathcal{E}(L) = | f_{I_a}(I_a) - \Hat{f}_{I_a}(I_a)|$. In Appendix A, we show that the truncation error $\mathcal{E}(L)$ is upper bounded as
\begin{equation}\label{errorbound}
\mathcal{E}(L) < \Xi e^{\alpha \xi I_a} \max_{p>L}\{c_p(\alpha,n)\} \sum_{n=1}^\beta \frac{\pi \omega_n}{2\sin[\pi(\alpha-n)]}     
\end{equation}
This bound illustrates that the truncation error can be made arbitrarily small by increasing $L$. In Fig. \ref{truncerror}, the truncation error in (\ref{errorbound}) is considered for different values of the upper limit $L$. In Fig. \ref{comp}, the exact PDF is compared with the truncated version for different values of the upper limit $L$.
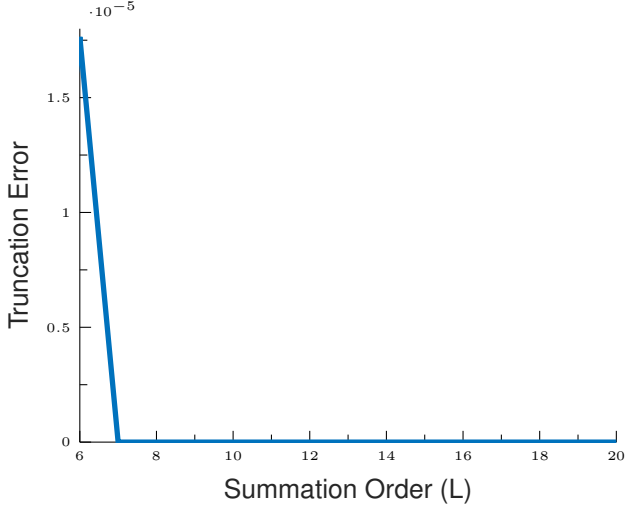
\begin{figure}[t]
\centering
\setlength\fheight{5.5cm}
\setlength\fwidth{7.5cm}
%
%
\definecolor{mycolor1}{rgb}{0.00000,0.44700,0.74100}%
\begin{tikzpicture}

\begin{axis}[%
width=0.951\fwidth,
height=\fheight,
at={(0\fwidth,0\fheight)},
scale only axis,
xmin=6,
xmax=20,
xlabel style={font=\color{white!15!black}},
xlabel={\textsf{Summation Order (L)}},
ymin=0,
ymax=1.8e-05,
ylabel style={font=\color{white!15!black}},
ylabel={\textsf{Truncation Error}},
axis background/.style={fill=white},
axis x line*=bottom,
axis y line*=left,
legend style={legend cell align=left, align=left, draw=white!15!black}
]
\addplot [color=mycolor1, line width=2.0pt]
  table[row sep=crcr]{%
4	0\\
5	0\\
6	1.76457737859903e-05\\
7	0\\
8	0\\
9	0\\
10	0\\
11	0\\
12	0\\
13	0\\
14	0\\
15	0\\
16	0\\
17	0\\
18	0\\
19	0\\
20	0\\
};

\end{axis}
\end{tikzpicture}%
    \caption{Truncation error with respect to the number of summation terms used in the series representation of the M\'alaga distribution. The results are generated with $\rho = 0.95, b_0 = 0.596, \Omega = 1.32, g = 0.001, \alpha = 2.1, \beta = 2$.  }
    \label{comp}
\end{figure}
\begin{figure}[b]
\centering
\setlength\fheight{5.5cm}
\setlength\fwidth{7.5cm}
%
%
\definecolor{mycolor1}{rgb}{0.00000,0.44700,0.74100}%
\definecolor{mycolor2}{rgb}{0.85000,0.32500,0.09800}%
\definecolor{mycolor3}{rgb}{0.92900,0.69400,0.12500}%
\definecolor{mycolor4}{rgb}{0.49400,0.18400,0.55600}%
\begin{tikzpicture}

\begin{axis}[%
width=0.951\fwidth,
height=\fheight,
at={(0\fwidth,0\fheight)},
scale only axis,
unbounded coords=jump,
xmin=0,
xmax=3.5,
xlabel style={font=\color{white!15!black}},
xlabel={\textsf{Threshold Argument} ($I_a$)},
ymin=-0.3,
ymax=0.4,
ylabel style={font=\color{white!15!black}},
ylabel={\textsf{PDF}},
axis background/.style={fill=white},
axis x line*=bottom,
axis y line*=left,
legend style={at={(0.6,0.35)}, anchor=north east, legend cell align=left, align=left, draw=white!15!black,draw=none,fill=none}
]
\addplot [color=mycolor1, line width=2.0pt]
  table[row sep=crcr]{%
0	nan\\
0.1	0.279743418735741\\
0.2	0.353604158754246\\
0.3	0.380210471404182\\
0.4	0.386155758866594\\
0.5	0.381634613742919\\
0.6	0.371506252993437\\
0.7	0.358364029560434\\
0.8	0.343688683493665\\
0.9	0.328361632966114\\
1	0.312920181044968\\
1.1	0.297694763759\\
1.2	0.28288720592575\\
1.3	0.268617395476124\\
1.4	0.254952124644942\\
1.5	0.241923420927666\\
1.6	0.229540463065524\\
1.7	0.217797466464698\\
1.8	0.206678974086468\\
1.9	0.196163442875445\\
2	0.186225691140309\\
2.1	0.176838573772793\\
2.2	0.16797412779041\\
2.3	0.159604351051399\\
2.4	0.151701725037647\\
2.5	0.144239558137405\\
2.6	0.137192202655144\\
2.7	0.13053518294467\\
2.8	0.124245261132996\\
2.9	0.118300459279163\\
3	0.112680051445127\\
3.1	0.107364535346457\\
3.2	0.102335590527597\\
3.3	0.0975760280482919\\
3.4	0.0930697352523869\\
3.5	0.0888016181631428\\
};
\addlegendentry{\textsf{Exact}}

\addplot [color=mycolor2, line width=2.0pt]
  table[row sep=crcr]{%
0	0.00126763994629109\\
0.1	0.279743418735502\\
0.2	0.353604158696245\\
0.3	0.380210469988673\\
0.4	0.386155745290757\\
0.5	0.381634535588234\\
0.6	0.371505927019483\\
0.7	0.358362940746536\\
0.8	0.343685591778532\\
0.9	0.328353877122564\\
1	0.312902535271182\\
1.1	0.297657665099503\\
1.2	0.282814129749235\\
1.3	0.268481112101766\\
1.4	0.254709522970562\\
1.5	0.241508534704256\\
1.6	0.228855282398621\\
1.7	0.216700061311841\\
1.8	0.204968398055727\\
1.9	0.193560824603731\\
2	0.182350857019525\\
2.1	0.171181479876293\\
2.2	0.15986031059749\\
2.3	0.148153536040059\\
2.4	0.13577865947841\\
2.5	0.122396059526297\\
2.6	0.107599337229757\\
2.7	0.0909044096798273\\
2.8	0.0717372955651439\\
2.9	0.0494205284993706\\
3	0.0231581266773746\\
3.1	-0.00798095826851602\\
3.2	-0.045080994904047\\
3.3	-0.0893993135660491\\
3.4	-0.14238792450068\\
3.5	-0.205717003949748\\
};
\addlegendentry{\textsf{Truncated $L$ = 6 Terms}}

\addplot [color=mycolor3, line width=2.0pt]
  table[row sep=crcr]{%
0	0.00126763994629109\\
0.1	0.279743418735741\\
0.2	0.353604158753924\\
0.3	0.380210471392271\\
0.4	0.386155758713745\\
0.5	0.381634612640091\\
0.6	0.371506247461654\\
0.7	0.358364007963051\\
0.8	0.343688613289329\\
0.9	0.328361434539662\\
1	0.312919678742347\\
1.1	0.297693600635163\\
1.2	0.282884703600661\\
1.3	0.268612334297793\\
1.4	0.254942411996655\\
1.5	0.241905606912053\\
1.6	0.229509052873253\\
1.7	0.217743967419539\\
1.8	0.206590602578781\\
1.9	0.196021402087398\\
2	0.186002914558902\\
2.1	0.176496811921513\\
2.2	0.167460236320892\\
2.3	0.158845617264022\\
2.4	0.150600046987668\\
2.5	0.142664265669552\\
2.6	0.134971282942851\\
2.7	0.127444644331485\\
2.8	0.119996338235462\\
2.9	0.112524329353427\\
3	0.104909696897778\\
3.1	0.0970133499337343\\
3.2	0.088672287181487\\
3.3	0.0796953643528042\\
3.4	0.0698585283151966\\
3.5	0.0588994739455785\\
};
\addlegendentry{\textsf{Truncated $L$ = 7 Terms}}

\addplot [color=mycolor4, line width=2.0pt, only marks, mark=asterisk, mark options={solid, mycolor4}]
  table[row sep=crcr]{%
0	0.00126763994629109\\
0.2	0.353604158754246\\
0.4	0.386155758865248\\
0.6	0.371506252919894\\
0.8	0.343688682245543\\
1	0.312920169855238\\
1.2	0.282887138895683\\
1.4	0.25495182056453\\
1.6	0.229539337404386\\
1.8	0.206675406037143\\
2	0.186215683724275\\
2.2	0.167948703466141\\
2.4	0.151642197544515\\
2.6	0.137062059032604\\
2.8	0.123976855602567\\
3	0.112153633395019\\
3.2	0.101347339176526\\
3.4	0.0912844294312563\\
};
\addlegendentry{\textsf{Truncated $L$ = 8 Terms}}
\end{axis}
\end{tikzpicture}%
    \caption{Exact PDF of the irradiance fluctuations compared to the truncated PDF. The results are generated with $\rho = 0.95, b_0 = 0.596, \Omega = 1.32, g = 0.001, \alpha = 2.1, \beta = 2$.}
    \label{truncerror}
\end{figure}
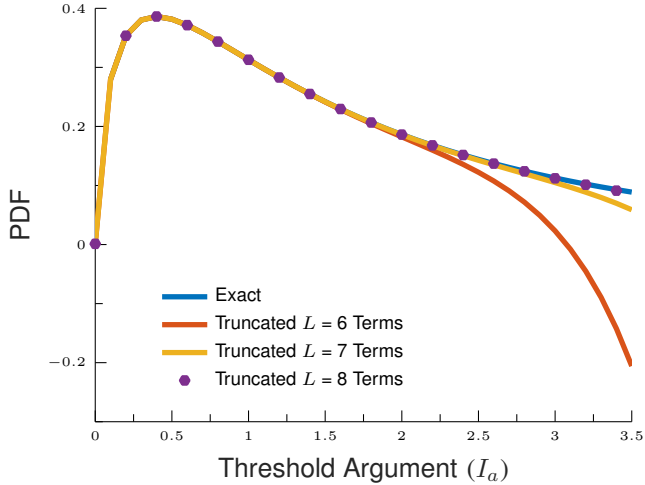}

\begin{remark}
\textcolor{black}{Proving the convergence of the truncated PDF of the atmospheric turbulences is necessary and sufficient to guarantee the convergence of the truncated CDF and PDF in (\ref{fso-cdf}) and (\ref{fso-pdf}), respectively.}
\end{remark}

\subsubsection{\textcolor{black}{Aggregate FSO Channel Model}}
Note that the optical signal suffers from the weather attenuation or also called optical pathloss during the propagation. Hence, the pathloss is given by \cite[Eq.~(4)]{mmwavec}
\begin{equation}\label{pathloss}
    I_\ell = \frac{\pi a^2}{(\theta L_{\text{RD}})^2}\exp\left(-\sigma L_{\text{RD}}\right)
\end{equation}
where $a$ is the radius of the receiver, $\theta$ is the receive beam divergence, $L_{\text{RD}}$ is the optical link distance between $R$ and $D$, and $\sigma$ is the weather attenuation coefficient.
Furthermore, the optical signal is subject to some fluctuations introduced by seismic activities or buildings swaying. This perturbation is translated into a misalignment between $R$ and $D$ or also called the pointing errors which causes additional power losses. Thereby, the pointing errors made by Jitter is modeled as follows \cite{j3}
\begin{equation}\label{pointing}
    I_p = A_0 \exp\left(-\frac{2X^2}{\omega^2_{z_{eq}}}\right)
\end{equation}
where $X$ is the radial displacement at the receiver aperture, $\omega^2_{z_{eq}} = \frac{\sqrt{\pi} \text{erf}(v)}{2v\exp(-v^2)}\omega^2_z$ is the equivalent beam waist, $\omega_z$ is the beam waist at distance $z$, $A_0 = [\text{erf}(v)]^2$, $v = \sqrt{\frac{\pi}{2}}\frac{a}{\omega_z}$, $a$ is the radius of the receiver aperture, and $\text{erf}(\cdot)$ is the error function.\\
The beam waist $\omega_z$ is related to the Rytov variance $\sigma^2_{\text{R}}$ \cite[Eq.~(15)]{djordjevic} which is given by
\begin{align}
\sigma^2_{\text{R}} = 1.23 C_n^2 k^{7/6} L_{\text{RD}}^{11/6}    
\end{align}
where $C_n^2$ is the refractive index, $k$ is the wave number given by $k = 2\pi/\lambda$, and $\lambda$ is the wavelength. If the radial displacement $X$ at the receiver is modeled following the Rayleigh distribution, the PDF of $I_p$ is given by
\begin{align}
    f_{I_p}(I_p) = \frac{\xi^2}{A_0^{\xi^2}}I_p^{\xi^2-1},~0\leq I_p\leq A_0
\end{align}
where $\xi^2 = \frac{\omega_{z_{eq}}}{2\sigma_s^2}$ is the pointing errors coefficient, and $\sigma_s^2$ is the Jitter variance at the receiver. 
\begin{table}[H]
\renewcommand{\arraystretch}{.8}
\caption{FSO Sub-System \cite{mmwavej,mmwavec}}
\label{table2}
\centering
\begin{tabular}{|l|c|c|}
\hline\hline
\bfseries Parameter & \bfseries Symbol & \bfseries Value\\
\hline
Wavelength & $\lambda$& 1550 nm\\
\hline
Modulation Index& $\mu$& 1\\
\hline
Receiver Radius & $a$& 5 cm\\
\hline
Divergence Angle& $\theta$ & 10 mrad\\
\hline
Responsivity& $\eta$& 0.5 A/W\\
\hline
Noise Variance& $\sigma^2_{\text{RD}}$& $\text{10}^{\text{-7}}$ A/Hz\\
\hline\hline
\end{tabular}
\end{table}

\begin{table}[H]
\renewcommand{\arraystretch}{.8}
\caption{Weather Dependent Parameters of FSO and MmWaves Channels \cite{mmwavej,mmwavec}}
\label{table3}
\centering
\begin{tabular}{|l|c|c|c|}
\hline\hline
\bfseries Weather Conditions & \bfseries $\sigma$ (dB/km) & \bfseries $\alpha_{\text{rain}}$ (dB/km)& \bfseries $C^2_n$\\
\hline
Clear Air & 0.43 & 0 & 5$\cdot$\text{10}$^{\text{-14}}$\\
\hline
Moderate Fog & 42.2 & 0 & 2$\cdot$\text{10}$^{\text{-15}}$\\
\hline
Moderate Rain (12.5 mm/h) & 5.8 & 5.6 & 5$\cdot$\text{10}$^{\text{-15}}$\\
\hline\hline
\end{tabular}
\end{table}
Since selection combining is applied at the receiver, the PDF of $I_{\tt{max}}$ is given by \cite[Eq.~(12)]{59} 
\begin{equation}
f_{I_{\text{max}}}(I) = M[F_{I_m}(I)]^{M-1}f_{I_m}(I).    
\end{equation}
In case of the heterodyne detection, the average SNR $\mu_1$ is given by $\mu_1 = \frac{\mu\mathbb{E}[I_{\tt{max}}]}{\sigma^2_{\text{RD}}}$. Regarding the IM/DD detection, the average electrical SNR $\mu_2$ is given by $\mu_2 = \frac{(\mu\mathbb{E}[I_{\tt{max}}])^2}{\sigma^2_{\text{RD}}}$ while the instantaneous optical SNR is $\gamma_{\text{RD}} = \frac{(\mu I_{\text{max}})^2}{\sigma^2_{\text{RD}}}$. Unifying the two detection schemes gives $\gamma_{\text{RD}} = \frac{(\mu I_{\tt{max}})^r}{\sigma_{\text{RD}}^2}$.

The average SNR $\overline{\gamma}_r$\footnote[1]{The average SNR $\overline{\gamma}_r$ is defined as $\overline{\gamma}_r = \mu^r\mathbb{E}[I_{\tt{max}}^r]/\sigma_{\text{RD}}^2$, while the average electrical SNR $\mu_r$ is given by $\mu_r = \mu^r\mathbb{E}[I_{\tt{max}}]^r/\sigma_{\text{RD}}^2$. Therefore, the relation between the average SNR and the average electrical SNR is trivial given that $ \frac{\mathbb{E}[I^2_{\text{max}}]}{\mathbb{E}[I_{\tt{max}}]^2} = \sigma^2_{\text{si}} + 1$, where $\sigma^2_{\text{si}}$ is the scintillation index \cite{scin}.} can be expressed as 
\begin{align}
 \overline{\gamma}_r = \frac{\mathbb{E}[I^r_{m}]}{\mathbb{E}[I_{m}]^r}\mu_r   
\end{align}
where $\mu_r$ is the average electrical SNR given by
\begin{equation}
    \mu_r = \frac{\mu^r\mathbb{E}[I_{m}]^r}{\sigma_{\text{RD}}^2}.
\end{equation}
Following the reasoning steps given by \cite{59}, and given that $\gamma_{\text{RD}} = \mu_rI_{\tt{max}}^r$, the CDF of $\gamma_{\text{RD}}$ is given by
\begin{equation}\label{fso-cdf}
\begin{split}
F_{\gamma_{\text{RD}}}(\gamma) =& \left[ F_{I_m}\left(\left(\frac{\gamma}{\mu_r} \right)^{\frac{1}{r}}\right) \right]^M = A^M\sum_{i=0}^{M}\sum_{j=M-i}^{(M-i)(L+\beta)}\sum_{t=0}^{iL}\\&(-1)^i{M \choose i}\chi_j\chi_t \left(\frac{\gamma}{\mu_r}\right)^{\frac{j+t+i\alpha}{r}}.    
\end{split}
\end{equation}
where $L$, $\chi_i$, and $\chi_t$ are given by \cite{59}.
\begin{IEEEproof}
The proof of (\ref{fso-cdf}) is reported in Appendix A.
\end{IEEEproof}

Differentiating the CDF (\ref{fso-cdf}) gives
\begin{equation}\label{fso-pdf}
\begin{split}
f_{\gamma_{\text{RD}}}(\gamma) =&  A^M\sum_{i=0}^{M}\sum_{j=M-i}^{(M-i)(L+\beta)}\sum_{t=0}^{iL}(-1)^i{M \choose i}\\&\times\frac{(j+t+i \alpha)\chi_j\chi_t}{r\mu_r}\left(\frac{\gamma}{\mu_r}\right)^{\frac{j+t+i\alpha}{r}-1}.      
\end{split}
\end{equation}

\subsection{Interference Model}
The instantaneous SNR of each interferer $\gamma_{\rm{R,k}}$ $\backsim \mathcal{G} (m_{\rm{ R,k}},1/\beta_{\rm{R}})$, where $\beta_{\rm R} \triangleq \frac{m_{\rm{R,k}}\sigma_{\text{SR}}^2}{\Omega_{\rm{R,k}}P_{\rm{R_k}}}$, ($m_{\rm{R,k}}, \Omega_{\rm{R,k}}$) are Nakagami-$m$ parameters between the $k$-th interferer and the relay. It has been shown in \cite{64} that the sum of $L$ i.i.d Gamma random variables with shape parameter $\sigma$ and scale parameter $\alpha$ is a Gamma random variable with parameters $\sigma L$ and $\alpha$. The PDF of the total Interference-to-Noise Ratio (INR) $\gamma_{\rm R} \triangleq \sum\limits_{k=1}^{M_{\rm R}}\gamma_{\rm{R,k}}$ can be expressed as follows
\begin{equation}
f_{\gamma_{\rm R}}(\gamma) = \frac{\beta_R^{m_{\rm R}}}{\Gamma(m_{\rm R})}\gamma^{m_{\rm R}-1}e^{-\beta_{\rm R}\gamma}
\end{equation}
where $m_{\rm R} \triangleq  \sum\limits_{k=1}^{M_{\rm R}}m_{\rm{R,k}}$.

\subsection{High Power Amplifier Nonlinearities}
The HPA nonlinearities impairment is assumed at the relay node. The amplification of the signal happens in two phases. In the first phase, the relay gain $G$ is applied to the received signal as $\phi = G y_{\text{SR}}$. The gain $G$ is given by
\textcolor{black}{
\begin{equation}
    G = \sqrt{\frac{\varrho^2}{\frac{P_sA_{\text{SR}}}{N}\mathbb{E}[\| \mathbf{h} \|^2] + P_r  \mathbb{E}[\| \mathbf{f} \|^2]     + \sigma^2_{\text{SR}}   }}
\end{equation}}
where $\varrho^2$ is the mean power of the signal at the output of the gain block. In the second phase, the signal goes through a nonlinear circuit $\psi = f(\phi)$.

We assume that the HPA of the relay is memoryless. A memoryless HPA is determined by both Amplitude to Amplitude (AM/AM) and Amplitude to Phase (AM/PM) characteristics. The functions AM/AM and AM/PM transform the signal distortion respectively as $A_m(|\phi|)$ and $A_p(|\phi|)$. The output signal of the nonlinear HPA circuit is given by
\begin{equation}
\psi = A_m(|\phi|)~\text{exp}\left(j(\text{arg}(\phi)+A_p(|\phi|))\right),  
\end{equation}
where $\text{arg}(z)$ is the angle of the complex signal $z$. The characteristic functions of the SEL and TWTA models are respectively given by \cite{26}
\begin{equation}
   A_m(|\phi|)=%
   \begin{cases}
     |\phi| & \text{if $|\phi|$ $\leq$ $A_{\text{sat}}$ }  \\
     A_{\text{sat}} & \text{otherwise}
   \end{cases}
\end{equation}
\begin{equation}
    A_p(|\phi|) = 0
\end{equation}
and
\begin{equation}
A_m(|\phi|) = \frac{A^2_{\text{sat}}|\phi|}{A^2_{\text{sat}} + |\phi|^2}
\end{equation}
\begin{equation}
A_p(|\phi|)=\frac{\Phi~|\phi|^2}{A^2_{\text{sat}} + |\phi|^2}    
\end{equation}
where $A_{\text{sat}}$ is the input saturation amplitude, and $\Phi$ controls the maximum phase distortion. From a given saturation level $A_{\text{sat}}$, the relay's power amplifier operates at an input back-off (IBO), which is defined as IBO = $\frac{A_{\text{sat}}^2}{\varrho^2}$. 

\begin{figure}[t]
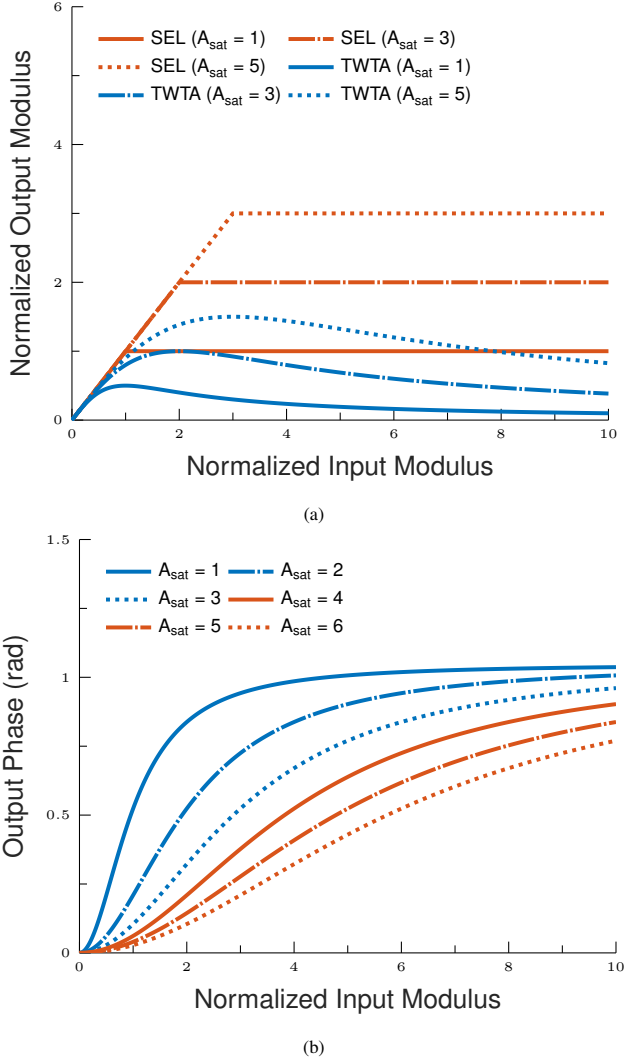

\begin{subfigure}[b]{0.5\textwidth}
\centering
\setlength\fheight{5.5cm}
\setlength\fwidth{7.5cm}
\input{figures/am.tikz}
    \caption{}
    \label{am}
    \end{subfigure}
    \begin{subfigure}[b]{0.5\textwidth}
\centering
\setlength\fheight{5.5cm}
\setlength\fwidth{7.5cm}
\input{figures/phase.tikz}
    \caption{}
    \label{phase}
    \end{subfigure}
    \caption{(a). AM/AM characteristics of SEL and TWTA for different values of $A_{\text{sat}}$. Note that for a given input saturation level, the amplifer output saturation level is higher for SEL compared to TWTA. Therefore, the signal amplified through TWTA amplifier experiences severe distortion and clipping. While the SEL amplifier amplifies the signal with less distortion compared to TWTA. (b). AM/PM characteristics of TWTA for different values of $A_{\text{sat}}$. For this scenario, we assume that $\Phi = \frac{\pi}{3}$. By increasing the input saturation level, the output phase diminishes. The phase shifting range that the incoming signal may experience enlarges with the TWTA output phase. If this range is lower, then the signal phase is subject to smaller rotation or shifting which eventually limits the distortion. }
\end{figure}

The characteristics of AM/AM and AM/PM of SEL and TWTA are shown in Figures.~\ref{am} and \ref{phase} for different values of $A_{\text{sat}}$.
\subsection{Bussgang Linearization Theory}
According to Bussgang Linearization Theory \cite{busg}, the output of the nonlinear HPA circuit is a function of the linear scale parameter $\epsilon$ of the input signal and a non linear distortion $d$ uncorrelated with the input signal and modeled as a complex Gaussian random variable $d \backsim \mathcal{N} (0,~\sigma^2_{d})$. In this case, the AM/AM characteristic $A_m(|\phi|)$ can be written as
\begin{equation}
A_m(|\phi|) = \epsilon x + d.    
\end{equation}
For the SEL power amplifier model, $\epsilon$ and $\sigma^2_{d}$ can be expressed as follows
\begin{equation}
    \epsilon = 1 - \exp\left(-\frac{A_{\text{sat}}^2}{\varrho^2}\right) + \frac{\sqrt{\pi} A_{\text{sat}}}{2\varrho^2}~\text{erfc}\left(\frac{A_{\text{sat}}}{\varrho}\right).
\end{equation}
\begin{equation}
\sigma^2_{d} = \varrho^2~\left[1 - \exp\left(-\frac{A_{\text{sat}}^2}{\varrho^2}\right) - \epsilon^2 \right].   
\end{equation}

Regarding the TWTA power amplifier model, if the AM/PM effect of the characteristic $A_p(|\phi|)$ is negligible (i.e., $\Phi \approx 0$), $\epsilon$ and $\sigma^2_{d}$ can be expressed as follows
\begin{equation}
\epsilon = \frac{A_{\text{sat}}^2}{\varrho^2}\left[1 + \frac{A_{\text{sat}}^2}{\varrho^2}\exp\left(\frac{A_{\text{sat}}^2}{\varrho^2}\right)  + \text{Ei}\left(-\frac{A_{\text{sat}}^2}{\varrho^2} \right)   \right]
\end{equation} 
\begin{equation}
\sigma^2_{d} = -\frac{A_{\text{sat}}^4}{\varrho^2}\left[\left(1 + \frac{A_{\text{sat}}^2}{\varrho^2}\right)\exp\left(\frac{A_{\text{sat}}^2}{\varrho^2}\right)\text{Ei}\left(-\frac{A_{\text{sat}}^2}{\varrho^2} \right) + 1 \right]-\varrho^2\epsilon^2
\end{equation}

where \text{Ei}($\cdot$) and \text{erfc}($\cdot$) are the exponential integral function and the complementary error function, respectively. Note that the ratio between the received SNR and the average transmitted Signal-to-noise-plus-distortion-ratio (SNDR) at the relay is given by
\begin{equation}
\rho_1 = 1 + \frac{\sigma_{d}^2}{\epsilon^2G^2\sigma_{\text{SR}}^2}.
\end{equation}

\subsection{In-Phase and Quadrature-Phase Imbalance}
For the proposed system, the IQ imbalance is modeled as the phase and/or magnitude imbalance between the In-phase and the Quadrature components at the destination. We assume an asymmetric IQ imbalance model, where the In-phase component is ideal and the quadrature branch is impaired \cite{imb}. Hence, the received signal at the destination is given by
\begin{equation}
\hat{y}_{\text{RD}} = \nu_1 y_{\text{RD}} + \nu_2 y^*_{\text{RD}}
\end{equation}
where $y^*_{\text{RD}}$ is the mirror signal introduced by the IQ imbalance at the destination and the coefficients $\nu_1$ and $\nu_2$ are respectively given by
\begin{equation}
\nu_1 = \frac{1 + \zeta e^{-j\theta}}{2}
\end{equation}
\begin{equation}
 \nu_2 = \frac{1 - \zeta e^{j\theta}}{2}   
\end{equation}
Where $\zeta$ and $\theta$ are respectively the amplitude and the phase imbalance. This type of impairment is modeled by the Image-Leakage Ratio (ILR) $\nu_2$, which is given by
\begin{equation}
\rho_2 = \left|\frac{\nu_2}{\nu_1}\right|^2.    
\end{equation}
Note that for an ideal receiver, $\theta$ = 0, and $\zeta$ = 1.
\subsection{End-to-End Signal-to-Interference-plus-Noise-and-Distortion Ratio (SINDR) Analysis}
\textcolor{black}{
The receive SINR at the relay is expressed as 
\begin{equation}
  \gamma_{\text{SR}}^{\text{eff}} = \frac{\gamma_{\text{SR}}}{\gamma_{\text{R}}+1}  
\end{equation}
}
Given that the system suffers from the joint effects of HPA nonlinearities, IQ imbalance, and interference, the SINDR can be written as follows \cite[Eq.~(10)]{ehsan} and \cite[Eq.~(16)]{22}
\begin{equation}\label{sindr0}
\begin{split}
{\scriptsize{\textsf{SINDR}}} = \frac{\gamma_{\text{SR}}^{\text{eff}}\gamma_{\text{RD}}}{\rho_2\gamma_{\text{SR}}^{\text{eff}}\gamma_{\text{RD}}+\rho_1(1+\rho_2)\gamma_{\text{RD}}+(1+\rho_2)(\mathbb{E}[\gamma_{\text{SR}}^{\text{eff}}] + \rho_1)}
\end{split}
\end{equation}
By expanding $\gamma_{\text{SR}}^{\text{eff}}$, the {\scriptsize{\textsf{SINDR}}} can be expressed as
\begin{strip}
\noindent\rule{18.5cm}{1pt}
\begin{equation}\label{sindr}
\begin{split}
{\scriptsize{\textsf{SINDR}}} = \frac{\gamma_{\text{SR}}\gamma_{\text{RD}}}{\rho_2\gamma_{\text{SR}}\gamma_{\text{RD}}+\rho_1(1+\rho_2)(1+\gamma_{\text{R}})\gamma_{\text{RD}}+(1+\rho_2)(1+\gamma_{\text{R}})(\mathbb{E}[\gamma_{\text{SR}}^{\text{eff}}] + \rho_1)}.
\end{split}
\end{equation}
\noindent\rule{18.5cm}{1pt}
\end{strip}

\section{Outage Probability Analysis}
The end-to-end outage probability is defined as the probability that the overall SINDR falls below a given threshold $\Gamma$. It can be generally written as
\begin{equation}\label{opdef}
P_{\text{out}}({\scriptsize{\textsf{SINDR}}},\Gamma) = \mathbb{P}[{\scriptsize{\textsf{SINDR}}} < \Gamma] = F_{{\scriptsize{\textsf{SINDR}}}}(\Gamma).
\end{equation} 
After replacing the expression of {\scriptsize{\textsf{SINDR}}} (\ref{sindr}) in (\ref{opdef}), the CDF can be written as follows
\newpage
\begin{strip}
\begin{equation}
\begin{split}
F_{{\scriptsize{\textsf{SINDR}}}}(\Gamma) =& \int\limits_{0}^{+\infty} \int\limits_{0}^{+\infty} \mathbb{P}\left(\gamma_{\text{SR}} \leq  \frac{\Gamma(\rho_1(1+\rho_2)(1+\gamma_{\text{R}})\gamma_{\text{RD}}+(1+\rho_2)(1+\gamma_{\text{R}})(\mathbb{E}[\gamma_{\text{SR}}^{\text{eff}}] + \rho_1))}{\gamma_{\text{RD}}(1-\rho_2\Gamma)} \Bigg|\gamma_{\text{RD}},~\gamma_{\text{R}}\right)f_{\gamma_{\text{RD}}}(\gamma_{\text{RD}})d\gamma_{\text{RD}}f_{\gamma_{\text{R}}}(\gamma_{\text{R}})d\gamma_{\text{R}}     
\end{split}
\end{equation}
\noindent\rule{18.5cm}{1pt}
\begin{equation}\label{cdf}
\begin{split}
F_{{\scriptsize{\textsf{SINDR}}}}(\Gamma) =& 1 - \frac{A^M\beta_{\text{R}}^{m_{\text{R}}}}{\Gamma(m_{\text{R}})}\exp\left(-\frac{Nm_{\text{SR}}\rho_1(1+\rho_2)\Gamma}{\overline{\gamma}_{\text{SR}}(1-\rho_2\Gamma)}  \right)\sum_{i=0}^M\sum_{j=0}^{(M-i)(L+\beta)}\sum_{t=0}^{iL}\sum_{n=1}^{Nm_{\text{SR}}-1}\sum_{k=0}^n\sum_{m=0}^k {n \choose k} {k \choose m} {M \choose i}\\&\times~\frac{(-1)^i\Gamma(m_{\text{R}}+m)(j+t+i\alpha)\chi_j\chi_t}{rn!\mu_r^{\frac{j+t+i\alpha}{r}}} \left(\frac{(1+\rho_2)(\mathbb{E}[\gamma_{\text{SR}}^{\text{eff}}]+\rho_1)\Gamma}{1-\rho_2\Gamma}  \right)^{n-k}\left( \frac{\rho_1(1+\rho_2)\Gamma}{1-\rho_2\Gamma}\right)^k\\&\times~\left( \frac{\overline{\gamma}_{\text{SR}}(1-\rho_2\Gamma)}{Nm_{\text{SR}}\rho_1(1+\rho_2)\Gamma+\beta_{\text{R}}\overline{\gamma}_{\text{SR}}(1-\rho_2\Gamma)} \right)^{m_{\text{R}}+m}G_{2,0}^{0,2}\Bigg( \frac{\overline{\gamma}_{\text{SR}}(1-\rho_2\Gamma)}{Nm_{\text{SR}}(1+\rho_2)(\mathbb{E}[\gamma_{\text{SR}}^{\text{eff}}]+\rho_1)\Gamma} ~\bigg|~\begin{matrix} n+1-k-\frac{j+t+i\alpha}{r},~1 \\ - \end{matrix} \Bigg). 
\end{split}
\end{equation}
\noindent\rule{18.5cm}{1pt}
\end{strip}
After applying the Binomial expansion $(x+y)^n=\sum_{k=0}^n{n \choose k}x^ky^{n-k}$, and using the identities \cite[Eq.~(3.351.3)]{64}, and \cite[Eq.~(2.24.3.1)]{prudnikov}, the CDF can be written as follows 

Note that expression (\ref{cdf}) is valid only if $\Gamma < \frac{1}{\rho_2}$, otherwise, the cdf is equal to a unity.
\newtheorem{thm}{Theorem}
\newtheorem{cor}[thm]{Corollary}
\begin{cor}
For large values of average SNR, we derive the high SNR approximation by expanding the Meijer-G function \cite[Eq.~(07.34.06.0044.01]{wolfram}.
\end{cor}

The joint effect of HPA nonlinearities and IQ imbalance introduces an irreducible outage floor that saturates the system performance at high SNR and hence, the diversity gain $G_{\text{d}}=0$. In fact, at high SNR, the outage floor is not perfectly horizontal, and practically, it is slowly lowered which is evidence of diversity gain. Although, the system still achieves a diversity gain, this particular gain is fairly negligible and could be considered null $G_{\text{d}} \cong 0$ due to the hardware impairments.

The system can achieve a better diversity gain as far as the joint impacts of HPA nonlinearities and IQ imbalance is mitigated or compensated. The best case scenario wherein the hardwares are ideal ($\rho_1 \longrightarrow 1, \rho_2 \longrightarrow 0$), the diversity gain can be given by
\begin{equation}\label{Gd}
    G_{\text{d}} = \min\left(Nm_{\text{SR}},~M\min\left(\frac{\xi^2}{r},~\frac{\alpha}{r},~\frac{\beta}{r}\right)\right). 
\end{equation}
\subsection*{Special Case:} If we consider the case of interference-free ($M_{\text{R}}=0$), $m_{\text{SR}} = 1, \rho_1 \longrightarrow 1, \rho_2 \longrightarrow 0$, we retrieve the results of \cite{59}.

\section{Bit Error Probability Analysis}
In this section, we invoke the study of the probability of error for OOK, M-PSK, and M-QAM modulations. The error performance expression can be given by
\begin{equation}\label{bep}
P_e = \frac{\delta}{2\Gamma(\tau)}\sum_{\ell=1}^{v}\int\limits_{0}^{+\infty}\Gamma(\tau,q_\ell\gamma)f_{{\scriptsize{\textsf{SINDR}}}}(\gamma)d\gamma,
\end{equation}
where $v,~\delta,~\tau$~, and $q_\ell$ vary depending on the type of detection (heterodyne technique or IM/DD) and modulation
being assumed. It is worth accentuating that this expression is
general enough to be used for both heterodyne and IM/DD
techniques and can be applicable to different modulation
schemes. The parameters $v,~\delta,~\tau$~, and $q_\ell$ are summarized in Table below.
 \begin{table*}[t]\label{beptab}
\renewcommand{\arraystretch}{1}
\caption{\textsc{Parameters for Different Modulations$^\dag$}}
\label{tab:example}
\centering
\begin{tabular}{|l|c|c|c|c|c|}
    \hline\hline
    \textbf{Modulation}  &  $\delta$ & $\tau$ & $q_\ell$ & $v$ & \textbf{Detection}\\
    \hline
   \textbf{OOK}    &  1& 0.5 & 0.5 & 1 & IM/DD\\
    \hline
   \textbf{BPSK}    &  1& 0.5 & 1 & 1 & Heterodyne\\
    \hline
     \textbf{M-PSK}    &  $\frac{2}{\max(\log_2(M), 2)}$ & 0.5 & $\text{sin}^2\left(\frac{(2\ell-1)\pi}{M}\right)$ & $\max\left(\frac{M}{4},1\right)$ & Heterodyne\\
    \hline
      \textbf{M-QAM}    & $\frac{4}{\log_2(M)}\left(1-\frac{1}{\sqrt{M}}  \right)$  & 0.5 & $\frac{3(2\ell-1)^2}{2(M-1)}$& $\frac{\sqrt{M}}{2}$& Heterodyne \\
    \hline
    \hline
\end{tabular}
\\ 
\rule{0in}{1.2em}$^\dag$\scriptsize In case of OOK modulation, the parameters $v,~\delta,~\tau$~, and $q_\ell$ are given by \cite[Eq.~(26)]{mod1}. For M-PSK and M-QAM modulations, these parameters are provided by \cite[Eqs.~(30,~31)]{zedini}.
\end{table*}
The average bit error probability expression in (\ref{bep}) can be rewritten in terms of the CDF (\ref{cdf}) by using integration by parts as
\begin{equation}\label{bepcdf}
P_e = \frac{\delta}{2\Gamma(\tau)}\sum_{\ell=1}^{v}q_\ell^{\tau}\int\limits_{0}^{+\infty}\gamma^{\tau-1}e^{-q_\ell\gamma}F_{{\scriptsize{\textsf{SINDR}}}}(\gamma)d\gamma.    
\end{equation}

As we mentioned earlier, the presence of the terms related to the impairments makes the integral calculus mathematically difficult if not impossible. Thus, deriving a closed-form of the probability of error is not tractable. In this case, we will evaluate the error performance by applying the numerical integration method.

\subsection*{Special case:} When the system is susceptible only to the HPA nonlinearities by keeping $\rho_1$ and setting $\rho_2 \longrightarrow 0$, it is possible to derive a closed-form expression of the error performance. First, we need to invert the argument of the Meijer-G function in Eq.~(\ref{cdf}). After applying the identities \cite[Eqs.~(07.34.03.0271.01,~07.34.03.0046.01)]{wolfram}, the integral in Eq.~(\ref{bepcdf}) involves three Meijer-G functions. After transforming each Meijer-G function into Fox-H function as $G_{p,q}^{m,n}\Bigg(z^C~\bigg|~\begin{matrix} a_1,\ldots,a_p \\ b_1,\ldots,b_q \end{matrix} \Bigg) = \frac{1}{C}~H_{p,q}^{m,n}\Bigg(z~\bigg|~\begin{matrix} (a_1,C^{-1}),\ldots,(a_p,C^{-1}) \\ (b_1,C^{-1}),\ldots,(b_q,C^{-1}) \end{matrix} \Bigg)$, and applying \cite[Eq.~(2.3)]{mittal}, the probability of error is derived as follows 
\newpage
\begin{strip}
\noindent\rule{18.5cm}{1pt}
\begin{equation}\label{bepexact}
\begin{split}
P_e =& \frac{v\delta}{2} - \frac{A^M\delta}{2\Gamma(m_{\text{R}})\Gamma(\tau)}\sum_{i=0}^M\sum_{j=0}^{(M-i)(L+\beta)}\sum_{t=0}^{iL}\sum_{n=1}^{Nm_{\text{SR}}-1}\sum_{k=0}^n\sum_{m=0}^k\sum_{\ell=1}^v{n \choose k} {k \choose m}{M \choose i}\left( \frac{\overline{\gamma}_{\text{SR}}}{Nm_{\text{SR}}\rho_1}\right)^{n+\tau}\\&\times~\frac{(-1)^i(j+t+i\alpha)\chi_j\chi_t\rho_1^kq_\ell^{\tau}(\mathbb{E}[\gamma_{\text{SR}}^{\text{eff}}]+\rho_1)^{n-k}}{n!r\beta_{\text{R}}^m\mu_r^{\frac{j+t+i\alpha}{r}}}
H_{1,0:1,1:0,2}^{0,1:1,1:2,0}\left(\begin{matrix} (\vartheta_1;1,1) \\ - \end{matrix} ~\bigg|~  \begin{matrix} (\vartheta_2,1) \\ (0,1) \end{matrix} ~\bigg|~ \begin{matrix} - \\ (\vartheta_3,1),~(0,1)\end{matrix}~\bigg|~ \frac{1}{\beta_{\text{R}}},~ \frac{\mathbb{E}[\gamma_{\text{SR}}^{\text{eff}}]+\rho_1}{\rho_1}\right)
\end{split}
\end{equation}
\noindent\rule{18.5cm}{1pt}
\end{strip}
where $H_{p_1,q_1:p_2,q_2:p_3,q_3}^{m_1,n_1:m_2,n_2:m_3:n_3}(-|(\cdot,\cdot))$ is the bivariate Fox H-function, $\vartheta_1 = 1-n-\tau, \vartheta_2 = 1-m-m_{\text{R}}, \vartheta_3 = k-n+\frac{j+t+i\alpha}{r}$. An efficient MATLAB implementation of the bivariate Fox-H function is provided by \cite{bfox}.
\section{Ergodic Capacity Analysis}
The channel capacity, expressed in nats/sec/Hz, is defined as the maximum error-free data rate transferred by the system. It can be defined as
\begin{equation}\label{cap}
\mathcal{I}({\scriptsize{\textsf{SINDR}}},\varpi) = \mathbb{E}[\log\left(1+\varpi{\scriptsize{\textsf{SINDR}}}\right)]
\end{equation}
where $\varpi$ can take the values 1 or $e/2\pi$ for heterodyne or IM/DD, respectively. This formula can be computed by numerical integration using the PDF of the end-to-end {\scriptsize{\textsf{SINDR}}}. However, deriving a closed form of the channel capacity in our case is very complex.
\begin{corollary}
To analytically characterize the capacity, most related work in the literature referred to the approximation given by \cite{22,32}
\begin{equation}\label{capapp}
\mathbb{E}\left[\log\left(1+\frac{\phi}{\psi}\right)\right] \cong \log\left(1 + \frac{\mathbb{E}[\phi]}{\mathbb{E}[\psi]}\right),
\end{equation}
\end{corollary}
Although Eq.~(\ref{capapp}) has no fundamental basis, it provides a very tight approximation to the exact performance. We can also provide further characterization of the capacity by deriving a tight upper bound using Jensen's inequality. This upper bound is stated by the following Theorem.
\begin{theorem}
For asymmetric (Nakagami-m/M\'alaga) fading channels, the ergodic capacity for AF variable relaying gain and non-ideal hardware is tightly upper bounded by
\end{theorem}
\begin{equation}\label{upper}
    \mathcal{I}({\scriptsize{\textsf{SINDR}}},\varpi) \leq \log\left(1 + \frac{\varpi\mathcal{J}}{\rho_2\mathcal{J} + 1}\right).
\end{equation}
The term $\mathcal{J}$ is given by
\begin{equation}\label{jensen}
\mathcal{J} = \frac{\mathbb{E}[\gamma_{\text{SR}}]\beta_{\text{R}}}{\rho_1(1+\rho_2)\Gamma(m_{\text{R}})}G_{2,1}^{1,2}\Bigg(\frac{\gamma_{\text{R}}}{\beta_{\text{R}}} ~\bigg|~\begin{matrix} 2-m_{\text{R}},~1 \\ 1 \end{matrix} \Bigg).   
\end{equation}
\begin{IEEEproof}
The capacity upper bound derivation including Eqs.~(\ref{upper},~\ref{jensen}) is reported in Appendix C.
\end{IEEEproof}
Although the capacity indefinitely increases mainly with the average transmitted power and the number of transmit and receive antennas, the joint effect of HPA nonlinearities and IQ imbalance limits the capacity at high SNR. According to \cite{22}, the capacity ceiling introduced by the joint effect of hardware impairments is given by
\begin{equation}\label{inf}
    \mathcal{I}_{\infty}({\scriptsize{\textsf{SINDR}}},\varpi) = \log\left(1 + \frac{\varpi}{\frac{(1+\rho_2)\iota}{\epsilon^2}-1} \right).
\end{equation}
Assuming perfect relay node and for high SNR, the capacity under the effect of IQ imbalance is saturated by the following limit
\begin{equation}\label{max}
\mathcal{I}_{\text{max}}({\scriptsize{\textsf{SINDR}}},\varpi) = \log\left(1 + \frac{\varpi}{\rho_2}  \right).
\end{equation}
The last ceiling only depends on ILR $\rho_2$. If $\rho_2 \longrightarrow 0$, then $\mathcal{I}_{\text{max}}({\scriptsize{\textsf{SINDR}}})\longrightarrow +\infty$. Thereby, the capacity is not upper bounded at high SNR which is an expected result for ideal hardwares.

We clearly observe the importance of considering the hardware impairments along with the analysis of such system and most importantly for high rate systems. Note that the ceilings expressions Eqs.~(\ref{inf},~\ref{max}) only depend on the hardware parameters and hence the capacity cannot be improved by the regular factors such as the transmitted power or the number of antennas.
The term $\iota$ is called the clipping factor which characterizes the relay amplifier. This term is defined for SEL and TWTA, respectively by \cite{22}
\begin{equation}
\iota = 1 - \exp\left( -\frac{A_{\text{sat}}^2}{\varrho^2} \right).    
\end{equation}
\begin{equation}
\iota = -\frac{A^4_{\text{sat}}}{\varrho^4}\left[\left(1+\frac{A^2_{\text{sat}}}{\varrho^2}\right)\exp\left(\frac{A^2_{\text{sat}}}{\varrho^2}\right)\text{Ei}\left(-\frac{A^2_{\text{sat}}}{\varrho^2} \right) + 1 \right].
\end{equation}
\subsection*{Special Case:}
The derivation of a closed-form expression of the capacity is possible not only for ideal hardware, but also for HPA nonlinearities similar to the special case derived for the probability of error ($\rho_1 \neq 1, \rho_2 \longrightarrow 0$).
The expression (\ref{cap}) can be expanded as
\begin{equation}
\mathcal{I}({\scriptsize{\textsf{SINDR}}},\varpi) = \int\limits_{0}^{+\infty} \log(1+\varpi\gamma) f_{{\scriptsize{\textsf{SINDR}}}}(\gamma)d\gamma
\end{equation}
Using integration by parts, the capacity can be reformulated as follows
\begin{equation}
\mathcal{I}({\scriptsize{\textsf{SINDR}}},\varpi) = \int\limits_{0}^{+\infty} \frac{\varpi\overline{F}_{{\scriptsize{\textsf{SINDR}}}}(\gamma)}{1+\varpi\gamma}d\gamma
\end{equation}
where $\overline{F}_{{\scriptsize{\textsf{SINDR}}}}(\cdot)$ is the complementary CDF of ${\scriptsize{\textsf{SINDR}}}$.
Using the following identities \cite[Eqs.~(28,~29a,~29b,~30)]{python}, and \cite[Appendix A]{j3}, the closed-form expression of the capacity is derived in terms of the multivariate Fox-H function.
\begin{strip}
\noindent\rule{18.5cm}{1pt}
\begin{equation}
\begin{split}
\mathcal{I}({\scriptsize{\textsf{SINDR}}},\varpi) =& \frac{\varpi A^M}{\Gamma(m_{\text{R}})}\sum_{i=0}^M\sum_{j=0}^{(M-i)(L+\beta)}\sum_{t=0}^{iL}\sum_{n=1}^{Nm_{\text{SR}}-1}\sum_{k=0}^n\sum_{m=0}^k {n \choose k} {k \choose m}{M \choose i}\frac{(-1)^i\Gamma(m_{\text{R}}+m)}{rn!\mu_r^{\frac{j+t+i\alpha}{r}}}\left( \frac{\overline{\gamma}_{\text{SR}}}{Nm_{\text{SR}}\rho_1}\right)^{n}(\mathbb{E}[\gamma_{\text{SR}}^{\text{eff}}]+\rho_1)^{n-k}\\&\times~\frac{(j+t+i\alpha)\chi_j\chi_t\rho_1^k}{\beta_{\text{R}}^{m}}
H_{1,0:0,1:1,1:0,2}^{0,1:1,0:1,1:2,0}\left(\begin{matrix} (-n;-1,-1,-1) \\ - \end{matrix} ~\bigg|~  \begin{matrix} - \\ \kappa_1 \end{matrix} ~\bigg|~ \begin{matrix} - \\ \kappa_1,~\kappa_1\end{matrix}~\bigg|~\begin{matrix} - \\ \kappa_1,~\kappa_2 \end{matrix} ~\bigg|~ \frac{Nm_{m_{\text{SR}}}\rho_1}{\beta_{\text{R}}\overline{\gamma}_{\text{SR}}},~\varpi,~ \frac{Nm_{\text{SR}}(\mathbb{E}[\gamma_{\text{SR}}^{\text{eff}}]+\rho_1)}{\overline{\gamma}_{\text{SR}}}\right)
\end{split}
\end{equation}
\noindent\rule{18.5cm}{1pt}
\end{strip}
where $\kappa_1 = (0,-1), \kappa_2 = (\vartheta_3,-1)$. An efficient PYTHON implementation of the multivariate Fox-H function is provided by \cite[Appendix A]{python}.

\section{Numerical Results and Discussion}
\begin{figure}[t]
    \centering
    \includegraphics[width=\linewidth]{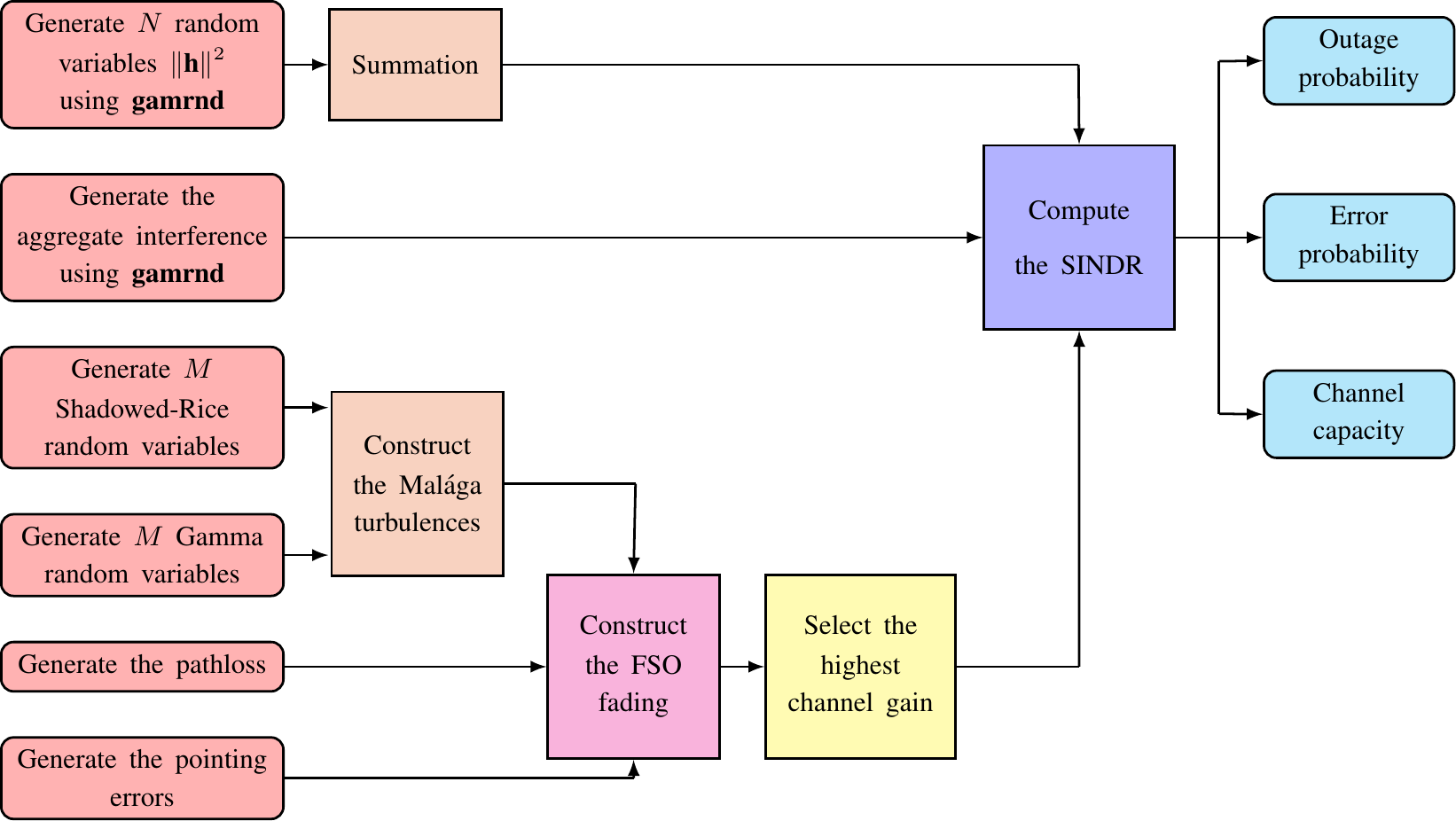}
    \caption{Block diagram of the simulation setup.}
    \label{sim}
\end{figure}
In this section, we will present the numerical results of the performance metrics depending on different combinations of the system parameters. We also check the accuracy of the analytical expressions with Monte Carlo simulations \footnote[2]{It is important to note here that these values for the parameters are selected from \cite{f1,f2,f3} subject to the standards to prove the validity of the obtained results. For all cases, $10^6$ realizations of the random variables were generated to perform the Monte Carlo simulation in MATLAB.}. The mmWave and the interference SNRs are generated following the Gamma distribution. The large-scale atmospheric turbulences can be generated using the Gamma distribution while the small-scale turbulences are generated by a compound of shadowed-Rice distribution. By multiplying the large-scale and small-scale turbulences, we construct the Mal\'aga distribution that describes the aggregate atmospheric turbulences. More details about the simulation process are described by the block diagram in Fig.~\ref{sim}.

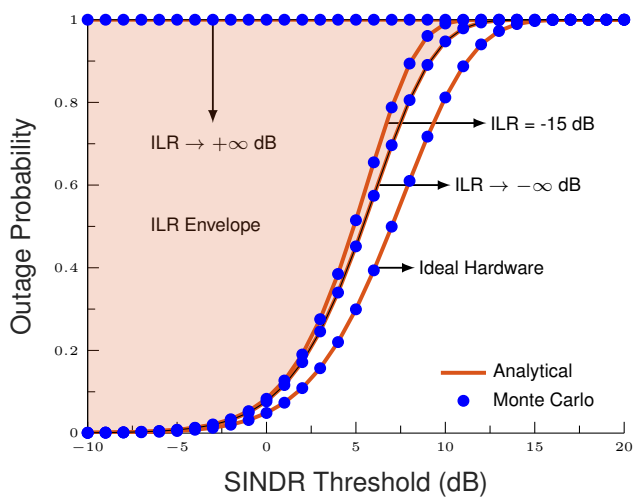
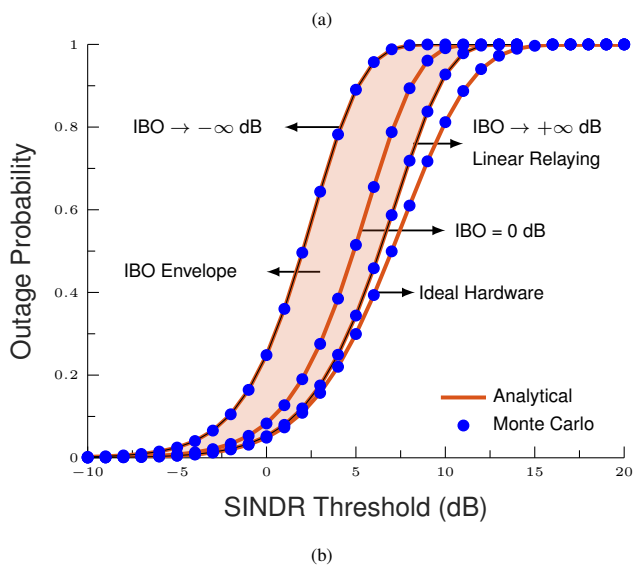
\begin{figure}[t]
\begin{subfigure}[b]{0.5\textwidth}
\centering
\setlength\fheight{5.5cm}
\setlength\fwidth{7.5cm}
%
%
\definecolor{mycolor1}{rgb}{0.85000,0.33000,0.10000}%
\definecolor{mycolor2}{rgb}{0.85000,0.32500,0.09800}%
\definecolor{mycolor3}{rgb}{0.00000,0.45000,0.74000}%
\definecolor{mycolor4}{rgb}{0.49400,0.18400,0.55600}%
\definecolor{mycolor5}{rgb}{0.30100,0.74500,0.93300}%
\definecolor{mycolor6}{rgb}{0.00000,0.44700,0.74100}%
\begin{tikzpicture}

\begin{axis}[%
width=0.951\fwidth,
height=\fheight,
at={(0\fwidth,0\fheight)},
scale only axis,
unbounded coords=jump,
xmin=-10,
xmax=20,
xlabel style={font=\color{white!15!black}},
xlabel={\textsf{SINDR Threshold (dB)}},
ymin=0,
ymax=1,
ylabel style={font=\color{white!15!black}},
ylabel={\textsf{Outage Probability}},
axis background/.style={fill=white},
axis x line*=bottom,
axis y line*=left,
legend style={at={(0.97,0.03)}, anchor=south east, legend cell align=left, align=left, fill=none, draw=none}
]

\node[right, align=left, rotate=0]
at (axis cs:-7,0.5) {\scriptsize\sffamily{ILR Envelope}};

\node[right, align=left, rotate=0]
at (axis cs:-7,0.7) {\scriptsize\sffamily{ILR}~$\rightarrow +\infty$ \scriptsize\sffamily{dB}};
\draw [latex-,black,line width=.7pt] (-3,.75) to (-3,1);

\node[right, align=left, rotate=0]
at (axis cs:8,0.4) {\scriptsize\sffamily{Ideal Hardware}};
\draw [-latex,black,line width=.7pt] (6,.4) to (8.3,.4);

\node[right, align=left, rotate=0]
at (axis cs:12,0.75) {\scriptsize\sffamily{ILR = -15 dB}};
\draw [-latex,black,line width=.7pt] (6.7,.75) to (12.3,.75);

\node[right, align=left, rotate=0]
at (axis cs:10,0.6) {\scriptsize\sffamily{ILR}~$\rightarrow -\infty$ \scriptsize\sffamily{dB}};
\draw [-latex,black,line width=.7pt] (6.3,.6) to (10.3,.6);

\addplot [color=mycolor1, line width=1.5pt]
  table[row sep=crcr]{%
-10	0.000842411397408038\\
-9	0.00127411656256027\\
-8	0.0019629098922409\\
-7	0.00307194139841926\\
-6	0.00486604688958425\\
-5	0.0077685964213412\\
-4	0.0124422454982781\\
-3	0.0198957224998223\\
-2	0.0316103396528477\\
-1	0.049662406965176\\
0	0.0767880679520815\\
1	0.116296412494784\\
2	0.171697772410537\\
3	0.245911096758234\\
4	0.340007657966498\\
5	0.451703273210399\\
6	0.574220651641021\\
7	0.696498008230801\\
8	0.805551774616096\\
9	0.890682349737651\\
10	0.947509348027423\\
11	0.979118248929762\\
12	0.993353286931255\\
13	0.998375959264676\\
14	0.999710681205034\\
15	0.999964812876288\\
16	0.999997314987043\\
17	0.999999884552032\\
18	0.999999997557402\\
19	0.999999999978547\\
20	0.999999999999937\\
};
\addlegendentry{\textsf{Analytical}}

\addplot [color=mycolor2,only marks, line width=.5pt, draw=none, mark=*, mark options={ fill=mycolor3, blue}]
  table[row sep=crcr]{%
-10	0.000842411397408038\\
-9	0.00127411656256027\\
-8	0.0019629098922409\\
-7	0.00307194139841926\\
-6	0.00486604688958425\\
-5	0.0077685964213412\\
-4	0.0124422454982781\\
-3	0.0198957224998223\\
-2	0.0316103396528477\\
-1	0.049662406965176\\
0	0.0767880679520815\\
1	0.116296412494784\\
2	0.171697772410537\\
3	0.245911096758234\\
4	0.340007657966498\\
5	0.451703273210399\\
6	0.574220651641021\\
7	0.696498008230801\\
8	0.805551774616096\\
9	0.890682349737651\\
10	0.947509348027423\\
11	0.979118248929762\\
12	0.993353286931255\\
13	0.998375959264676\\
14	0.999710681205034\\
15	0.999964812876288\\
16	0.999997314987043\\
17	0.999999884552032\\
18	0.999999997557402\\
19	0.999999999978547\\
20	0.999999999999937\\
};
\addlegendentry{\textsf{Monte Carlo}}

\addplot [color=mycolor1, line width=1.5pt, forget plot]
  table[row sep=crcr]{%
-10	1\\
-9	1\\
-8	1\\
-7	1\\
-6	1\\
-5	1\\
-4	1\\
-3	1\\
-2	1\\
-1	1\\
0	1\\
1	1\\
2	1\\
3	1\\
4	1\\
5	1\\
6	1\\
7	1\\
8	1\\
9	1\\
10	1\\
11	1\\
12	1\\
13	1\\
14	1\\
15	1\\
16	1\\
17	1\\
18	1\\
19	1\\
20	1\\
};
\addplot [color=mycolor4, only marks, line width=.5pt, draw=none, mark=*, mark options={solid, fill=mycolor3, blue}, forget plot]
  table[row sep=crcr]{%
-10	1\\
-9	1\\
-8	1\\
-7	1\\
-6	1\\
-5	1\\
-4	1\\
-3	1\\
-2	1\\
-1	1\\
0	1\\
1	1\\
2	1\\
3	1\\
4	1\\
5	1\\
6	1\\
7	1\\
8	1\\
9	1\\
10	1\\
11	1\\
12	1\\
13	1\\
14	1\\
15	1\\
16	1\\
17	1\\
18	1\\
19	1\\
20	1\\
};
\addplot [color=mycolor1, line width=1.5pt, forget plot]
  table[row sep=crcr]{%
-10	0.000859726173744191\\
-9	0.00130303200122373\\
-8	0.00201271558730687\\
-7	0.00316012596331394\\
-6	0.00502575754156287\\
-5	0.00806267874094824\\
-4	0.012989080963557\\
-3	0.0209152707652119\\
-2	0.0335025838397154\\
-1	0.053132810014098\\
0	0.0830302191313337\\
1	0.127219359117319\\
2	0.190126465282074\\
3	0.275576383514541\\
4	0.384994248450533\\
5	0.51495853707327\\
6	0.655032749534944\\
7	0.787875786801107\\
8	0.893960544421346\\
9	0.960803889643138\\
10	0.990851513668436\\
11	0.998998198599061\\
12	0.999973628662059\\
13	0.999999976516186\\
14	1\\
15	nan\\
16	1\\
17	1\\
18	1\\
19	1\\
20	1\\
};
\addplot [color=mycolor5, only marks, line width=.5pt, draw=none, mark=*, mark options={solid, fill=mycolor3, blue}, forget plot]
  table[row sep=crcr]{%
-10	0.000859726173744191\\
-9	0.00130303200122373\\
-8	0.00201271558730687\\
-7	0.00316012596331394\\
-6	0.00502575754156287\\
-5	0.00806267874094824\\
-4	0.012989080963557\\
-3	0.0209152707652119\\
-2	0.0335025838397154\\
-1	0.053132810014098\\
0	0.0830302191313337\\
1	0.127219359117319\\
2	0.190126465282074\\
3	0.275576383514541\\
4	0.384994248450533\\
5	0.51495853707327\\
6	0.655032749534944\\
7	0.787875786801107\\
8	0.893960544421346\\
9	0.960803889643138\\
10	0.990851513668436\\
11	0.998998198599061\\
12	0.999973628662059\\
13	0.999999976516186\\
14	1\\
15	nan\\
16	1\\
17	1\\
18	1\\
19	1\\
20	1\\
};
\addplot [color=mycolor1, line width=1.5pt, forget plot]
  table[row sep=crcr]{%
-10	0.000576970291472745\\
-9	0.000864833996873537\\
-8	0.0013206192309354\\
-7	0.00204893725855226\\
-6	0.00321822030858177\\
-5	0.00509545150903978\\
-4	0.00809509328813018\\
-3	0.0128437556242681\\
-2	0.0202582977150715\\
-1	0.0316275232547759\\
0	0.0486755276688826\\
1	0.0735686776317052\\
2	0.108811869368536\\
3	0.156971731530151\\
4	0.220178915413479\\
5	0.29941532336856\\
6	0.393695405793138\\
7	0.499390650294424\\
8	0.610067165687127\\
9	0.717202648236842\\
10	0.811904717387887\\
11	0.887247412680424\\
12	0.940290026856521\\
13	0.972695833576302\\
14	0.989496834651675\\
15	0.996702797464466\\
16	0.999185247866507\\
17	0.999848397687897\\
18	0.999979911783006\\
19	0.999998234426502\\
20	0.999999905943003\\
};
\addplot [color=mycolor6, only marks, line width=.5pt, draw=none, mark=*, mark options={solid, fill=mycolor3, blue}, forget plot]
  table[row sep=crcr]{%
-10	0.000576970291472745\\
-9	0.000864833996873537\\
-8	0.0013206192309354\\
-7	0.00204893725855226\\
-6	0.00321822030858177\\
-5	0.00509545150903978\\
-4	0.00809509328813018\\
-3	0.0128437556242681\\
-2	0.0202582977150715\\
-1	0.0316275232547759\\
0	0.0486755276688826\\
1	0.0735686776317052\\
2	0.108811869368536\\
3	0.156971731530151\\
4	0.220178915413479\\
5	0.29941532336856\\
6	0.393695405793138\\
7	0.499390650294424\\
8	0.610067165687127\\
9	0.717202648236842\\
10	0.811904717387887\\
11	0.887247412680424\\
12	0.940290026856521\\
13	0.972695833576302\\
14	0.989496834651675\\
15	0.996702797464466\\
16	0.999185247866507\\
17	0.999848397687897\\
18	0.999979911783006\\
19	0.999998234426502\\
20	0.999999905943003\\
};

\addplot[area legend, draw=black, fill=mycolor1, fill opacity=0.2, forget plot]
table[row sep=crcr] {%
x	y\\
-10	0.000842411397408038\\
-9	0.00127411656256027\\
-8	0.0019629098922409\\
-7	0.00307194139841926\\
-6	0.00486604688958425\\
-5	0.0077685964213412\\
-4	0.0124422454982781\\
-3	0.0198957224998223\\
-2	0.0316103396528477\\
-1	0.049662406965176\\
0	0.0767880679520815\\
1	0.116296412494784\\
2	0.171697772410537\\
3	0.245911096758234\\
4	0.340007657966498\\
5	0.451703273210399\\
6	0.574220651641021\\
7	0.696498008230801\\
8	0.805551774616096\\
9	0.890682349737651\\
10	0.947509348027423\\
11	0.979118248929762\\
12	0.993353286931255\\
13	0.998375959264676\\
14	0.999710681205034\\
15	0.999964812876288\\
16	0.999997314987043\\
17	0.999999884552032\\
18	0.999999997557402\\
19	0.999999999978547\\
20	0.999999999999937\\
20	1\\
19	1\\
18	1\\
17	1\\
16	1\\
15	1\\
14	1\\
13	1\\
12	1\\
11	1\\
10	1\\
9	1\\
8	1\\
7	1\\
6	1\\
5	1\\
4	1\\
3	1\\
2	1\\
1	1\\
0	1\\
-1	1\\
-2	1\\
-3	1\\
-4	1\\
-5	1\\
-6	1\\
-7	1\\
-8	1\\
-9	1\\
-10	1\\
}--cycle;
\end{axis}
\end{tikzpicture}%
    \caption{}
    \label{e1}
    \end{subfigure}
    \begin{subfigure}[b]{0.5\textwidth}
\centering
\setlength\fheight{5.5cm}
\setlength\fwidth{7.5cm}
%
%
\definecolor{mycolor1}{rgb}{0.85000,0.33000,0.10000}%
\definecolor{mycolor2}{rgb}{0.85000,0.32500,0.09800}%
\definecolor{mycolor3}{rgb}{0.00000,0.45000,0.74000}%
\definecolor{mycolor4}{rgb}{0.49400,0.18400,0.55600}%
\definecolor{mycolor5}{rgb}{0.30100,0.74500,0.93300}%
\definecolor{mycolor6}{rgb}{0.00000,0.44700,0.74100}%
\begin{tikzpicture}

\begin{axis}[%
width=0.951\fwidth,
height=\fheight,
at={(0\fwidth,0\fheight)},
scale only axis,
unbounded coords=jump,
xmin=-10,
xmax=20,
xlabel style={font=\color{white!15!black}},
xlabel={\textsf{SINDR Threshold (dB)}},
ymin=0,
ymax=1,
ylabel style={font=\color{white!15!black}},
ylabel={\textsf{Outage Probability}},
axis background/.style={fill=white},
axis x line*=bottom,
axis y line*=left,
legend style={at={(0.97,0.03)}, anchor=south east, legend cell align=left, align=left, fill=none, draw=none}
]

\node[right, align=left, rotate=0]
at (axis cs:-8.5,0.45) {\scriptsize\sffamily{IBO Envelope}};
\draw [latex-,black,line width=.7pt] (0,.45) to (3,.45);

\node[right, align=left, rotate=0]
at (axis cs:-8,0.8) {\scriptsize\sffamily{IBO}~$\rightarrow -\infty$~\scriptsize\sffamily{dB}};
\draw [latex-,black,line width=.7pt] (1,.8) to (4.1,.8);

\node[right, align=left, rotate=0]
at (axis cs:8,0.4) {\scriptsize\sffamily{Ideal Hardware}};
\draw [-latex,black,line width=.7pt] (6,.4) to (8.3,.4);

\node[right, align=left, rotate=0]
at (axis cs:10,0.55) {\scriptsize\sffamily{IBO = 0 dB}};
\draw [-latex,black,line width=.7pt] (5.2,.55) to (10,.55);

\node[right, align=left, rotate=0]
at (axis cs:11,0.8) {\scriptsize\sffamily{IBO}~$\rightarrow +\infty$~\scriptsize\sffamily{dB}};
\node[right, align=left, rotate=0]
at (axis cs:11,0.72) {\scriptsize\sffamily{Linear Relaying}};
\draw [-latex,black,line width=.7pt] (8.2,.76) to (11,.76);

\addplot [color=mycolor1, line width=1.5pt]
  table[row sep=crcr]{%
-10	0.00228225204087684\\
-9	0.00358155861940379\\
-8	0.00571708147022532\\
-7	0.00925036988673578\\
-6	0.0151056417776475\\
-5	0.0247683170238947\\
-4	0.0405424653269586\\
-3	0.0658203210736885\\
-2	0.105222185831587\\
-1	0.164311345579336\\
0	0.248430884739729\\
1	0.360251691132251\\
2	0.496249094326802\\
3	0.643809533329829\\
4	0.782267092588541\\
5	0.890374987666466\\
6	0.957319925634003\\
7	0.988219006557005\\
8	0.997975445456377\\
9	0.999825305095256\\
10	0.999994877806344\\
11	0.9999999766634\\
12	0.999999999997299\\
13	1\\
14	1\\
15	nan\\
16	1\\
17	1\\
18	1\\
19	1\\
20	1\\
};
\addlegendentry{\textsf{Analytical}}

\addplot [color=mycolor2,only marks, line width=.5pt, draw=none, mark=*, mark options={solid, fill=mycolor3, blue}]
  table[row sep=crcr]{%
-10	0.00228225204087684\\
-9	0.00358155861940379\\
-8	0.00571708147022532\\
-7	0.00925036988673578\\
-6	0.0151056417776475\\
-5	0.0247683170238947\\
-4	0.0405424653269586\\
-3	0.0658203210736885\\
-2	0.105222185831587\\
-1	0.164311345579336\\
0	0.248430884739729\\
1	0.360251691132251\\
2	0.496249094326802\\
3	0.643809533329829\\
4	0.782267092588541\\
5	0.890374987666466\\
6	0.957319925634003\\
7	0.988219006557005\\
8	0.997975445456377\\
9	0.999825305095256\\
10	0.999994877806344\\
11	0.9999999766634\\
12	0.999999999997299\\
13	1\\
14	1\\
15	nan\\
16	1\\
17	1\\
18	1\\
19	1\\
20	1\\
};
\addlegendentry{\textsf{Monte Carlo}}

\addplot [color=mycolor1, line width=1.5pt, forget plot]
  table[row sep=crcr]{%
-10	0.000584338478844648\\
-9	0.000877454236036246\\
-8	0.00134298546754619\\
-7	0.00208973821481884\\
-6	0.00329432156521203\\
-5	0.00523950890165403\\
-4	0.00836978569200963\\
-3	0.0133676210029603\\
-2	0.0212509108277424\\
-1	0.0334848427731325\\
0	0.0520881140574652\\
1	0.0796924518365517\\
2	0.119486190460359\\
3	0.174942351294061\\
4	0.249212932003378\\
5	0.344091261001478\\
6	0.458548869660638\\
7	0.587104477193004\\
8	0.718732341587998\\
9	0.837606598500579\\
10	0.927271605964426\\
11	0.978574732773759\\
12	0.997089241055431\\
13	0.999931566668283\\
14	0.999999996940629\\
15	nan\\
16	1\\
17	1\\
18	1\\
19	1\\
20	1\\
};
\addplot [color=mycolor4,only marks, line width=.5pt, draw=none, mark=*, mark options={solid, fill=mycolor3, blue}, forget plot]
  table[row sep=crcr]{%
-10	0.000584338478844648\\
-9	0.000877454236036246\\
-8	0.00134298546754619\\
-7	0.00208973821481884\\
-6	0.00329432156521203\\
-5	0.00523950890165403\\
-4	0.00836978569200963\\
-3	0.0133676210029603\\
-2	0.0212509108277424\\
-1	0.0334848427731325\\
0	0.0520881140574652\\
1	0.0796924518365517\\
2	0.119486190460359\\
3	0.174942351294061\\
4	0.249212932003378\\
5	0.344091261001478\\
6	0.458548869660638\\
7	0.587104477193004\\
8	0.718732341587998\\
9	0.837606598500579\\
10	0.927271605964426\\
11	0.978574732773759\\
12	0.997089241055431\\
13	0.999931566668283\\
14	0.999999996940629\\
15	nan\\
16	1\\
17	1\\
18	1\\
19	1\\
20	1\\
};
\addplot [color=mycolor1, line width=1.5pt, forget plot]
  table[row sep=crcr]{%
-10	0.000859726173744191\\
-9	0.00130303200122373\\
-8	0.00201271558730687\\
-7	0.00316012596331394\\
-6	0.00502575754156287\\
-5	0.00806267874094824\\
-4	0.012989080963557\\
-3	0.0209152707652119\\
-2	0.0335025838397154\\
-1	0.053132810014098\\
0	0.0830302191313337\\
1	0.127219359117319\\
2	0.190126465282074\\
3	0.275576383514541\\
4	0.384994248450533\\
5	0.51495853707327\\
6	0.655032749534944\\
7	0.787875786801107\\
8	0.893960544421346\\
9	0.960803889643138\\
10	0.990851513668436\\
11	0.998998198599061\\
12	0.999973628662059\\
13	0.999999976516186\\
14	1\\
15	nan\\
16	1\\
17	1\\
18	1\\
19	1\\
20	1\\
};
\addplot [color=mycolor5,only marks, line width=.5pt, draw=none, mark=*, mark options={solid, fill=mycolor3, blue}, forget plot]
  table[row sep=crcr]{%
-10	0.000859726173744191\\
-9	0.00130303200122373\\
-8	0.00201271558730687\\
-7	0.00316012596331394\\
-6	0.00502575754156287\\
-5	0.00806267874094824\\
-4	0.012989080963557\\
-3	0.0209152707652119\\
-2	0.0335025838397154\\
-1	0.053132810014098\\
0	0.0830302191313337\\
1	0.127219359117319\\
2	0.190126465282074\\
3	0.275576383514541\\
4	0.384994248450533\\
5	0.51495853707327\\
6	0.655032749534944\\
7	0.787875786801107\\
8	0.893960544421346\\
9	0.960803889643138\\
10	0.990851513668436\\
11	0.998998198599061\\
12	0.999973628662059\\
13	0.999999976516186\\
14	1\\
15	nan\\
16	1\\
17	1\\
18	1\\
19	1\\
20	1\\
};
\addplot [color=mycolor1, line width=1.5pt, forget plot]
  table[row sep=crcr]{%
-10	0.000576970291472745\\
-9	0.000864833996873537\\
-8	0.0013206192309354\\
-7	0.00204893725855226\\
-6	0.00321822030858177\\
-5	0.00509545150903978\\
-4	0.00809509328813018\\
-3	0.0128437556242681\\
-2	0.0202582977150715\\
-1	0.0316275232547759\\
0	0.0486755276688826\\
1	0.0735686776317052\\
2	0.108811869368536\\
3	0.156971731530151\\
4	0.220178915413479\\
5	0.29941532336856\\
6	0.393695405793138\\
7	0.499390650294424\\
8	0.610067165687127\\
9	0.717202648236842\\
10	0.811904717387887\\
11	0.887247412680424\\
12	0.940290026856521\\
13	0.972695833576302\\
14	0.989496834651675\\
15	0.996702797464466\\
16	0.999185247866507\\
17	0.999848397687897\\
18	0.999979911783006\\
19	0.999998234426502\\
20	0.999999905943003\\
};
\addplot [color=mycolor6,only marks, line width=.5pt, draw=none, mark=*, mark options={solid, fill=mycolor3, blue}, forget plot]
  table[row sep=crcr]{%
-10	0.000576970291472745\\
-9	0.000864833996873537\\
-8	0.0013206192309354\\
-7	0.00204893725855226\\
-6	0.00321822030858177\\
-5	0.00509545150903978\\
-4	0.00809509328813018\\
-3	0.0128437556242681\\
-2	0.0202582977150715\\
-1	0.0316275232547759\\
0	0.0486755276688826\\
1	0.0735686776317052\\
2	0.108811869368536\\
3	0.156971731530151\\
4	0.220178915413479\\
5	0.29941532336856\\
6	0.393695405793138\\
7	0.499390650294424\\
8	0.610067165687127\\
9	0.717202648236842\\
10	0.811904717387887\\
11	0.887247412680424\\
12	0.940290026856521\\
13	0.972695833576302\\
14	0.989496834651675\\
15	0.996702797464466\\
16	0.999185247866507\\
17	0.999848397687897\\
18	0.999979911783006\\
19	0.999998234426502\\
20	0.999999905943003\\
};

\addplot[area legend, draw=black, fill=mycolor1, fill opacity=0.2, forget plot]
table[row sep=crcr] {%
x	y\\
-10	0.00228225204087684\\
-9	0.00358155861940379\\
-8	0.00571708147022532\\
-7	0.00925036988673578\\
-6	0.0151056417776475\\
-5	0.0247683170238947\\
-4	0.0405424653269586\\
-3	0.0658203210736885\\
-2	0.105222185831587\\
-1	0.164311345579336\\
0	0.248430884739729\\
1	0.360251691132251\\
2	0.496249094326802\\
3	0.643809533329829\\
4	0.782267092588541\\
5	0.890374987666466\\
6	0.957319925634003\\
7	0.988219006557005\\
8	0.997975445456377\\
9	0.999825305095256\\
10	0.999994877806344\\
11	0.9999999766634\\
12	0.999999999997299\\
13	1\\
14	1\\
14	0.999999996940629\\
13	0.999931566668283\\
12	0.997089241055431\\
11	0.978574732773759\\
10	0.927271605964426\\
9	0.837606598500579\\
8	0.718732341587998\\
7	0.587104477193004\\
6	0.458548869660638\\
5	0.344091261001478\\
4	0.249212932003378\\
3	0.174942351294061\\
2	0.119486190460359\\
1	0.0796924518365517\\
0	0.0520881140574652\\
-1	0.0334848427731325\\
-2	0.0212509108277424\\
-3	0.0133676210029603\\
-4	0.00836978569200963\\
-5	0.00523950890165403\\
-6	0.00329432156521203\\
-7	0.00208973821481884\\
-8	0.00134298546754619\\
-9	0.000877454236036246\\
-10	0.000584338478844648\\
}--cycle;
\end{axis}
\end{tikzpicture}%
    \caption{}
    \label{e2}
    \end{subfigure}
    \caption{\textcolor{black}{Performance results of the outage probability at {\scriptsize{\textsf{SNR}}} = -10 dB: Fig.~\ref{e1} illustrates the impacts of the IBO range on the outage while the {\scriptsize{\textsf{ILR}}} is fixed at -15 dB. Fig.~\ref{e2} provides the effects of the {\scriptsize{\textsf{ILR}}} range on the outage while the IBO is fixed at 0 dB. .}}
    \label{fig1}
\end{figure}
\textcolor{black}{The dependence of the outage on the joint effects of the HPA nonlinearities and IQ imbalance is illustrated by Fig.~\ref{fig1}. In Fig.~\ref{e1}, we observe that the system is vulnerable to the HPA nonlinearities distortion wherein the outage saturates when the IBO, which is the maximum power delivered by the PA, is very low and vice versa. In fact, when the IBO is lower, the PA is unable to provide sufficient power to the relay to amplify the signal which in turn causes signal clipping resulting in information losses and higher outage. In other terms, the PA devices are nonlinear devices and high power amplification may introduce clipping which is translated in frequency domain as spectral regrowth which in turn causes interference between the adjacent subcarriers or aliasing resulting in information losses. The same analysis follows for the effect of the {\scriptsize{\textsf{ILR}}} on the outage which is illustrated by Fig.~\ref{e2}. We observe that the system performance gets much better for low {\scriptsize{\textsf{ILR}}} and vice versa. In fact, the {\scriptsize{\textsf{ILR}}} measures the severity of the mismatch at the receiver and as long as the mismatch is pronounced, the outage is higher.}

\begin{figure}[H]
\centering
\setlength\fheight{5.5cm}
\setlength\fwidth{7.5cm}
\input{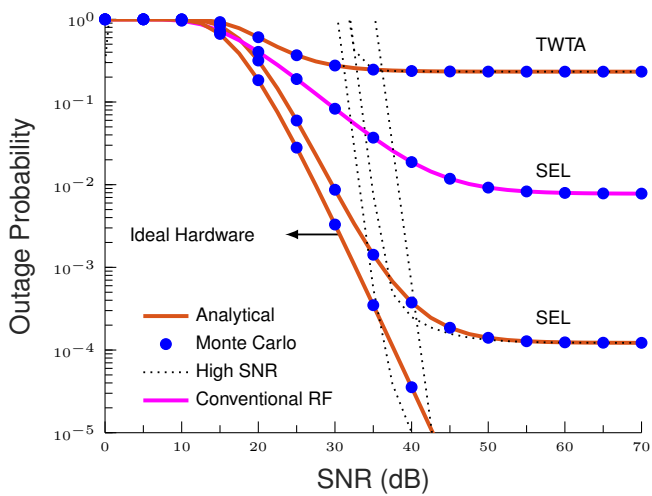}
    \caption{\textcolor{black}{Performance results: Fig.~\ref{e3} illustrates a comparison between the SEL and TWTA and their effects on the outage. We also introduce a comparison with a conventional RF relaying system \cite{22}}.}
    \label{e3}
\end{figure}
\textcolor{black}{In Fig.~\ref{e3}, we observe that the impact of the hardware impairments is not significant at low {\scriptsize{\textsf{SNR}}}, however, this impact becomes remarkable at high {\scriptsize{\textsf{SNR}}} wherein the distortion creates an irreducible outage floor. Besides, we note that the introduction of the wireless optical signaling and multiple relays decreases the outage much lower than the conventional RF relaying systems \cite{22} at low {\scriptsize{\textsf{SNR}}}. At high {\scriptsize{\textsf{SNR}}}, even though both systems are saturated by the outage floor, the proposed system achieves better outage due to the higher diversity order compared to the conventional RF system.}

\begin{figure}[t]
\centering
\setlength\fheight{5.5cm}
\setlength\fwidth{7.5cm}
\input{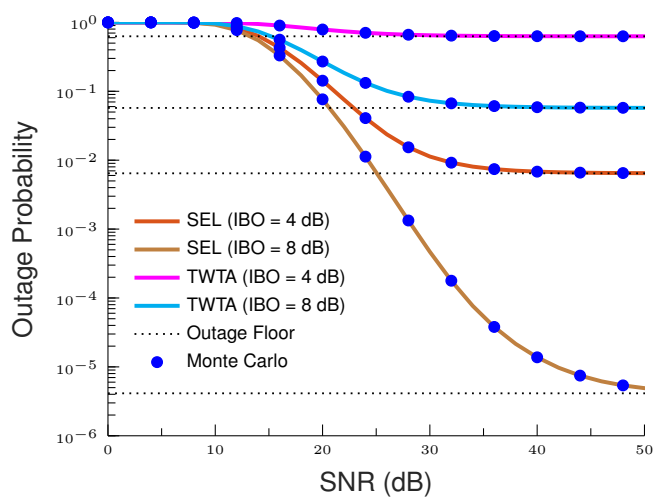}
    \caption{\textcolor{black}{Outage probability versus the average SNR for different values of IBO, and $\rho_2$ = -15 dB. The impairment model is SEL and IM/DD is the detection technique.}}
    \label{op1}
\end{figure}

\begin{figure}[b]
\centering
\setlength\fheight{5.5cm}
\setlength\fwidth{7.5cm}
\input{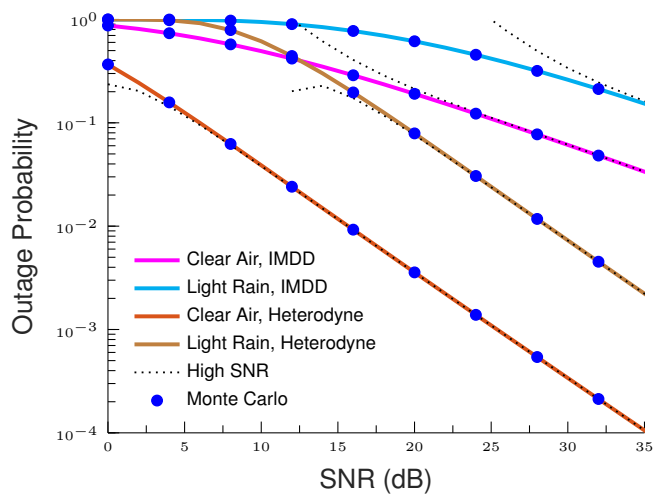}
    \caption{\textcolor{black}{Outage probability versus the average SNR for different weather conditions, and detection techniques under ideal hardware scenario.}}
    \label{op2}
\end{figure}


Fig.~\ref{op1} illustrates the impact the IBO on the probability of outage under SEL and TWTA hardware impairments. We observe that the impairments create an outage floor that degrades the outage performance mainly at high SNR. For a given IBO value, we note that the system performance is more vulnerable to TWTA than to SEL impairments. For an IBO = 8 dB, the performance under SEL model yields an outage of roughly $10^{-5}$ while the outage is around 0.01 for TWTA which is relatively higher. On the other hand, we observe that by increasing the IBO from 4 to 8 dB, the outage performance significantly improves for TWTA and SEL. In fact, when the IBO value is relatively lower, the amplifier cannot deliver enough power for the signal amplification which introduces the clipping to the signal peaks. Thereby, this distortion leads to the saturation of the outage performance.

Fig.~\ref{op2} illustrates the joint effects of the FSO atmospheric pathloss and the optical receiver detection technique on the outage probability. As expected, the system performs better in clear than rainy weather. In fact, when the signal hits the rain droplets, it is significantly scattered in different directions leading to decrease the received power. Hence, the outage performance gets much worse. On the other side, the performance can be further improved under heterodyne detection than IMDD mode. This result can be interpreted by the diversity gain (\ref{Gd}) which is better for heterodyne than IMDD method.


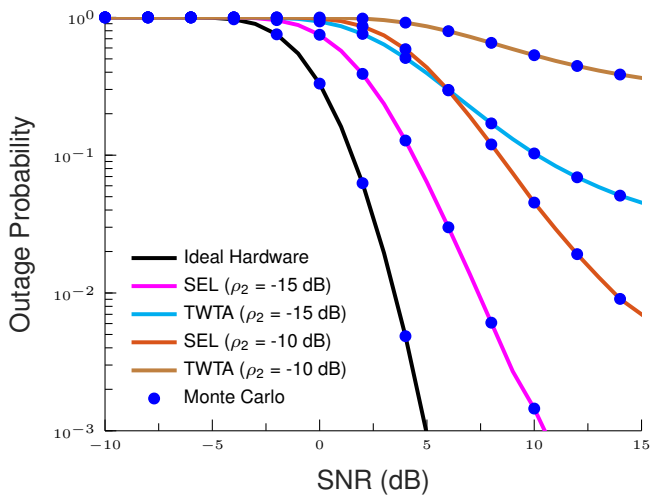
\begin{figure}[t]
\centering
\setlength\fheight{5.5cm}
\setlength\fwidth{7.5cm}
%
%
\definecolor{mycolor1}{rgb}{0.00000,0.44700,0.74100}%
\definecolor{mycolor2}{rgb}{0.85000,0.33000,0.10000}%
\definecolor{mycolor3}{rgb}{0.85000,0.32500,0.09800}%
\definecolor{mycolor4}{rgb}{0.00000,0.45000,0.74000}%
\definecolor{mycolor5}{rgb}{1.00000,0.00000,1.00000}%
\begin{tikzpicture}

\begin{axis}[%
width=0.951\fwidth,
height=\fheight,
at={(0\fwidth,0\fheight)},
scale only axis,
xmin=-10,
xmax=15,
axis x line*=bottom,
axis y line*=left,
xlabel style={font=\color{white!15!black}},
xlabel={\sffamily{SNR (dB)}},
ymode=log,
ymin=0.001,
ymax=1,
yminorticks=true,
ylabel style={font=\color{white!15!black}},
ylabel={\sffamily{Outage Probability}},
axis background/.style={fill=white},
legend style={at={(0.03,0.03)}, anchor=south west, legend cell align=left, align=left, draw=white!15!black,draw=none,fill=none}
]
\addplot [color=black, line width=1.5pt]
  table[row sep=crcr]{%
-10	1\\
-9	1\\
-8	0.99999\\
-7	0.99983\\
-6	0.99884\\
-5	0.9931\\
-4	0.96834\\
-3	0.89695\\
-2	0.75353\\
-1	0.5468\\
0	0.33079\\
1	0.16099\\
2	0.06267\\
3	0.01951\\
4	0.00486\\
5	0.00094\\
6	0.00024\\
7	3e-05\\
8	0\\
9	0\\
10	0\\
11	0\\
12	0\\
13	0\\
14	0\\
15	0\\
16	0\\
17	0\\
18	0\\
19	0\\
20	0\\
};
\addlegendentry{\sffamily{Ideal Hardware}}

\addplot [color=mycolor5, solid , line width=1.5pt]
  table[row sep=crcr]{%
-10	1\\
-9	1\\
-8	1\\
-7	0.99999\\
-6	0.99993\\
-5	0.9995\\
-4	0.99744\\
-3	0.98791\\
-2	0.95527\\
-1	0.8798\\
0	0.74795\\
1	0.57198\\
2	0.38925\\
3	0.23583\\
4	0.12778\\
5	0.06359\\
6	0.02997\\
7	0.01371\\
8	0.00609\\
9	0.0027\\
10	0.00145\\
11	0.00068\\
12	0.00051\\
13	0.00029\\
14	0.00021\\
15	0.00014\\
16	0.00011\\
17	0.00011\\
18	8e-05\\
19	8e-05\\
20	6e-05\\
};
\addlegendentry{\sffamily{$\text{SEL (}\rho{}_\text{2}\text{ = -15 dB)}$}}

\addplot [color=cyan, line width=1.5pt]
  table[row sep=crcr]{%
-10	1\\
-9	1\\
-8	1\\
-7	1\\
-6	0.99999\\
-5	0.99994\\
-4	0.99962\\
-3	0.99818\\
-2	0.99263\\
-1	0.97504\\
0	0.93479\\
1	0.86287\\
2	0.75906\\
3	0.63563\\
4	0.50762\\
5	0.39339\\
6	0.29772\\
7	0.22476\\
8	0.17016\\
9	0.13083\\
10	0.10293\\
11	0.08334\\
12	0.06912\\
13	0.0588\\
14	0.05089\\
15	0.04528\\
16	0.04127\\
17	0.03824\\
18	0.03637\\
19	0.03461\\
20	0.03329\\
};
\addlegendentry{\sffamily{$\text{TWTA (}\rho{}_\text{2}\text{ = -15 dB)}$}}

\addplot [color=mycolor2, line width=1.5pt]
  table[row sep=crcr]{%
-10	1\\
-9	1\\
-8	1\\
-7	1\\
-6	1\\
-5	1\\
-4	0.99994\\
-3	0.99978\\
-2	0.99882\\
-1	0.99467\\
0	0.97974\\
1	0.94043\\
2	0.86198\\
3	0.74018\\
4	0.58781\\
5	0.4324\\
6	0.29551\\
7	0.19092\\
8	0.1197\\
9	0.07397\\
10	0.04531\\
11	0.0291\\
12	0.01914\\
13	0.01304\\
14	0.00906\\
15	0.007\\
16	0.00537\\
17	0.00428\\
18	0.00351\\
19	0.00305\\
20	0.00266\\
};
\addlegendentry{\sffamily{$\text{SEL (}\rho{}_\text{2}\text{ = -10 dB)}$}}

\addplot [color=brown, line width=1.5pt]
  table[row sep=crcr]{%
-10	1\\
-9	1\\
-8	1\\
-7	1\\
-6	1\\
-5	1\\
-4	1\\
-3	0.99997\\
-2	0.9999\\
-1	0.9994\\
0	0.9979\\
1	0.99297\\
2	0.98072\\
3	0.95597\\
4	0.9162\\
5	0.86117\\
6	0.79462\\
7	0.72339\\
8	0.6534\\
9	0.58846\\
10	0.53166\\
11	0.48309\\
12	0.44411\\
13	0.4113\\
14	0.38493\\
15	0.36363\\
16	0.34732\\
17	0.33408\\
18	0.32336\\
19	0.31501\\
20	0.30805\\
};
\addlegendentry{\sffamily{$\text{TWTA (}\rho{}_\text{2}\text{ = -10 dB)}$}}

\addplot [color=mycolor3, line width=.5pt, only marks,draw=none, mark=*, mark options={solid, fill=mycolor4, blue}]
  table[row sep=crcr]{%
-10	1\\
-8	0.99999\\
-6	0.99884\\
-4	0.96834\\
-2	0.75353\\
0	0.33079\\
2	0.06267\\
4	0.00486\\
6	0.00024\\
8	0\\
10	0\\
12	0\\
14	0\\
16	0\\
18	0\\
20	0\\
};
\addlegendentry{\sffamily{Monte Carlo}}

\addplot [color=mycolor3, line width=.5pt, only marks,draw=none, mark=*, mark options={solid, fill=mycolor4, blue}]
  table[row sep=crcr]{%
-10	1\\
-8	1\\
-6	0.99993\\
-4	0.99744\\
-2	0.95527\\
0	0.74795\\
2	0.38925\\
4	0.12778\\
6	0.02997\\
8	0.00609\\
10	0.00145\\
12	0.00051\\
14	0.00021\\
16	0.00011\\
18	8e-05\\
20	6e-05\\
};

\addplot [color=mycolor3, line width=.5pt, only marks,draw=none, mark=*, mark options={solid, fill=mycolor4, blue}]
  table[row sep=crcr]{%
-10	1\\
-8	1\\
-6	0.99999\\
-4	0.99962\\
-2	0.99263\\
0	0.93479\\
2	0.75906\\
4	0.50762\\
6	0.29772\\
8	0.17016\\
10	0.10293\\
12	0.06912\\
14	0.05089\\
16	0.04127\\
18	0.03637\\
20	0.03329\\
};

\addplot [color=mycolor3, line width=.5pt, only marks,draw=none, mark=*, mark options={solid, fill=mycolor4, blue}]
  table[row sep=crcr]{%
-10	1\\
-8	1\\
-6	1\\
-4	0.99994\\
-2	0.99882\\
0	0.97974\\
2	0.86198\\
4	0.58781\\
6	0.29551\\
8	0.1197\\
10	0.04531\\
12	0.01914\\
14	0.00906\\
16	0.00537\\
18	0.00351\\
20	0.00266\\
};

\addplot [color=mycolor3, line width=.5pt, only marks,draw=none, mark=*, mark options={solid, fill=mycolor4, blue}]
  table[row sep=crcr]{%
-10	1\\
-8	1\\
-6	1\\
-4	1\\
-2	0.9999\\
0	0.9979\\
2	0.98072\\
4	0.9162\\
6	0.79462\\
8	0.6534\\
10	0.53166\\
12	0.44411\\
14	0.38493\\
16	0.34732\\
18	0.32336\\
20	0.30805\\
};

\end{axis}
\end{tikzpicture}%
    \caption{Probability of outage for SEL and TWTA for different values of $\rho_2$. The IBO is -3 dB while the impairment model is SEL and IM/DD is the detection technique.}
    \label{op3}
\end{figure}

Fig.~\ref{op3} illustrates a comparison between the joint effects of HPA and IQI, and the  ideal hardware. For a given $\rho_2 = -15$ dB, we observe that the system is more susceptible to TWTA rather than SEL. On the other hand, if we consider both SEL and TWTA, as the ILR $\rho_2$ increases, the performance gets much worse. The severity of joint effects of impairments is measured in comparison with the ideal hardware in two ways. The first way states that for a given $10^{-3}$ outage, the system experiences a power loss of roughly 5 dB in comparison with the best case scenario of hardware impairment (SEL, $\rho_2$ = - 15 dB). This power loss becomes more pronounced for the worst case scenario (TWTA, $\rho_2$ = -10 dB). The second way shows that the performance experiences an irreducible outage floor that degrades the diversity gain, most importantly at high SNR.

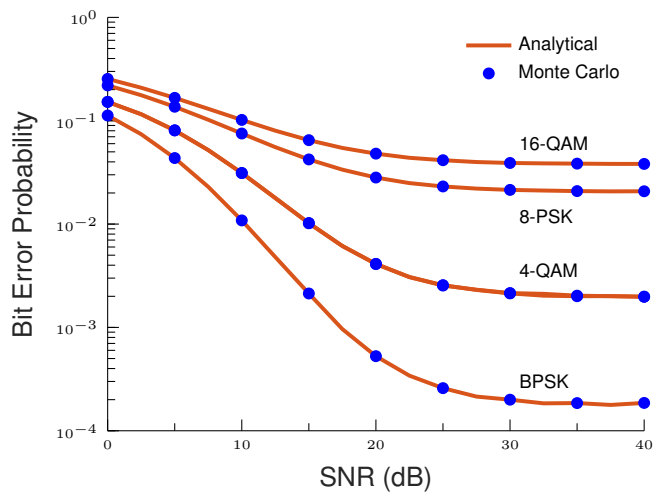
\begin{figure}[t]
\centering
\setlength\fheight{5.5cm}
\setlength\fwidth{7.5cm}
%
%
\definecolor{mycolor1}{rgb}{0.85000,0.33000,0.10000}%
\definecolor{mycolor2}{rgb}{0.85000,0.32500,0.09800}%
\definecolor{mycolor3}{rgb}{0.00000,0.45000,0.74000}%
\definecolor{mycolor4}{rgb}{0.49400,0.18400,0.55600}%
\definecolor{mycolor5}{rgb}{0.30100,0.74500,0.93300}%
\definecolor{mycolor6}{rgb}{0.00000,0.44700,0.74100}%
\definecolor{mycolor7}{rgb}{0.92900,0.69400,0.12500}%
\begin{tikzpicture}

\begin{axis}[%
width=0.951\fwidth,
height=\fheight,
at={(0\fwidth,0\fheight)},
scale only axis,
xmin=0,
xmax=40,
xlabel style={font=\color{white!15!black}},
xlabel={\textsf{SNR (dB)}},
ymode=log,
ymin=0.0001,
ymax=1,
yminorticks=true,
ylabel style={font=\color{white!15!black}},
ylabel={\textsf{Bit Error Probability}},
axis background/.style={fill=white},
axis x line*=bottom,
axis y line*=left,
legend style={legend cell align=left, align=left, fill=none, draw=none}
]

\node[right, align=left, rotate=0]
at (axis cs:30,0.0003) {\scriptsize\sffamily{BPSK}};

\node[right, align=left, rotate=0]
at (axis cs:30,0.0035) {\scriptsize\sffamily{4-QAM}};

\node[right, align=left, rotate=0]
at (axis cs:30,0.012) {\scriptsize\sffamily{8-PSK}};

\node[right, align=left, rotate=0]
at (axis cs:30,0.06) {\scriptsize\sffamily{16-QAM}};

\addplot [color=mycolor1, line width=1.5pt]
  table[row sep=crcr]{%
0	0.152199036949685\\
2.5	0.11569141392303\\
5	0.0806706798485238\\
7.5	0.0520025114136063\\
10	0.0312160854852828\\
12.5	0.0178150092273838\\
15	0.0102254695109798\\
17.5	0.00612527866062733\\
20	0.0041190146411927\\
22.5	0.00306256519294709\\
25	0.00254425063925581\\
27.5	0.00231784973945192\\
30	0.00214862761196182\\
32.5	0.00211323587085332\\
35	0.00202922221536588\\
37.5	0.00200539526124202\\
40	0.00198511494626164\\
42.5	0.0019739366119415\\
45	0.00196498433509391\\
47.5	0.00200744525082026\\
50	0.00198830085919059\\
};
\addlegendentry{\textsf{Analytical}}

\addplot [color=mycolor2, line width=.5pt, only marks,draw=none, mark=*, mark options={solid, fill=mycolor3, blue}]
  table[row sep=crcr]{%
0	0.152199036949685\\
5	0.0806706798485238\\
10	0.0312160854852828\\
15	0.0102254695109798\\
20	0.0041190146411927\\
25	0.00254425063925581\\
30	0.00214862761196182\\
35	0.00202922221536588\\
40	0.00198511494626164\\
45	0.00196498433509391\\
50	0.00198830085919059\\
};
\addlegendentry{\textsf{Monte Carlo}}

\addplot [color=mycolor1, line width=1.5pt, forget plot]
  table[row sep=crcr]{%
0	0.112348383374564\\
2.5	0.0740533672157973\\
5	0.0435448106028303\\
7.5	0.0229335159389658\\
10	0.0108532477856869\\
12.5	0.00479706320655515\\
15	0.00212954877350865\\
17.5	0.00096314868619067\\
20	0.000526823313556448\\
22.5	0.000341324743429108\\
25	0.000258448842476855\\
27.5	0.00021417815771794\\
30	0.000199866656620026\\
32.5	0.000184235144975151\\
35	0.000185308368657196\\
37.5	0.000177837604479449\\
40	0.000185970794702728\\
42.5	0.00017602847029697\\
45	0.000178611339108022\\
47.5	0.000177031362834737\\
50	0.000176177705743509\\
};
\addplot [color=mycolor4, line width=.5pt, draw=none, mark=*, mark options={solid, fill=mycolor3, blue}, forget plot]
  table[row sep=crcr]{%
0	0.112348383374564\\
5	0.0435448106028303\\
10	0.0108532477856869\\
15	0.00212954877350865\\
20	0.000526823313556448\\
25	0.000258448842476855\\
30	0.000199866656620026\\
35	0.000185308368657196\\
40	0.000185970794702728\\
45	0.000178611339108022\\
50	0.000176177705743509\\
};
\addplot [color=mycolor1, line width=1.5pt, forget plot]
  table[row sep=crcr]{%
0	0.152180909860942\\
2.5	0.115471655914997\\
5	0.0807219799936904\\
7.5	0.0520134155965917\\
10	0.0311503581743319\\
12.5	0.0179241114070667\\
15	0.0101856718081159\\
17.5	0.00614490682058392\\
20	0.0041052167624965\\
22.5	0.00307774027384492\\
25	0.00256326383871362\\
27.5	0.00230376661277888\\
30	0.00213476850899\\
32.5	0.00202932420130619\\
35	0.00200423128859504\\
37.5	0.00200894580156195\\
40	0.0019796365380819\\
42.5	0.00199222518576472\\
45	0.00197619450121683\\
47.5	0.00197723537951647\\
50	0.00201149135358415\\
};
\addplot [color=mycolor5, line width=.5pt, draw=none, mark=*, mark options={solid, fill=mycolor3, blue}, forget plot]
  table[row sep=crcr]{%
0	0.152180909860942\\
5	0.0807219799936904\\
10	0.0311503581743319\\
15	0.0101856718081159\\
20	0.0041052167624965\\
25	0.00256326383871362\\
30	0.00213476850899\\
35	0.00200423128859504\\
40	0.0019796365380819\\
45	0.00197619450121683\\
50	0.00201149135358415\\
};
\addplot [color=mycolor1, line width=1.5pt, forget plot]
  table[row sep=crcr]{%
0	0.253869175824832\\
2.5	0.208631372980032\\
5	0.166504272638459\\
7.5	0.130384429555244\\
10	0.101862686742682\\
12.5	0.0801492946556792\\
15	0.0648780518920593\\
17.5	0.05454445710948\\
20	0.0480344906642187\\
22.5	0.0437904699961869\\
25	0.0414947996981467\\
27.5	0.0398565001379345\\
30	0.039019112378106\\
32.5	0.0386725352819835\\
35	0.0384668887676191\\
37.5	0.0382282848773659\\
40	0.0382082184489361\\
42.5	0.0380056703105435\\
45	0.0380009269694525\\
47.5	0.0380917402905122\\
50	0.0379610170746011\\
};
\addplot [color=mycolor6, line width=.5pt, draw=none, mark=*, mark options={solid, fill=mycolor3, blue}, forget plot]
  table[row sep=crcr]{%
0	0.253869175824832\\
5	0.166504272638459\\
10	0.101862686742682\\
15	0.0648780518920593\\
20	0.0480344906642187\\
25	0.0414947996981467\\
30	0.039019112378106\\
35	0.0384668887676191\\
40	0.0382082184489361\\
45	0.0380009269694525\\
50	0.0379610170746011\\
};
\addplot [color=mycolor1, line width=1.5pt, forget plot]
  table[row sep=crcr]{%
0	0.219813561826516\\
2.5	0.176666587139582\\
5	0.136585618136172\\
7.5	0.102179880929612\\
10	0.0752363491030439\\
12.5	0.0555092380837796\\
15	0.042226009675176\\
17.5	0.0336092481610284\\
20	0.0282056725756495\\
22.5	0.0249156205520523\\
25	0.0231222050149038\\
27.5	0.0220477598809385\\
30	0.0214159580390635\\
32.5	0.0211092549168048\\
35	0.0208522213558318\\
37.5	0.0207227419168144\\
40	0.0207717616867842\\
42.5	0.0206486506176844\\
45	0.0206195559341811\\
47.5	0.0206294757052846\\
50	0.0206928646302216\\
};
\addplot [color=mycolor7, line width=.5pt, draw=none, mark=*, mark options={solid, fill=mycolor3, blue}, forget plot]
  table[row sep=crcr]{%
0	0.219813561826516\\
5	0.136585618136172\\
10	0.0752363491030439\\
15	0.042226009675176\\
20	0.0282056725756495\\
25	0.0231222050149038\\
30	0.0214159580390635\\
35	0.0208522213558318\\
40	0.0207717616867842\\
45	0.0206195559341811\\
50	0.0206928646302216\\
};
\end{axis}
\end{tikzpicture}%
    \caption{\textcolor{black}{Performance results of the bit error probability for various modulation schemes with hardware impairment: IBO = 0 dB and ILR = -15 dB.}}
    \label{bep1}
\end{figure}

Fig.~\ref{bep1} provides the error performance for different modulation schemes and for ideal hardware. Although the results are expected, this plot is presented for the purpose to check the accuracy of the analytical derivation. 


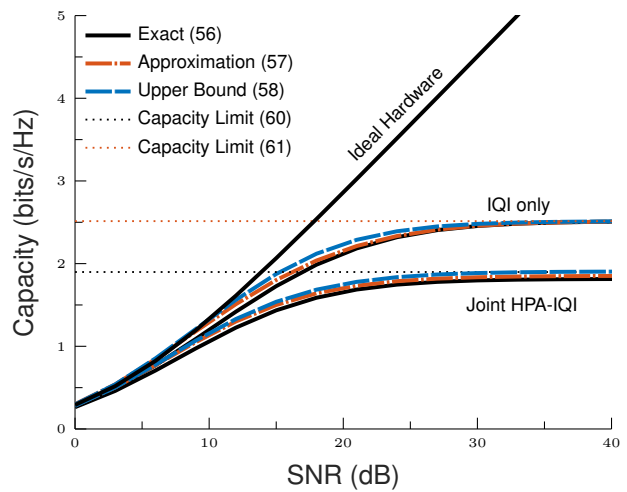
\begin{figure}[b]
\centering
\setlength\fheight{5.5cm}
\setlength\fwidth{7.5cm}
%
%
\definecolor{mycolor1}{rgb}{0.00000,0.44700,0.74100}%
\definecolor{mycolor2}{rgb}{0.85000,0.33000,0.10000}%
\begin{tikzpicture}

\begin{axis}[%
width=0.951\fwidth,
height=\fheight,
at={(0\fwidth,0\fheight)},
scale only axis,
xmin=0,
xmax=40,
axis x line*=bottom,
axis y line*=left,
xlabel style={font=\color{white!15!black}},
xlabel={\sffamily{SNR (dB)}},
ymin=0,
ymax=5,
ylabel style={font=\color{white!15!black}},
ylabel={\sffamily{Capacity (bits/s/Hz)}},
axis background/.style={fill=white},
legend style={at={(0,1)}, anchor=north west, legend cell align=left, align=left, draw=white!15!black,draw=none,fill=none}
]
\node[right, align=left, rotate=45]
at (axis cs:20.1,3.2) {\scriptsize\sffamily{Ideal Hardware}};
\node[right, align=left]
at (axis cs:30,2.7) {\scriptsize\sffamily{IQI only}};
\node[right, align=left]
at (axis cs:28.5,1.5) {\scriptsize\sffamily{Joint HPA-IQI}};
\addplot [color=black, line width=1.5pt]
  table[row sep=crcr]{%
0	0.278235511249795\\
3	0.492278997431635\\
6	0.769770773234962\\
9	1.08901454632495\\
12	1.41921791853267\\
15	1.72741946782181\\
18	1.98684052124083\\
21	2.18381716218016\\
24	2.31942300515215\\
27	2.40499722669329\\
30	2.45525325034468\\
33	2.48318298044839\\
36	2.49810512827053\\
39	2.50587097803469\\
42	2.50984658966005\\
45	2.51186214953533\\
48	2.51287844136441\\
51	2.51338937950447\\
54	2.51364586012726\\
57	2.51377450753515\\
60	2.51383900985685\\
63	2.51387134411291\\
66	2.51388755126784\\
69	2.51389567449877\\
72	2.5138997458619\\
};
\addlegendentry{\sffamily{Exact (\ref{cap})}}

\addplot [color=mycolor2, dash pattern={on 10pt off 1pt on 1pt off 1pt}, line width=1.5pt]
  table[row sep=crcr]{%
0	0.300342104828943\\
3	0.539857085897619\\
6	0.844891933114302\\
9	1.18096730027252\\
12	1.51004200614107\\
15	1.801536117596\\
18	2.03747631811369\\
21	2.21306510308001\\
24	2.33386144084717\\
27	2.41117312131782\\
30	2.45758176804948\\
33	2.48397174794812\\
36	2.49834959625858\\
39	2.50594148801945\\
42	2.50986585293354\\
45	2.51186722368615\\
48	2.51287974887455\\
51	2.51338971232491\\
54	2.51364594429699\\
57	2.51377452875036\\
60	2.51383901519508\\
63	2.513871345455\\
66	2.51388755160509\\
69	2.51389567458351\\
72	2.51389974588318\\
};
\addlegendentry{\sffamily{Approximation (\ref{capapp})}}

\addplot [color=mycolor1, dash pattern={on 10pt off 1pt on 0pt off 0pt} , line width=1.5pt]
  table[row sep=crcr]{%
0	0.297010141462029\\
3	0.539524851845297\\
6	0.854422542141943\\
9	1.2088942847604\\
12	1.56212178543713\\
15	1.87501417206849\\
18	2.11958646520247\\
21	2.28784595903571\\
24	2.39125282506822\\
27	2.44968230151995\\
30	2.48096368863051\\
33	2.49719668960147\\
36	2.5054795268082\\
39	2.50966874189771\\
42	2.51177799181125\\
45	2.51283756665194\\
48	2.51336922853103\\
51	2.5136358458074\\
54	2.51376950998393\\
57	2.51383651056383\\
60	2.51387009286159\\
63	2.51388692449911\\
66	2.51389536045635\\
69	2.51389958848946\\
72	2.51390170753548\\
};
\addlegendentry{\sffamily{Upper Bound (\ref{upper})}}

\addplot [color=black, dotted, line width=.7pt]
  table[row sep=crcr]{%
0	1.89841810879489\\
3	1.89841810879489\\
6	1.89841810879489\\
9	1.89841810879489\\
12	1.89841810879489\\
15	1.89841810879489\\
18	1.89841810879489\\
21	1.89841810879489\\
24	1.89841810879489\\
27	1.89841810879489\\
30	1.89841810879489\\
33	1.89841810879489\\
36	1.89841810879489\\
39	1.89841810879489\\
42	1.89841810879489\\
45	1.89841810879489\\
48	1.89841810879489\\
51	1.89841810879489\\
54	1.89841810879489\\
57	1.89841810879489\\
60	1.89841810879489\\
63	1.89841810879489\\
66	1.89841810879489\\
69	1.89841810879489\\
72	1.89841810879489\\
};
\addlegendentry{\sffamily{Capacity Limit (\ref{inf})}}

\addplot [color=mycolor2,dotted, line width=.7pt ]
  table[row sep=crcr]{%
0	2.51390383667526\\
3	2.51390383667526\\
6	2.51390383667526\\
9	2.51390383667526\\
12	2.51390383667526\\
15	2.51390383667526\\
18	2.51390383667526\\
21	2.51390383667526\\
24	2.51390383667526\\
27	2.51390383667526\\
30	2.51390383667526\\
33	2.51390383667526\\
36	2.51390383667526\\
39	2.51390383667526\\
42	2.51390383667526\\
45	2.51390383667526\\
48	2.51390383667526\\
51	2.51390383667526\\
54	2.51390383667526\\
57	2.51390383667526\\
60	2.51390383667526\\
63	2.51390383667526\\
66	2.51390383667526\\
69	2.51390383667526\\
72	2.51390383667526\\
};
\addlegendentry{\sffamily{Capacity Limit (\ref{max})}}

\addplot [color=black, line width=1.5pt, forget plot]
  table[row sep=crcr]{%
0	0.263429662839151\\
3	0.460765654934606\\
6	0.707269279389116\\
9	0.974127103423097\\
12	1.22630476797452\\
15	1.4349830284731\\
18	1.58725849092396\\
21	1.68669250199317\\
24	1.746038421886\\
27	1.77918392327909\\
30	1.79688509728022\\
33	1.80607795327956\\
36	1.81077479395809\\
39	1.81315276456666\\
42	1.81435082756658\\
45	1.81495288602306\\
48	1.81525503765161\\
51	1.81540657517882\\
54	1.81548254979963\\
57	1.81552063383699\\
60	1.81553972271118\\
63	1.81554929022354\\
66	1.81555408544218\\
69	1.81555648877057\\
72	1.81555769329462\\
};
\addplot [color=mycolor2,dash pattern={on 10pt off 1pt on 1pt off 1pt} , line width=1.5pt, forget plot]
  table[row sep=crcr]{%
0	0.283228524723548\\
3	0.502482357096826\\
6	0.771198186603308\\
9	1.0499972291879\\
12	1.30063007344215\\
15	1.49952459226227\\
18	1.64105747161213\\
21	1.73273684056618\\
24	1.78765160688154\\
27	1.81858080356187\\
30	1.83523768222955\\
33	1.8439439427095\\
36	1.84841120881002\\
39	1.85067882060357\\
42	1.85182296369056\\
45	1.85239838179593\\
48	1.85268728302133\\
51	1.8528322057761\\
54	1.8529048718193\\
57	1.85294129932491\\
60	1.85295955839127\\
63	1.85296871012144\\
66	1.85297329698222\\
69	1.85297559589106\\
72	1.85297674808306\\
};
\addplot [color=mycolor1,dash pattern={on 10pt off 1pt on 0pt off 0pt} , line width=1.5pt, forget plot]
  table[row sep=crcr]{%
0	0.280022837800204\\
3	0.501700258338007\\
6	0.777735714692569\\
9	1.06806715773773\\
12	1.3307451997063\\
15	1.538490648593\\
18	1.68502908156447\\
21	1.77944122846167\\
24	1.83618463640232\\
27	1.86848994341459\\
30	1.88611822846686\\
33	1.8954368978381\\
36	1.90025604195088\\
39	1.90271391732261\\
42	1.90395734957081\\
45	1.90458358445528\\
48	1.90489822879325\\
51	1.90505612389901\\
54	1.90513530933345\\
57	1.90517500877197\\
60	1.90519490882187\\
63	1.90520488327687\\
66	1.90520988254843\\
69	1.90521238817028\\
72	1.90521364396872\\
};

\addplot [color=black, line width=1.5pt, forget plot]
  table[row sep=crcr]{%
0	0.290177930542475\\
3	0.518423340673136\\
6	0.823506632027024\\
9	1.19383810563462\\
12	1.61323619707209\\
15	2.06579868933258\\
18	2.53885438759142\\
21	3.02371143451099\\
24	3.51502326612334\\
27	4.00974387333656\\
30	4.50622472962263\\
33	5.0036023463098\\
36	5.50143331248485\\
39	5.99949250941261\\
42	6.4976663534169\\
45	6.99589772323698\\
48	7.49415794096779\\
51	7.99243262111062\\
54	8.49071455068793\\
57	8.98900011385531\\
60	9.48728749819844\\
63	9.98557579530846\\
66	10.4838645498897\\
69	10.9821535337508\\
72	11.4804426325243\\
};
\end{axis}
\end{tikzpicture}%
    \caption{Exact, approximate, and upper bound of the capacity versus the average SNR.}
    \label{cap1}
\end{figure}

Fig.~\ref{cap1} illustrates the channel capacity performance for ideal and non-ideal hardware. Basically, the capacity increases linearly with the average SNR for ideal hardware. However, the capacity saturates at high SNR by a ceiling created by the hadrware impairments. Most importantly, the capacity is more vulnerable to the joint effect of HPA and IQI while the performance is further improved by removing the HPA nonlinearities. At high SNR, a perfect compensation for the HPA nonlinearities leads to an improvement of the spectral efficiency of around 1 bits/sec/Hz.

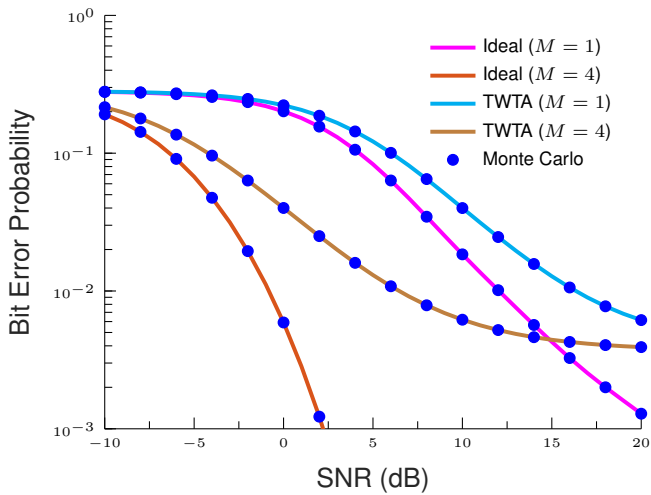
\begin{figure}[H]
\centering
\setlength\fheight{5.5cm}
\setlength\fwidth{7.5cm}
%
%
\definecolor{mycolor1}{rgb}{0.00000,0.44700,0.74100}%
\definecolor{mycolor2}{rgb}{0.85000,0.33000,0.10000}%
\definecolor{mycolor3}{rgb}{0.85000,0.32500,0.09800}%
\definecolor{mycolor4}{rgb}{0.00000,0.45000,0.74000}%
\definecolor{mycolor5}{rgb}{1.00000,0.00000,1.00000}%

\begin{tikzpicture}

\begin{axis}[%
width=0.951\fwidth,
height=\fheight,
at={(0\fwidth,0\fheight)},
scale only axis,
xmin=-10,
xmax=20,
xlabel style={font=\color{white!15!black}},
xlabel={\sffamily{SNR (dB)}},
ymode=log,
ymin=0.001,
axis x line*=bottom,
axis y line*=left,
ymax=1,
yminorticks=true,
ylabel style={font=\color{white!15!black}},
ylabel={\sffamily{Bit Error Probability}},
axis background/.style={fill=white},
legend style={at={(0.97,0.97)}, anchor=north east, legend cell align=left, align=left, draw=white!15!black,draw=none,fill=none}
]

\addplot [color=mycolor5, line width=1.5pt]
  table[row sep=crcr]{%
-10	0.278309792654462\\
-9	0.276813480692253\\
-8	0.274747476241704\\
-7	0.271910629868152\\
-6	0.26804218891796\\
-5	0.262812859604769\\
-4	0.255822290361503\\
-3	0.246611390485529\\
-2	0.234702614794027\\
-1	0.21968405798467\\
0	0.201346582394593\\
1	0.179857561409103\\
2	0.155910085982974\\
3	0.130747195638264\\
4	0.105970277770094\\
5	0.0831423957410862\\
6	0.0633806397931723\\
7	0.0471952277628659\\
8	0.0345899495998616\\
9	0.0252084061693219\\
10	0.0184491840947201\\
11	0.0136254847567753\\
12	0.0101377279681511\\
13	0.00756923075299349\\
14	0.00566899699070823\\
15	0.00427434640285683\\
16	0.00326230623421377\\
17	0.00253252448652181\\
18	0.00200079480865657\\
19	0.00160248563438324\\
20	0.00128676297443349\\
};
\addlegendentry{\sffamily{Ideal ($M = 1$)}}

\addplot [color=mycolor2 , line width=1.5pt]
  table[row sep=crcr]{%
-10	0.19135601043798\\
-9	0.168118496956815\\
-8	0.142744907211833\\
-7	0.116507834584556\\
-6	0.0909357633726592\\
-5	0.0675343034065465\\
-4	0.0474851956846939\\
-3	0.0314382942189531\\
-2	0.0194700453107774\\
-1	0.0111885700233124\\
0	0.00590927463093036\\
1	0.00283779000923562\\
2	0.00122468477648483\\
3	0.000468718726807188\\
4	0.000156412894375439\\
5	4.44093496182864e-05\\
6	1.03500583992643e-05\\
7	1.88403381355029e-06\\
8	2.50937749038788e-07\\
9	2.24958124469285e-08\\
10	1.22066295708222e-09\\
11	3.50473039363639e-11\\
12	4.4986557264978e-13\\
13	2.09612669169794e-15\\
14	2.74409303370329e-18\\
15	7.33526169638428e-22\\
16	2.66354725420162e-26\\
17	7.74412091451147e-32\\
18	9.13158608420587e-39\\
19	1.84955014858761e-47\\
20	2.19951463981691e-58\\
};
\addlegendentry{\sffamily{Ideal ($M = 4$)}}

\addplot [color=cyan, line width=1.5pt]
  table[row sep=crcr]{%
-10	0.279355278864522\\
-9	0.278272141783187\\
-8	0.276776395623803\\
-7	0.27472198595719\\
-6	0.271918993623015\\
-5	0.268126125934642\\
-4	0.263046533907745\\
-3	0.256331697475302\\
-2	0.247600772313804\\
-1	0.236484626918831\\
0	0.222701480486067\\
1	0.206160104735258\\
2	0.187066092256701\\
3	0.165985708763609\\
4	0.143818121411972\\
5	0.121656638310641\\
6	0.100580841153762\\
7	0.081468396971345\\
8	0.0648901614054771\\
9	0.0510850796831941\\
10	0.0399887227954032\\
11	0.0313091824916037\\
12	0.0246383385727381\\
13	0.0195581224184216\\
14	0.015705766110575\\
15	0.0127940414877527\\
16	0.0106032852720171\\
17	0.00896452395786105\\
18	0.00774449862964325\\
19	0.0068354439888554\\
20	0.00615006735007524\\
};
\addlegendentry{\sffamily{TWTA ($M = 1$)}}

\addplot [color=brown, line width=1.5pt]
  table[row sep=crcr]{%
-10	0.215685354717957\\
-9	0.198167237866649\\
-8	0.178604142301993\\
-7	0.157704097651475\\
-6	0.136366228551539\\
-5	0.115538969792924\\
-4	0.0960629264125031\\
-3	0.0785513850289567\\
-2	0.0633450429455814\\
-1	0.0505370332234975\\
0	0.0400337968717984\\
1	0.0316196567975824\\
2	0.0250119660928224\\
3	0.0199055032968386\\
4	0.0160060058978944\\
5	0.0130516739327871\\
6	0.0108230039136692\\
7	0.00914379063106315\\
8	0.00787698108385599\\
9	0.00691834329854533\\
10	0.00618980077558319\\
11	0.00563339922457518\\
12	0.00520629482125541\\
13	0.00487681559649512\\
14	0.00462147869726611\\
15	0.00442278539727667\\
16	0.0042676145629809\\
17	0.00414606058077625\\
18	0.00405059422411515\\
19	0.00397545528582241\\
20	0.00391621065283529\\
};
\addlegendentry{\sffamily{TWTA ($M = 4$)}}

\addplot [color=mycolor3, line width=.5pt, only marks,draw=none, mark=*, mark options={solid, fill=mycolor4, blue}]
  table[row sep=crcr]{%
-10	0.278309792654462\\
-8	0.274747476241704\\
-6	0.26804218891796\\
-4	0.255822290361503\\
-2	0.234702614794027\\
0	0.201346582394593\\
2	0.155910085982974\\
4	0.105970277770094\\
6	0.0633806397931723\\
8	0.0345899495998616\\
10	0.0184491840947201\\
12	0.0101377279681511\\
14	0.00566899699070823\\
16	0.00326230623421377\\
18	0.00200079480865657\\
20	0.00128676297443349\\
};
\addlegendentry{\sffamily{Monte Carlo}}

\addplot [color=mycolor3, line width=.5pt, only marks,draw=none, mark=*, mark options={solid, fill=mycolor4, blue}]
  table[row sep=crcr]{%
-10	0.19135601043798\\
-8	0.142744907211833\\
-6	0.0909357633726592\\
-4	0.0474851956846939\\
-2	0.0194700453107774\\
0	0.00590927463093036\\
2	0.00122468477648483\\
4	0.000156412894375439\\
6	1.03500583992643e-05\\
8	2.50937749038788e-07\\
10	1.22066295708222e-09\\
12	4.4986557264978e-13\\
14	2.74409303370329e-18\\
16	2.66354725420162e-26\\
18	9.13158608420587e-39\\
20	2.19951463981691e-58\\
};

\addplot [color=mycolor3, line width=.5pt, only marks,draw=none, mark=*, mark options={solid, fill=mycolor4, blue}]
  table[row sep=crcr]{%
-10	0.279355278864522\\
-8	0.276776395623803\\
-6	0.271918993623015\\
-4	0.263046533907745\\
-2	0.247600772313804\\
0	0.222701480486067\\
2	0.187066092256701\\
4	0.143818121411972\\
6	0.100580841153762\\
8	0.0648901614054771\\
10	0.0399887227954032\\
12	0.0246383385727381\\
14	0.015705766110575\\
16	0.0106032852720171\\
18	0.00774449862964325\\
20	0.00615006735007524\\
};

\addplot [color=mycolor3, line width=.5pt, only marks,draw=none, mark=*, mark options={solid, fill=mycolor4, blue}]
  table[row sep=crcr]{%
-10	0.215685354717957\\
-8	0.178604142301993\\
-6	0.136366228551539\\
-4	0.0960629264125031\\
-2	0.0633450429455814\\
0	0.0400337968717984\\
2	0.0250119660928224\\
4	0.0160060058978944\\
6	0.0108230039136692\\
8	0.00787698108385599\\
10	0.00618980077558319\\
12	0.00520629482125541\\
14	0.00462147869726611\\
16	0.0042676145629809\\
18	0.00405059422411515\\
20	0.00391621065283529\\
};

\end{axis}
\end{tikzpicture}%
    \caption{Probability of error of ideal hardware and HPA-TWTA for different receive apertures for OOK modulation.}
    \label{bep2}
\end{figure}

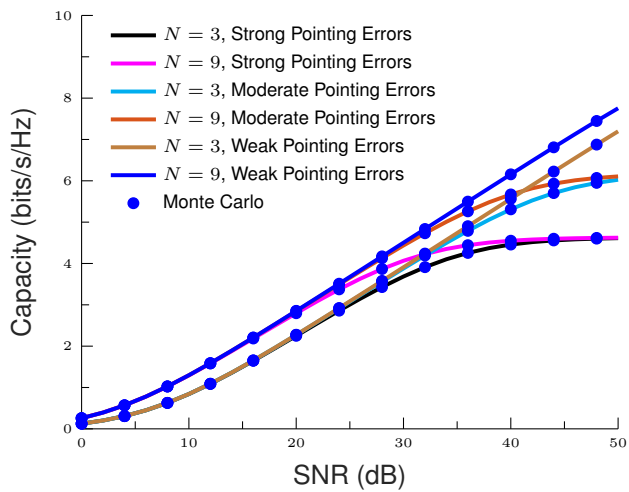
\begin{figure}[H]
\centering
\setlength\fheight{5.5cm}
\setlength\fwidth{7.5cm}
%
%
\definecolor{mycolor1}{rgb}{0.00000,0.44700,0.74100}%
\definecolor{mycolor2}{rgb}{0.85000,0.33000,0.10000}%
\definecolor{mycolor3}{rgb}{0.85000,0.32500,0.09800}%
\definecolor{mycolor4}{rgb}{0.00000,0.45000,0.74000}%
\definecolor{mycolor5}{rgb}{1.00000,0.00000,1.00000}%
\begin{tikzpicture}

\begin{axis}[%
width=0.951\fwidth,
height=\fheight,
at={(0\fwidth,0\fheight)},
scale only axis,
xmin=0,
xmax=50,
axis x line*=bottom,
axis y line*=left,
xlabel style={font=\color{white!15!black}},
xlabel={\sffamily{SNR (dB)}},
ymin=0,
ymax=10,
ylabel style={font=\color{white!15!black}},
ylabel={\sffamily{Capacity (bits/s/Hz)}},
axis background/.style={fill=white},
legend style={at={(.035,1)}, anchor=north west, legend cell align=left, align=left, draw=white!15!black,draw=none,fill=none}
]
\addplot [color=black, line width=1.5pt]
  table[row sep=crcr]{%
0	0.124882327274002\\
2	0.201141246541767\\
4	0.307429901781736\\
6	0.448618616016886\\
8	0.627012761759168\\
10	0.841313846787107\\
12	1.08690990747315\\
14	1.35725379739502\\
16	1.64539832815354\\
18	1.94496326263322\\
20	2.25039123718011\\
22	2.55670339140674\\
24	2.85901470615644\\
26	3.15201325063192\\
28	3.4295930886249\\
30	3.68488330694908\\
32	3.91093519949238\\
34	4.10211512824562\\
36	4.25571664141429\\
38	4.37285232190739\\
40	4.45798316990161\\
42	4.51739991093358\\
44	4.55758565498023\\
46	4.58414811964872\\
48	4.60142738642613\\
50	4.61254749049678\\
};
\addlegendentry{\sffamily{$N=3$, Strong Pointing Errors}}

\addplot [color=mycolor5, line width=1.5pt]
  table[row sep=crcr]{%
0	0.257908979513062\\
2	0.393626918870782\\
4	0.567995134136903\\
6	0.779483705969026\\
8	1.02322203234514\\
10	1.29244311854739\\
12	1.58004565353305\\
14	1.87956976401413\\
16	2.18545572880234\\
18	2.49280124331669\\
20	2.79687627092779\\
22	3.0925897614109\\
24	3.37407887327174\\
26	3.63464194804514\\
28	3.86728195112729\\
30	4.06597902224752\\
32	4.22733030573594\\
34	4.35166738286848\\
36	4.44287355567832\\
38	4.50701179104591\\
40	4.5506382452702\\
42	4.5795922620549\\
44	4.59847971227888\\
46	4.61065729653447\\
48	4.61844830484195\\
50	4.62340788369367\\
};
\addlegendentry{\sffamily{$N=9$, Strong Pointing Errors}}

\addplot [color=cyan, line width=1.5pt]
  table[row sep=crcr]{%
0	0.124987174711732\\
2	0.201346307730887\\
4	0.307807815481506\\
6	0.449290045781918\\
8	0.628175841542559\\
10	0.84328764222638\\
12	1.09020013270711\\
14	1.36265520164658\\
16	1.65415508081737\\
18	1.95901657006795\\
20	2.27275222474971\\
22	2.59199484246689\\
24	2.91421505706478\\
26	3.23738444493483\\
28	3.55964130554812\\
30	3.8789624943791\\
32	4.19282822706185\\
34	4.49788218933243\\
36	4.78964128631749\\
38	5.06240418256768\\
40	5.30960899388051\\
42	5.52487004947833\\
44	5.70359274462446\\
46	5.84448427474435\\
48	5.9500240207093\\
50	6.02555792936155\\
};
\addlegendentry{\sffamily{$N=3$, Moderate Pointing Errors}}

\addplot [color=mycolor2 , line width=1.5pt]
  table[row sep=crcr]{%
0	0.258186888884087\\
2	0.394159678443225\\
4	0.568965309813432\\
6	0.781184085630185\\
8	1.02611530815412\\
10	1.29725525379514\\
12	1.58791259656645\\
14	1.89226475291879\\
16	2.20573332521021\\
18	2.52490124909957\\
20	2.84722535529243\\
22	3.17069820040458\\
24	3.49351702076194\\
26	3.81376285004804\\
28	4.12907429009351\\
30	4.43631149323462\\
32	4.73125085356489\\
34	5.00843754768315\\
36	5.26143017426307\\
38	5.48369150882083\\
40	5.6701218814524\\
42	5.81867076197065\\
44	5.93107929050969\\
46	6.0122326682057\\
48	6.06857487565956\\
50	6.10653194123022\\
};
\addlegendentry{\sffamily{$N=9$, Moderate Pointing Errors}}

\addplot [color=brown, line width=1.5pt]
  table[row sep=crcr]{%
0	0.125000406053023\\
2	0.201372350042177\\
4	0.307856149373506\\
6	0.449376534107653\\
8	0.628326652204037\\
10	0.84354504475468\\
12	1.09063131808753\\
14	1.36336619764032\\
16	1.65531298961641\\
18	1.96088497866275\\
20	2.27574720450557\\
22	2.59677298389776\\
24	2.9218110162882\\
26	3.24942401807889\\
28	3.57866712345674\\
30	3.90892132068605\\
32	4.23977525910815\\
34	4.57094286441208\\
36	4.90220414723553\\
38	5.23335901490687\\
40	5.56418479259745\\
42	5.8943894158213\\
44	6.22355116549936\\
46	6.55103430339998\\
48	6.87586785344152\\
50	7.19657509219033\\
};
\addlegendentry{\sffamily{$N=3$, Weak Pointing Errors}}

\addplot [color=blue, line width=1.5pt]
  table[row sep=crcr]{%
0	0.258222373862965\\
2	0.394228196555025\\
4	0.569090933288443\\
6	0.781405571132968\\
8	1.02649408814139\\
10	1.29788807908951\\
12	1.58895177306015\\
14	1.89395039630623\\
16	2.20844440459551\\
18	2.52923589408479\\
20	2.8541264845459\\
22	3.18164852666677\\
24	3.51083809637284\\
26	3.84106484389261\\
28	4.1719116109133\\
30	4.50309079654201\\
32	4.83438491765185\\
34	5.16560084416447\\
36	5.4965288824497\\
38	5.82689856775259\\
40	6.15632255932646\\
42	6.48421853460767\\
44	6.8096969566604\\
46	7.13140170680059\\
48	7.44729499268399\\
50	7.75439652763779\\
};
\addlegendentry{\sffamily{$N=9$, Weak Pointing Errors}}

\addplot [color=mycolor3, line width=.5pt, only marks,draw=none, mark=*, mark options={solid, fill=mycolor4, blue}]
  table[row sep=crcr]{%
0	0.124882327274002\\
4	0.307429901781736\\
8	0.627012761759168\\
12	1.08690990747315\\
16	1.64539832815354\\
20	2.25039123718011\\
24	2.85901470615644\\
28	3.4295930886249\\
32	3.91093519949238\\
36	4.25571664141429\\
40	4.45798316990161\\
44	4.55758565498023\\
48	4.60142738642613\\
};
\addlegendentry{\sffamily{Monte Carlo}}

\addplot [color=mycolor3, line width=.5pt, only marks,draw=none, mark=*, mark options={solid, fill=mycolor4, blue}]
  table[row sep=crcr]{%
0	0.257908979513062\\
4	0.567995134136903\\
8	1.02322203234514\\
12	1.58004565353305\\
16	2.18545572880234\\
20	2.79687627092779\\
24	3.37407887327174\\
28	3.86728195112729\\
32	4.22733030573594\\
36	4.44287355567832\\
40	4.5506382452702\\
44	4.59847971227888\\
48	4.61844830484195\\
};

\addplot [color=mycolor3, line width=.5pt, only marks,draw=none, mark=*, mark options={solid, fill=mycolor4, blue}]
  table[row sep=crcr]{%
0	0.124987174711732\\
4	0.307807815481506\\
8	0.628175841542559\\
12	1.09020013270711\\
16	1.65415508081737\\
20	2.27275222474971\\
24	2.91421505706478\\
28	3.55964130554812\\
32	4.19282822706185\\
36	4.78964128631749\\
40	5.30960899388051\\
44	5.70359274462446\\
48	5.9500240207093\\
};

\addplot [color=mycolor3, line width=.5pt, only marks,draw=none, mark=*, mark options={solid, fill=mycolor4, blue}]
  table[row sep=crcr]{%
0	0.258186888884087\\
4	0.568965309813432\\
8	1.02611530815412\\
12	1.58791259656645\\
16	2.20573332521021\\
20	2.84722535529243\\
24	3.49351702076194\\
28	4.12907429009351\\
32	4.73125085356489\\
36	5.26143017426307\\
40	5.6701218814524\\
44	5.93107929050969\\
48	6.06857487565956\\
};

\addplot [color=mycolor3, line width=.5pt, only marks,draw=none, mark=*, mark options={solid, fill=mycolor4, blue}]
  table[row sep=crcr]{%
0	0.125000406053023\\
4	0.307856149373506\\
8	0.628326652204037\\
12	1.09063131808753\\
16	1.65531298961641\\
20	2.27574720450557\\
24	2.9218110162882\\
28	3.57866712345674\\
32	4.23977525910815\\
36	4.90220414723553\\
40	5.56418479259745\\
44	6.22355116549936\\
48	6.87586785344152\\
};

\addplot [color=mycolor3, line width=.5pt, only marks,draw=none, mark=*, mark options={solid, fill=mycolor4, blue}]
  table[row sep=crcr]{%
0	0.258222373862965\\
4	0.569090933288443\\
8	1.02649408814139\\
12	1.58895177306015\\
16	2.20844440459551\\
20	2.8541264845459\\
24	3.51083809637284\\
28	4.1719116109133\\
32	4.83438491765185\\
36	5.4965288824497\\
40	6.15632255932646\\
44	6.8096969566604\\
48	7.44729499268399\\
};

\end{axis}
\end{tikzpicture}%
    \caption{Capacity versus the average SNR for various pointing errors values and number of transmit antennas.}
    \label{cap2}
\end{figure}

Fig.~\ref{bep2} shows the error performance with respect to different number of receive apertures for ideal and joint TWTA-IQI scenarios. For both scenarios, the system shows an observable diversity gain in comparison to single receive antenna. For joint HPA-IQI, we note that the diversity improves the performance at low SNR. However, the receive diversity contribution becomes very limited by the error floor at high SNR. Most importantly, the selection receive diversity gain for joint HPA-IQI is worse than the ideal case with single receive antenna. 
Fig.~\ref{cap2} illustrates the joint impact of the number of receive antennas and the pointing error fading on the channel capacity. For low SNR, we observe that the capacity improves by increasing the number of antennas and vice versa. In fact, large number of antennas yields a higher array gain which increases the received power and hence the capacity. The impact of the number of antennas is still observable for high SNR but for weak pointing errors. However, the performance gets much worse for moderate and strong pointing errors as the capacity is saturated at high SNR. In fact, the pointing errors is a way to measure the degree of fluctuations of the laser beam. For moderate and severe pointing errors, the receiver cannot intercept the whole received power leading to additional losses. Most importantly, we observe that the impact of the number of antennas totally disappears for moderate and strong pointing errors at high SNR. This result eventually shows that the system depends to a large extent on the state of the optical channel.

\begin{figure}[H]
\centering
\setlength\fheight{5.5cm}
\setlength\fwidth{7.5cm}
\input{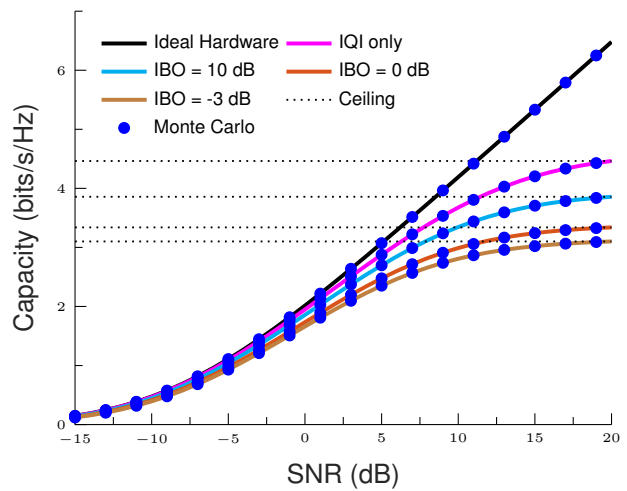}
    \caption{Capacity under joint effects of HPA-IQI for different values of the amplifier saturation levels, and for linear relaying (IQI only), $\rho_2$ = -13 dB.}
    \label{cap3}
\end{figure}

Fig.~\ref{cap3} shows the capacity performance with respect to ideal and hardware impairments scenarios. The effects of the impairments are negligible at low SNR as the performances are identical to the ideal hardware. As the SNR increases, however, the impairments effects become more pronounced as the capacity saturates by a finite ceiling. The capacity improves as the IBO increases, and gets much better for linear relaying (no HPA nonlinearities) which is expected, since the system suffers only from IQI. In fact, if the IBO increases, the amplifier saturation level also increases and the relay becomes able to deliver more output power. Consequently, the signal experiences less distortion and the effects of HPA nonlinearities are mitigated.


\section{Conclusion}
In this work, we present a generalized performance analysis of MIMO hybrid mmwave and FSO relaying systems for different scenarios. We conclude that for such system, it is important to consider the joint effects of HPA-IQI into the system model. Results show that the performance significantly deviates from the ideal case essentially at high SNR. Consequently, such imperfections have not to be neglected for high rate system to get consistent analytical results. In addition, the outage performance is better for SEL-IQI than TWTA-IQI as the last one is more severe. In addition, the diversity gain achieved by the system is acceptable at low SNR compared to the ideal case, however, it substantially degrades at high SNR mainly by TWTA-IQI and hence the MIMO architecture is very constrained by the hardware impairments. Consequently, the best strategy to leverage the diversity gain is to compensate for the hardware impairments. Although, previous algorithms have been developed for the impairment compensation, there are still residual impairments incorporated into the hardware.
\appendices
\textcolor{black}{\section{Proof of The Upper Bound of The Truncation Error (\ref{errorbound})}
In this appendix, we present the derivation of the truncation error in (\ref{errorbound}) which can be expressed as
\begin{equation}
\begin{split}
\mathcal{E}(L) =& \Xi \sum_{n=1}^\beta \frac{\pi\omega_n}{2\sin[\pi(\alpha-n)]}\sum_{p=L+1}^\infty \left[ \rho_p(\alpha,n)I_a^{p+n-1} \right.\\&\left.- \rho_p(n,\alpha)I_a^{p+\alpha-1}  \right]    
\end{split}
\end{equation}
Replacing $\rho_p(x,y)$ given by (\ref{rho}), the truncation error can be written as
\begin{equation}
\mathcal{E}(L) = \Xi \sum_{n=1}^\beta \frac{\pi\omega_n}{2\sin[\pi(\alpha-n)]}\sum_{p=L+1}^\infty \frac{ \left( \alpha \xi I_a \right)^p }{p!}c_p(\alpha,n)  
\end{equation}
where $c_p(x,y)$ and $b_p(x,y)$ are defined as
\begin{equation}\label{bp}
b_p(x,y) = \frac{\left( \alpha\xi \right)^yI_a^{y-1}}{\Gamma(p-x+y+1)}    
\end{equation}
\begin{equation}
c_p(x,y) = b_p(x,y) - b_p(y,x)
\end{equation}
Note that $b_q(x,y)$ in (\ref{bp}) decreases with the increase in p, the truncation error can be upper bounded by
\begin{equation}\label{errexp}
\mathcal{E}(L) < \Xi \max_{p>L}\{c_p(\alpha,n)\} \sum_{n=1}^\beta\frac{\pi\omega_n}{2\sin[\pi(\alpha-n)]}\sum_{p=L+1}^\infty\frac{ \left( \alpha \xi I_a \right)^p }{p!}
\end{equation}
Considering the Taylor series expansion of the exponential function, (\ref{errexp}) can be reduced to
\begin{equation}
\mathcal{E}(L) < \Xi e^{\alpha\xi I_a} \max_{p>L} \{c_p(\alpha,n)\}   \sum_{n=1}^\beta\frac{\pi\omega_n}{2\sin[\pi(\alpha-n)]}  
\end{equation}}
\section{Proof of The CDF Expression (\ref{fso-cdf})}
Since selection combining is assumed at the receiver, The PDF of $I_{\tt{max}}$ is given by
\begin{equation}
f_{I_{\tt{max}}}(x) = M[F_{I_{\tt{m}}}(x)]^{M-1} f_{I_{\tt{m}}}(x).   
\end{equation}
According to \cite{zf}, the Bessel function $K_{\nu}(\cdot)$ in (\ref{ird}) can be expanded in terms of power series as 
\begin{equation}\label{bessel}
K_{\nu}(x) = \frac{\pi}{2\sin(\pi\nu)}\sum_{p=0}^{\infty}\left[\frac{(0.5x)^{2p-\nu}}{\Gamma(p-\nu+1)p!} - \frac{(0.5x)^{2p+\nu}}{\Gamma(p+\nu+1)p!}  \right]    
\end{equation}
where $\nu \notin \mathbb{Z}$ and $|x| < \infty$. Injecting (\ref{bessel}) in (\ref{ird}), we get
\begin{equation}
\begin{split}
f_{I_{\tt{m}}}(x) =& A\sum_{n=1}^{\beta} \frac{a_n\pi}{2\sin(\pi(\alpha-n))}\sum_{p=0}^\infty \left[\rho_p(\alpha,n)x^{p+n-1} \right.\\&\left.- \rho_p(n,\alpha)x^{p+\alpha-1} \right]   
\end{split}
\end{equation}
where $\alpha \notin \mathbb{Z}$ is assumed and $\delta = \alpha\beta/(g\beta+\Omega^{'})$ and we express $\rho_p(x,y)$ as 
\begin{equation}
\rho_p(x,y) = \frac{\delta^{p-\frac{x-y}{2}}}{\Gamma(p-x+y+1)p!}    
\end{equation}

The CDF of $I_{\tt{m}}$ is then expressed as
\begin{equation}
\begin{split}
F_{I_{\tt{m}}}(x) =& A\sum_{n=1}^{\infty}\frac{a_n\pi}{2\sin(\pi(\alpha-n))}\sum_{p=0}^{\infty}\left[\frac{\rho_p(\alpha,n)}{p+n}x^{p+n} \right.\\&\left.- \frac{\rho_p(n,\alpha)}{p+\alpha} x^{p+\alpha} \right].  
\end{split}
\end{equation}
We introduce a finite large upper bound $L$ to approximate the infinite summation \cite{sim,sim2}. The CDF of the optical SNR $\gamma_r = \overline{\gamma}_rI_{\tt{max}}^r$ is expressed as
\begin{equation}
\begin{split}
F_{\gamma_{\text{RD}}}(x) =& \left[ F_{I_{\tt{m}}} \left(\frac{x}{\overline{\gamma}_r} \right)^{\frac{1}{r}} \right]^M = A^M\sum_{i=0}^{M}{M \choose i}(-1)^i\sum_{j=M-i}^{(M-i)(L+\beta)}\chi_j\\&\sum_{t=0}^{iL}\chi_t \left(\frac{x}{\overline{\gamma}_r} \right)^{\frac{j+t+i\alpha}{r}},
\end{split}
\end{equation}
where $\chi_i$ and $\chi_t$ are the coefficients of $\left(\frac{x}{\overline{\gamma}_r} \right)^{\frac{j}{r}}$ and $\left(\frac{x}{\overline{\gamma}_r} \right)^{\frac{t}{r}}$ in the expansions of
\begin{equation}
\begin{split}
    &\left[\sum_{n=1}^{\beta} \frac{a_n\pi}{2\sin(\pi\nu)}\sum_{p=0}^L \frac{\rho_p(\alpha,n)}{p+n}\left(  \frac{x}{\overline{\gamma}_r}\right)^{\frac{p+n}{r}}   \right]^{M-i} \\&\times  \left[\sum_{n=1}^{\beta} \frac{a_n\pi}{2\sin(\pi\nu)}\sum_{p=0}^L \frac{\rho_p(\alpha,n)}{p+n}\left(  \frac{x}{\overline{\gamma}_r}\right)^{\frac{p}{r}}   \right]^{i}.    
\end{split}
\end{equation}

\section{Proof of the Capacity Upper Bound}
The overall SINDR (\ref{sindr}) can be upper bounded by
\begin{equation}
    {\scriptsize\textsf{SINDR}} \leq \frac{\gamma_{\text{SR}}}{\rho_2\gamma_{\text{SR}}+\rho_1(1+\rho_2)(1+\gamma_{\text{R}})} = \frac{ \frac{\gamma_{\text{SR}}}{\rho_1(1+\rho_2)(1+\gamma_{\text{R}})}}{\rho_2\frac{\gamma_{\text{SR}}}{\rho_1(1+\rho_2)(1+\gamma_{\text{R}})}+1}.
\end{equation}
By defining $\psi = \frac{\gamma_{\text{SR}}}{\rho_1(1+\rho_2)(1+\gamma_{\text{R}})}$, the function $\log\left(1 + \frac{\varpi\psi}{\rho_2\psi + 1}\right)$ is concave of $\psi$ for $\psi > 0$, since its second derivative is given by
\begin{equation}
\frac{-(2\rho_2^2\psi+2\rho_2(\psi+1)+1)}{(\rho_2\psi+1)^2(\rho_2\psi + \psi + 1)^2} < 0.
\end{equation}
Then, we can apply Jensen's inequality to get
\begin{equation}
\mathbb{E}\left[\log\left(1 + \frac{\varpi\psi}{\rho_2\psi + 1}\right)\right] \leq \log\left(1 + \frac{\varpi\mathbb{E}[\psi]}{\rho_2\mathbb{E}[\psi] + 1}\right).
\end{equation}
Given that $\mathcal{J} = \mathbb{E}[\psi]$ and after applying the identity \cite[Eq.~(2.243.1)]{prudnikov}, the term $\mathcal{J}$ is derived (\ref{jensen}).

\bibliography{main.bib}

\begin{thebibliography}{10}
\providecommand{\url}[1]{#1}
\csname url@samestyle\endcsname
\providecommand{\newblock}{\relax}
\providecommand{\bibinfo}[2]{#2}
\providecommand{\BIBentrySTDinterwordspacing}{\spaceskip=0pt\relax}
\providecommand{\BIBentryALTinterwordstretchfactor}{4}
\providecommand{\BIBentryALTinterwordspacing}{\spaceskip=\fontdimen2\font plus
\BIBentryALTinterwordstretchfactor\fontdimen3\font minus
  \fontdimen4\font\relax}
\providecommand{\BIBforeignlanguage}[2]{{%
\expandafter\ifx\csname l@#1\endcsname\relax
\typeout{** WARNING: IEEEtran.bst: No hyphenation pattern has been}%
\typeout{** loaded for the language `#1'. Using the pattern for}%
\typeout{** the default language instead.}%
\else
\language=\csname l@#1\endcsname
\fi
#2}}
\providecommand{\BIBdecl}{\relax}
\BIBdecl

\bibitem{surv}
Y.~Niu, Y.~Li, D.~Jin, L.~Su, and A.~V. Vasilakos, ``{A survey of millimeter
  wave communications (mmWave) for 5G: opportunities and challenges},''
  \emph{Wireless Networks}, vol.~21, no.~8, pp. 2657--2676, Nov 2015.

\bibitem{rheath}
R.~W. {Heath}, N.~{González-Prelcic}, S.~{Rangan}, W.~{Roh}, and A.~M.
  {Sayeed}, ``{An Overview of Signal Processing Techniques for Millimeter Wave
  MIMO Systems},'' \emph{IEEE Journal of Selected Topics in Signal Processing},
  vol.~10, no.~3, pp. 436--453, April 2016.

\bibitem{5gnr}
E.~{Onggosanusi}, M.~S. {Rahman}, L.~{Guo}, Y.~{Kwak}, H.~{Noh}, Y.~{Kim},
  S.~{Faxer}, M.~{Harrison}, M.~{Frenne}, S.~{Grant}, R.~{Chen}, R.~{Tamrakar},
  and a.~Q.~{Gao}, ``{Modular and High-Resolution Channel State Information and
  Beam Management for 5G New Radio},'' \emph{IEEE Communications Magazine},
  vol.~56, no.~3, pp. 48--55, March 2018.

\bibitem{product}
A.~{El-Yamany} and M.~{Petri}, ``{An Adaptive IEEE 802.11ad Indoor mmWave
  Inner-Receiver Architecture},'' in \emph{2018 2nd URSI Atlantic Radio Science
  Meeting (AT-RASC)}, May 2018, pp. 1--4.

\bibitem{1}
S.~Arnon, J.~Barry, G.~Karagiannidis, R.~Schober, and M.~Uysal,
  \emph{{{A}dvanced {O}ptical {W}ireless {C}ommunication {S}ystems}},
  1st~ed.\hskip 1em plus 0.5em minus 0.4em\relax New York, NY, USA: Cambridge
  University Press, 2012.

\bibitem{c1}
E.~Balti, M.~Guizani, B.~Hamdaoui, and Y.~Maalej, ``{Partial Relay Selection
  for Hybrid RF/FSO Systems with Hardware Impairments},'' in \emph{2016 IEEE
  Global Communications Conference (GLOBECOM)}, Dec 2016, pp. 1--6.

\bibitem{c2}
E.~Balti, M.~Guizani, and B.~Hamdaoui, ``{Hybrid Rayleigh and Double-Weibull
  over impaired RF/FSO system with outdated CSI},'' in \emph{2017 IEEE
  International Conference on Communications (ICC)}, May 2017, pp. 1--6.

\bibitem{c3}
E.~Balti, M.~Guizani, B.~Hamdaoui, and B.~Khalfi, ``{Mixed RF/FSO Relaying
  Systems with Hardware Impairments},'' in \emph{GLOBECOM 2017 - 2017 IEEE
  Global Communications Conference}, Dec 2017, pp. 1--6.

\bibitem{j1}
E.~Balti and M.~Guizani, ``{Impact of Non-Linear High-Power Amplifiers on
  Cooperative Relaying Systems},'' \emph{IEEE Transactions on Communications},
  vol.~65, no.~10, pp. 4163--4175, Oct 2017.

\bibitem{j2}
E.~Balti, M.~Guizani, B.~Hamdaoui, and B.~Khalfi, ``{Aggregate Hardware
  Impairments Over Mixed RF/FSO Relaying Systems With Outdated CSI},''
  \emph{IEEE Transactions on Communications}, vol.~66, no.~3, pp. 1110--1123,
  March 2018.

\bibitem{j3}
E.~Balti and M.~Guizani, ``{Mixed RF/FSO Cooperative Relaying Systems with
  Co-Channel Interference},'' \emph{IEEE Transactions on Communications}, pp.
  1--1, 2018.

\bibitem{j4}
E.~{Balti} and B.~K. {Johnson}, ``{Tractable Approach to MmWaves Cellular
  Analysis with FSO Backhauling under Feedback Delay and Hardware
  Limitations},'' \emph{IEEE Transactions on Wireless Communications}, pp.
  1--1, 2019.

\bibitem{asym}
E.~Balti and B.~K. Johnson, ``Asymmetric rf/fso relaying with hpa
  non-linearities and feedback delay constraints,'' 2019.

\bibitem{eT}
E.~Balti, ``\BIBforeignlanguage{English}{Analysis of hybrid free space optics
  and radio frequency cooperative relaying systems},'' Master's thesis, 2018.

\bibitem{4}
M.~A. Al-Habash, L.~C. Andrews, and R.~L. Phillips, ``{Mathematical model for
  the irradiance probability density function of a laser beam propagating
  through turbulent media},'' \emph{Optical Engineering}, vol.~40, pp.
  1554--1562, Aug. 2001.

\bibitem{pathloss1}
M.~{Khatun}, H.~{Mehrpouyan}, D.~{Matolak}, and I.~{Guvenc}, ``{Millimeter wave
  systems for airports and short-range aviation communications: A survey of the
  current channel models at mmWave frequencies},'' in \emph{2017 IEEE/AIAA 36th
  Digital Avionics Systems Conference (DASC)}, Sep. 2017, pp. 1--8.

\bibitem{pathloss2}
P.~{Korrai} and D.~{Sen}, ``{Downlink SINR Coverage and Rate Analysis with Dual
  Slope Pathloss Model in mmWave Networks},'' in \emph{2017 IEEE Wireless
  Communications and Networking Conference (WCNC)}, March 2017, pp. 1--6.

\bibitem{cell}
A.~{Lopusina}, E.~{Kocan}, and M.~{Pejanovic-Djurisic}, ``{Macrodiversity for
  performance improvement of mmwave cellular communications},'' in \emph{2017
  40th International Conference on Telecommunications and Signal Processing
  (TSP)}, July 2017, pp. 174--177.

\bibitem{cell1}
X.~{Wang} and M.~C. {Gursoy}, ``Joint energy and sinr coverage in energy
  harvesting mmwave cellular networks with user-centric base station
  deployments,'' in \emph{2018 IEEE Global Conference on Signal and Information
  Processing (GlobalSIP)}, Nov 2018, pp. 1271--1275.

\bibitem{cell2}
N.~{Rupasinghe}, Y.~{Kakishima}, I.~{Guvenc}, K.~{Kitao}, and T.~{Imai},
  ``{Geometry performance for 5G mmWave cellular networks},'' in \emph{2016
  International Symposium on Antennas and Propagation (ISAP)}, Oct 2016, pp.
  874--875.

\bibitem{antenna}
E.~Balti and B.~K. Johnson, ``Sub-6 ghz microstrip antenna: Design and
  radiation modeling,'' 2019.

\bibitem{modulation1}
F.~{Kemmochi}, K.~{Fujisawa}, and H.~{Otsuka}, ``{}potential design for
  modulation and coding scheme in mmwave multicarrier hetnets,'' in \emph{2019
  IEEE 90th Vehicular Technology Conference (VTC2019-Fall)}.

\bibitem{modulation2}
J.~{Wang}, L.~{He}, and J.~{Song}, ``{An Overview of Spatial Modulation
  Techniques for Millimeter Wave MIMO Systems},'' in \emph{2017 IVth
  International Conference on Engineering and Telecommunication (EnT)}, Nov
  2017, pp. 51--56.

\bibitem{modulation3}
T.~Y. {Elganimi} and A.~A. {Elghariani}, ``{Space-Time block coded spatial
  modulation aided mmWave MIMO with hybrid precoding},'' in \emph{2018 26th
  Signal Processing and Communications Applications Conference (SIU)}, May
  2018, pp. 1--6.

\bibitem{macrocell}
G.~R. {MacCartney} and T.~S. {Rappaport}, ``{Study on 3GPP rural macrocell path
  loss models for millimeter wave wireless communications},'' in \emph{2017
  IEEE International Conference on Communications (ICC)}, May 2017, pp. 1--7.

\bibitem{microcell}
J.~{Lee}, J.~{Choi}, J.~{Lee}, and S.~{Kim}, ``{28 GHz Millimeter-Wave Channel
  Models in Urban Microcell Environment Using Three-Dimensional Ray Tracing},''
  \emph{IEEE Antennas and Wireless Propagation Letters}, vol.~17, no.~3, pp.
  426--429, March 2018.

\bibitem{v2xdiversity}
E.~Balti and B.~K. Johnson, ``Mmwaves cellular v2x for cooperative diversity
  relay fast fading channels,'' 2020.

\bibitem{muratbook}
Z.~Ghassemlooy, W.~Popoola, and S.~Rajbhandari, \emph{{O}ptical {W}ireless
  {C}ommunications: {S}ystem and {C}hannel {M}odelling with {MATLAB}},
  1st~ed.\hskip 1em plus 0.5em minus 0.4em\relax Boca Raton, FL, USA: CRC
  Press, Inc., 2012.

\bibitem{djordjevic}
G.~T. {Djordjevic}, M.~I. {Petkovic}, A.~M. {Cvetkovic}, and G.~K.
  {Karagiannidis}, ``Mixed rf/fso relaying with outdated channel state
  information,'' \emph{IEEE Journal on Selected Areas in Communications},
  vol.~33, no.~9, pp. 1935--1948, 2015.

\bibitem{zedini}
E.~Zedini, H.~Soury, and M.~S. Alouini, ``{On the Performance Analysis of
  Dual-Hop Mixed {FSO/RF} Systems},'' \emph{IEEE Transactions on Wireless
  Communications}, vol.~15, no.~5, pp. 3679--3689, May 2016.

\bibitem{10}
K.~Kumar and D.~K. Borah, ``{Quantize and Encode Relaying Through {FSO} and
  Hybrid {FSO/RF} Links},'' \emph{IEEE Transactions on Vehicular Technology},
  vol.~64, no.~6, pp. 2361--2374, June 2015.

\bibitem{12}
E.~Lee, J.~Park, D.~Han, and G.~Yoon, ``{Performance Analysis of the Asymmetric
  Dual-Hop Relay Transmission With Mixed {{RF/FSO}} Links},'' \emph{IEEE
  Photonics Technology Letters}, vol.~23, no.~21, pp. 1642--1644, Nov 2011.

\bibitem{prs}
I.~{Krikidis}, J.~{Thompson}, S.~{Mclaughlin}, and N.~{Goertz},
  ``Amplify-and-forward with partial relay selection,'' \emph{IEEE
  Communications Letters}, vol.~12, no.~4, pp. 235--237, 2008.

\bibitem{16}
N.~S. Ferdinand, N.~Rajatheva, and M.~Latva-aho, ``{Effects of Feedback Delay
  in Partial Relay Selection Over {Nakagami}- m Fading Channels},'' \emph{IEEE
  Transactions on Vehicular Technology}, vol.~61, no.~4, pp. 1620--1634, May
  2012.

\bibitem{fd1}
E.~Balti, N.~Mensi, and S.~Yan, ``A modified zero-forcing max-power design for
  hybrid beamforming full-duplex systems,'' 2020.

\bibitem{fd2}
E.~Balti, N.~Mensi, and D.~B. Rawat, ``Adaptive gradient search beamforming for
  full-duplex mmwave mimo systems,'' 2020.

\bibitem{pls1}
N.~Mensi, D.~B. Rawat, and E.~Balti, ``Physical layer security for v2i
  communications: Reflecting surfaces vs. relaying,'' 2020.

\bibitem{pls2}
------, ``Pls for v2i communications using friendly jammer and double kappa-mu
  shadowed fading,'' 2020.

\bibitem{neji1}
N.~{Mensi}, A.~{Makhlouf}, and M.~{Guizani}, ``Incentives for safe driving in
  vanet,'' in \emph{2016 4th International Conference on Control Engineering
  Information Technology (CEIT)}, 2016, pp. 1--6.

\bibitem{neji2}
N.~{Mensi}, M.~{Guizani}, and A.~{Makhlouf}, ``Study of vehicular cloud during
  traffic congestion,'' in \emph{2016 4th International Conference on Control
  Engineering Information Technology (CEIT)}, 2016, pp. 1--6.

\bibitem{maalej}
Y.~{Maalej}, A.~{Abderrahim}, M.~{Guizani}, B.~{Hamdaoui}, and E.~{Balti},
  ``Advanced activity-aware multi-channel operations1609.4 in vanets for
  vehicular clouds,'' in \emph{2016 IEEE Global Communications Conference
  (GLOBECOM)}, 2016, pp. 1--6.

\bibitem{17}
D.~Dardari, V.~Tralli, and A.~Vaccari, ``{A theoretical characterization of
  nonlinear distortion effects in {OFDM} systems},'' \emph{IEEE Transactions on
  Communications}, vol.~48, no.~10, pp. 1755--1764, Oct 2000.

\bibitem{19}
T.~Riihonen, S.~Werner, F.~Gregorio, R.~Wichman, and J.~Hamalainen, ``{BEP}
  analysis of {OFDM} relay links with nonlinear power amplifiers,'' in
  \emph{2010 IEEE Wireless Communication and Networking Conference}, April
  2010, pp. 1--6.

\bibitem{20}
J.~Qi, S.~Aissa, and M.~S. Alouini, ``{Analysis and compensation of {I/Q}
  imbalance in amplify-and-forward cooperative systems},'' in \emph{2012 IEEE
  Wireless Communications and Networking Conference (WCNC)}, April 2012, pp.
  215--220.

\bibitem{22}
N.~Maletic, M.~Cabarkapa, and N.~Neskovic, ``{Performance of fixed-gain
  amplify-and-forward nonlinear relaying with hardware impairments},''
  \emph{International Journal of Communication Systems}, pp. n/a--n/a, 2015.

\bibitem{23}
C.~{Zhang}, P.~{Ren}, J.~{Peng}, G.~{Wei}, Q.~{Du}, and Y.~{Wang}, ``{Optimal
  Relay Power Allocation for Amplify-and-Forward Relay Networks with Non-linear
  Power Amplifiers},'' \emph{ArXiv e-prints}, Apr. 2011.

\bibitem{31}
J.~Qi, S.~Aissa, and M.~S. Alouini, ``{Analysis and compensation of {I/Q}
  imbalance in amplify-and-forward cooperative systems},'' in \emph{2012 IEEE
  Wireless Communications and Networking Conference (WCNC)}, April 2012, pp.
  215--220.

\bibitem{32}
E.~Bjornson, M.~Matthaiou, and M.~Debbah, ``{A New Look at Dual-Hop Relaying:
  Performance Limits with Hardware Impairments},'' \emph{IEEE Transactions on
  Communications}, vol.~61, no.~11, pp. 4512--4525, November 2013.

\bibitem{33}
M.~Matthaiou, A.~Papadogiannis, E.~Bjornson, and M.~Debbah, ``{Two-Way Relaying
  Under the Presence of Relay Transceiver Hardware Impairments},'' \emph{IEEE
  Communications Letters}, vol.~17, no.~6, pp. 1136--1139, June 2013.

\bibitem{55}
E.~Zedini, I.~S. Ansari, and M.~S. Alouini, ``Performance analysis of mixed
  {N}akagami-m and {G}amma-{G}amma dual-hop {FSO} transmission systems,''
  \emph{IEEE Photonics Journal}, vol.~7, no.~1, pp. 1--20, Feb 2015.

\bibitem{56}
H.~AlQuwaiee, I.~S. Ansari, and M.~S. Alouini, ``On the performance of
  free-space optical communication systems over double generalized gamma
  channel,'' \emph{IEEE Journal on Selected Areas in Communications}, vol.~33,
  no.~9, pp. 1829--1840, Sept 2015.

\bibitem{58}
I.~S. Ansari, F.~Yilmaz, and M.~S. Alouini, ``On the performance of hybrid {RF}
  and {{RF/FSO}} dual-hop transmission systems,'' in \emph{2013 2nd
  International Workshop on Optical Wireless Communications (IWOW)}, Oct 2013,
  pp. 45--49.

\bibitem{59}
L.~Yang, M.~O. Hasna, and X.~Gao, ``{Performance of Mixed {{RF/FSO}} With
  Variable Gain over Generalized Atmospheric Turbulence Channels},'' \emph{IEEE
  Journal on Selected Areas in Communications}, vol.~33, no.~9, pp. 1913--1924,
  Sept 2015.

\bibitem{mmwavej}
B.~He and R.~Schober, ``{Bit-interleaved coded modulation for hybrid {RF/FSO}
  systems},'' \emph{IEEE Transactions on Communications}, vol.~57, no.~12, pp.
  3753--3763, December 2009.

\bibitem{mmwavec}
N.~D. Chatzidiamantis, G.~K. Karagiannidis, E.~E. Kriezis, and M.~Matthaiou,
  ``{Diversity Combining in Hybrid {RF/FSO} Systems with {PSK} Modulation},''
  in \emph{2011 IEEE International Conference on Communications (ICC)}, June
  2011, pp. 1--6.

\bibitem{scin}
M.~Niu, J.~Cheng, and J.~F. Holzman, ``{Error Rate Performance Comparison of
  Coherent and Subcarrier Intensity Modulated Optical Wireless
  Communications},'' \emph{J. Opt. Commun. Netw.}, vol.~5, no.~6, pp. 554--564,
  Jun 2013.

\bibitem{64}
N.~H. M~Evans and B.~Peacock, ``{Statistical Distributions, Third Edition},''
  \emph{Measurement Science and Technology}, vol.~12, no.~1, p. 117, 2001.

\bibitem{26}
D.~Dardari, V.~Tralli, and A.~Vaccari, ``{A theoretical characterization of
  nonlinear distortion effects in OFDM systems},'' \emph{IEEE Transactions on
  Communications}, vol.~48, no.~10, pp. 1755--1764, Oct 2000.

\bibitem{busg}
N.~Y. Ermolova and S.~G. Haggman, ``{An extension of Bussgang's theory to
  complex-valued signals},'' in \emph{Signal Processing Symposium, 2004. NORSIG
  2004. Proceedings of the 6th Nordic}, June 2004, pp. 45--48.

\bibitem{imb}
J.~Qi, S.~Aissa, and M.~S. Alouini, ``Impact of i/q imbalance on the
  performance of two-way csi-assisted af relaying,'' in \emph{2013 IEEE
  Wireless Communications and Networking Conference (WCNC)}, April 2013, pp.
  2507--2512.

\bibitem{ehsan}
E.~Soleimani-Nasab and M.~Uysal, ``{Generalized Performance Analysis of Mixed
  {{RF/FSO}} Cooperative Systems},'' \emph{IEEE Transactions on Wireless
  Communications}, vol.~15, no.~1, pp. 714--727, Jan 2016.

\bibitem{prudnikov}
A.~Prudnikov and Y.~A. Brychkov, \emph{{INTEGRAL AND SERIES, Volume 3, More
  Special Functions}}, Computing Center of the USSR Academy of Sciences,
  Moscow, 1990.

\bibitem{wolfram}
\BIBentryALTinterwordspacing
``{The Wolfram Functions Site}.'' [Online]. Available:
  \url{http://functions.wolfram.com}
\BIBentrySTDinterwordspacing

\bibitem{mod1}
E.~Zedini and M.~S. Alouini, ``{Multihop Relaying Over IM/DD FSO Systems With
  Pointing Errors},'' \emph{Journal of Lightwave Technology}, vol.~33, no.~23,
  pp. 5007--5015, Dec 2015.

\bibitem{mittal}
P.~K. Mittal and K.~C. Gupta, ``{An integral involving generalized function of
  two variables},'' \emph{Proceedings of the Indian Academy of Sciences -
  Section A}, vol.~75, no.~3, pp. 117--123, 1972.

\bibitem{bfox}
A.~Soulimani, M.~Benjillali, H.~Chergui, and D.~B. da~Costa, ``{Performance
  Analysis of {M-QAM} Multihop Relaying over mmWave Weibull Fading Channels},''
  \emph{CoRR}, vol. abs/1610.08535, 2016.

\bibitem{python}
{H. R. Alhennawi and M. M. H. El Ayadi and M. H. Ismail and H. A. M. Mourad},
  ``{Closed-Form Exact and Asymptotic Expressions for the Symbol Error Rate and
  Capacity of the $H$-Function Fading Channel},'' \emph{IEEE Transactions on
  Vehicular Technology}, vol.~65, no.~4, pp. 1957--1974, April 2016.

\bibitem{f1}
A.~Jurado-Navas, J.~Garrido-Balsellss, J.~Paris, and A.~Puerta-Notario,
  \emph{\BIBforeignlanguage{English}{{A Unifying Statistical Model for
  Atmospheric Optical Scintillation}}}.\hskip 1em plus 0.5em minus 0.4em\relax
  InTech, 2011, pp. 181--206.

\bibitem{f2}
A.~Jurado-Navas, J.~M. Garrido-Balsells, J.~F. Paris, M.~Castillo-V\'{a}zquez,
  and A.~Puerta-Notario, ``{Impact of pointing errors on the performance of
  generalized atmospheric optical channels},'' \emph{Opt. Express}, vol.~20,
  no.~11, pp. 12\,550--12\,562, May 2012.

\bibitem{f3}
H.~{Samimi} and M.~{Uysal}, ``{End-to-end performance of mixed RF/FSO
  transmission systems},'' \emph{IEEE/OSA Journal of Optical Communications and
  Networking}, vol.~5, no.~11, pp. 1139--1144, Nov 2013.

\bibitem{zf}
M.~{Matthaiou}, N.~D. {Chatzidiamantis}, G.~K. {Karagiannidis}, and J.~A.
  {Nossek}, ``{ZF Detectors over Correlated K Fading MIMO Channels},''
  \emph{IEEE Transactions on Communications}, vol.~59, no.~6, pp. 1591--1603,
  June 2011.

\bibitem{sim}
X.~{Song} and J.~{Cheng}, ``{Optical Communication Using Subcarrier Intensity
  Modulation in Strong Atmospheric Turbulence},'' \emph{Journal of Lightwave
  Technology}, vol.~30, no.~22, pp. 3484--3493, Nov 2012.

\bibitem{sim2}
H.~{Samimi} and M.~{Uysal}, ``{Performance of coherent differential phase-shift
  keying free-space optical communication systems in M-distributed
  turbulence},'' \emph{IEEE/OSA Journal of Optical Communications and
  Networking}, vol.~5, no.~7, pp. 704--710, July 2013.

\end{thebibliography}
\bibliographystyle{IEEEtran}
\end{document}